# Precision studies of neutron decay, left-right asymmetry model of weak interaction, CP-violation and baryon asymmetry of the Universe

A.P. Serebrov, O.L. Fedin, O.M. Zherebtsov, A.K. Fomin, R.M. Samoilov, N.S. Budanov, V.A. Kayaikin

*National Research Center "Kurchatov Institute" - Petersburg Institute of Nuclear Physics, 188300, Gatchina, Russia*

*e-mail: serebrov_ap@pnpi.nrcki.ru*

**Abstract**

An analysis of the latest, most accurate experimental data on neutron decay points to the need to extend the Standard Model by introducing a right vector boson admixture $W_R$. A left-right asymmetry model of the weak interaction is presented, in which the sign of the mixing angle $W_R$ and $W_L$ is reversed upon transition from $W^-$ to $W^+$. In this model, CP violation occurs due to the presence of a right vector boson admixture. The model parameter, the mixing angle, $\zeta = -(2.7 \pm 0.9) \cdot 10^{-3}$ is derived from neutron decay — lifetime and decay asymmetries — as well as from studies of proton decay in nuclei, the so-called superallowed $0^+ - 0^+$ transitions. An estimate for the mass $W_R$ $M_{W_R} = (340 \pm 60)$ GeV is given by the relation: from the parameter $\delta = (5.5 \pm 1.8) \times 10^{-2}$, which is the ratio of the squares of the mass states of vector bosons.

The possibility of describing CP-violation effects in neutral meson oscillations within the left-right asymmetry weak interaction model with parameters $\delta$ and $\zeta$ is investigated. It is shown that within this model, CP violation effects in the decays of $K^0$ -mesons, $D^0$-mesons, $B^0$-mesons, and $B_s^0$-mesons can be successfully described. Moreover, the sign of CP violation is opposite for particles and antiparticles, i.e., the CP violation phase is equal to $\pm \pi / 2$, which corresponds to 100% CP violation. The results of calculations within the left-right asymmetry model with parameters $\delta$ and $\zeta$ are confirmed by experimental results with neutral mesons.

The formation of baryon asymmetry in the Universe is considered in a left-right asymmetry model of weak interaction with an admixture of a right vector boson with different signs of the mixing angle with $W^-$ and $W^+$. This leads to a difference in the lifetimes of neutrons and antineutrons, which decay via $W^-$ and $W^+$. This difference gives rise to baryon asymmetry during the hadronization of quark-gluon plasma at temperatures below 150 MeV. During the phase transition from quark-gluon plasma to hadronic liquid, all three of A.D. Sakharov's conditions for the emergence of baryon asymmetry in the Universe are met: CP violation and non-stationarity of the process. As a result, baryon number violation occurs due to the difference in the decay probabilities of neutrons and antineutrons.

The emergence of lepton asymmetry in the Universe in the left-right model is associated with the presence of sterile (right) neutrinos, which do not thermalize and leave the cosmic plasma, carrying with them a lepton asymmetry with a sign opposite to the baryon asymmetry. Overall, a baryon-lepton asymmetry arises with the conservation of the difference between the baryon and lepton numbers, B - L . A mechanism for the formation of dark matter by sterile neutrinos and a mechanism for the generation of lepton asymmetry are presented.

The possibility of increasing the experimental accuracy in measuring neutron decay asymmetries is noted in order to increase the level of confidence in substantiating the left-right asymmetry model of weak interaction.

## 1. Introduction

Precision studies of neutron decay allow us to search for deviations from the Standard Model. Neutron decay has been studied for over half a century. The measurement accuracy has steadily increased and currently stands at $4 \cdot 10^{-4}$ for the neutron lifetime and $10^{-3}$ for the decay asymmetry. This research process involves numerous research groups at renowned research centers around the world. Research in Russia has made a significant contribution since the middle of the last century, particularly through ultracold neutron research and precise measurements of the neutron lifetime, as well as neutrino asymmetry measurements.

A brief overview of the current state of neutron decay research is presented in [1], where it is shown that there is a deviation from the Standard Model at the level of 3.7 sigma. This same paper discusses studies of proton decay in nuclei, the so-called superallowed $0^+ - 0^+$transitions, which have been thoroughly investigated for more than 30 years [2, 3]. These studies make it possible to independently determine $V_{ud}$ the element of the CKM matrix. Moreover, it was shown that the value $V_{ud}^{00}$ obtained from these studies differs from the value $V_{ud}^{unit}$ obtained from the unitarity of the first row of the CKM matrix by 2.4 sigma [4], where $V_{ud}^{unit}=\sqrt{1-V_{us}^2-V_{ub}^2}$ . Although this difference does not exceed 3 sigma, this is alarming, since the question of the unitarity of the CKM matrix is fundamentally important, and especially in conjunction with the deviation from the SM in neutron decay. An attempt to reconcile the value $V_{ud}^{n}$ obtained from neutron decay and $V_{ud}^{00}$ from proton decay in nuclei within the framework of the simplest left-right manifest model [5-9] failed, however, it turned out that better agreement can be obtained if the sign of the mixing angle is changed to the opposite sign when changing the sign of the charge of the vector boson [1]. Thus, the so-called asymmetry version of the left-right model arose, where the mixing scheme of left and right bosons has the following form:

In general:

$$\begin{pmatrix} W_L^{\pm} \\ W_R^{\pm} \end{pmatrix} = \begin{pmatrix} \cos\zeta & e^{i(-\pi/2+\vartheta^{\pm})}\sin\zeta \\ e^{i(-\pi/2+\vartheta^{\mp})}\sin\zeta & \cos\zeta \end{pmatrix} \begin{pmatrix} W_1^{\pm} \\ W_2^{\pm} \end{pmatrix} \ (1.1), \quad \begin{pmatrix} W_L^{\pm} \\ W_R^{\pm} \end{pmatrix} = \begin{pmatrix} \cos\zeta & \mp\sin\zeta \\ \pm\sin\zeta & \cos\zeta \end{pmatrix} \begin{pmatrix} W_1^{\pm} \\ W_2^{\pm} \end{pmatrix}. \ (1.2)$$

where $\zeta$ is the mixing angle of the flavor states $W_L$ and $W_R$, and $\delta$ is the ratio of the squares of the masses of the states $W_1$ and $W_2$, $\vartheta$ is the phase of CP violation.

Based on experience fitting experimental data, we have obtained an indication that the best agreement can be obtained for $\vartheta^{+}=-\pi/2$ and $\vartheta^{-}=\pi/2$. In this case, matrix (1.1) transforms into matrix (1.2). Both matrices for $W^-$and for $W^+$are unitary. However, the matrix for $W^+$does not transform into the matrix for $W^-$ under complex conjugation due to CP violation, which corresponds precisely to the fact that we are introducing CP violation, and 100%. This introduces an asymmetry between particles and antiparticles, which allows us to explain the asymmetry of the Universe.

The matrix for $W^+$and for $W^-$are related by $\delta$ and $\zeta$ . Each matrix is applied to the Hamiltonian corresponding to its problem (neutron decay via $W^-$, antineutron decay via $W^+$, proton decay in a nucleus via $W^+$, meson decay via $W^-$, antimeson decay via $W^+$). In principle, we have a system of independent equations from which we can independently determine the model parameters $\delta$ and $\zeta$ . However, given that the process of neutron and proton decay in a nucleus ( $0^+ - 0^+$ transitions) is based on a similar quark object, it should be expected that fitting the entire set of experimental data can be performed with a single set of $\delta$ and $\zeta$ . Further data analysis will show how fruitful this approach will be.

Thus, the left-right asymmetry model does not require a new theoretical justification, but is merely a special case of applying the traditional weak interaction Hamiltonian to the corresponding process, where a special approach is applied only to the selection of the corresponding CP-violation scheme. A distinctive feature of this CP-violation scheme in the left-right asymmetry model is the choice of the CP-violation phase for $W^-$ phase $\vartheta^{-}=\pi/2$, and for $W^+$ phase $\vartheta^{+}=-\pi/2$. This choice of phases is intuitive based on the set of experimental data, taking into account the main task – explaining the asymmetry of the Universe.

The actual parameters themselves $\delta$ and $\zeta$ may depend on the structure of the quark object and should be used to adjust the values of the CKM matrix elements to ensure CKM unitarity. In particular, after the definition of $V_{ud}$ the matrix element must be adjusted $V_{us}$ , etc. Thus, the main changes will relate to the nature of CP violation in connection with the possibility of the existence of a right vector boson.

Let us compare the symmetrical and asymmetrical left-right models.

**LRSM is a matrix with the same mixing angle signs for W + and W- ,** but the matrices are complex conjugate, as expected for particles and antiparticles. This mixing matrix was proposed in [7].

$$\begin{pmatrix} W_L^{-} \\ W_R^{-} \end{pmatrix} = \begin{pmatrix} \cos\zeta & +\sin\zeta \\ -e^{-i\omega}\sin\zeta & e^{-i\omega}\cos\zeta \end{pmatrix} \begin{pmatrix} W_1^{-} \\ W_2^{-} \end{pmatrix}, \qquad \begin{pmatrix} W_L^{+} \\ W_R^{+} \end{pmatrix} = \begin{pmatrix} \cos\zeta & +\sin\zeta \\ -e^{i\omega}\sin\zeta & e^{i\omega}\cos\zeta \end{pmatrix} \begin{pmatrix} W_1^{+} \\ W_2^{+} \end{pmatrix} \quad (1.3)$$

$$\begin{pmatrix} W_L^{-} \\ W_R^{-} \end{pmatrix} = (S^-) \begin{pmatrix} W_1^{-} \\ W_2^{-} \end{pmatrix}, \qquad \begin{pmatrix} W_L^{+} \\ W_R^{+} \end{pmatrix} = (S^+) \begin{pmatrix} W_1^{+} \\ W_2^{+} \end{pmatrix}$$

The above matrices $(S^-)$ and $(S^+)$ are unitary. The transition from a particle to an antiparticle is accomplished by complex conjugation of the matrix. Verifying the transition from a particle to an antiparticle via complex conjugation:

$$S_-^* = \begin{pmatrix} \cos\zeta & \sin\zeta \\ -e^{i\omega}\sin\zeta & e^{i\omega}\cos\zeta \end{pmatrix} = S_+,\ S_+^* = \begin{pmatrix} \cos\zeta & \sin\zeta \\ -e^{-i\omega}\sin\zeta & e^{-i\omega}\cos\zeta \end{pmatrix} = S_- \qquad (1.4)$$

shows that the matrix for particles is transformed into the matrix for antiparticles under complex conjugation.

Let's check the Hermitian property:

$S_-^\dagger = \begin{pmatrix} \cos\zeta & -e^{i\omega}\sin\zeta \\ \sin\zeta & e^{i\omega}\cos\zeta \end{pmatrix} \neq S_-$ , $S_-$ non-Hermitian, $S_+^\dagger = \begin{pmatrix} \cos\zeta & -e^{-i\omega}\sin\zeta \\ \sin\zeta & e^{-i\omega}\cos\zeta \end{pmatrix} \neq S_+$, $S_+$ non-Hermitian

Therefore, CP violation is introduced through the complex phase in the CKM matrix. However, CP violation (T violation) is asymmetric for particles and antiparticles, so it does not allow for the formation of an asymmetry in the Universe. After completing the comparative analysis, we will provide more detailed explanations.

**LRAM is a matrix with different mixing angle signs for W + and W-**

$$\begin{pmatrix} W_L^- \\ W_R^- \end{pmatrix} = \begin{pmatrix} \cos\zeta & +\sin\zeta \\ -\sin\zeta & \cos\zeta \end{pmatrix} \begin{pmatrix} W_1^- \\ W_2^- \end{pmatrix} \qquad \begin{pmatrix} W_L^+ \\ W_R^+ \end{pmatrix} = \begin{pmatrix} \cos\zeta & -\sin\zeta \\ +\sin\zeta & \cos\zeta \end{pmatrix} \begin{pmatrix} W_1^+ \\ W_2^+ \end{pmatrix} \qquad (1.5)$$

$$A_- = \begin{pmatrix} \cos\zeta & \sin\zeta \\ -\sin\zeta & \cos\zeta \end{pmatrix} \qquad A_+ = \begin{pmatrix} \cos\zeta & -\sin\zeta \\ \sin\zeta & \cos\zeta \end{pmatrix}$$

The above matrices A are unitary. The transition from particle to antiparticle is accomplished by complex conjugation of the matrix. Verifying the transition from particle to antiparticle via complex conjugation:

$$A_-^* = \begin{pmatrix} \cos\zeta & \sin\zeta \\ -\sin\zeta & \cos\zeta \end{pmatrix} \neq A_+,\ A_+^* = \begin{pmatrix} \cos\zeta & -\sin\zeta \\ \sin\zeta & \cos\zeta \end{pmatrix} \neq A_- \qquad (1.6)$$

shows that the matrix for particles does not transform into the matrix for antiparticles under complex conjugation

Let's check the Hermitian property:

$A_-^\dagger = \begin{pmatrix} \cos\zeta & -\sin\zeta \\ \sin\zeta & \cos\zeta \end{pmatrix} \neq A_-$, $A_-$ non- Hermitian, $A_+^\dagger = \begin{pmatrix} \cos\zeta & \sin\zeta \\ -\sin\zeta & \cos\zeta \end{pmatrix} \neq A_+$, $A_+$ non-Hermitian

In this case, T-invariance is not violated, but CP-invariance is, meaning different lifetimes become possible for neutral particles and antiparticles, such as neutron-antiteutron and neutral meson-antimeson. This explains the baryon and lepton asymmetry of the Universe. After completing the comparative analysis, we will provide more detailed explanations.

Finally, this analysis can be supplemented by the fact that unitarity is not satisfied for matrices containing an imaginary unit in off-diagonal elements (whether symmetrically or asymmetrically). Furthermore, the presence of an imaginary unit indicates non-stationarity or decay, which is not the case here.

In general, the Lagrangian of the charged current can be written in the form under the assumption of symmetry of the left and right currents $g_L = g_R \equiv g$

$$L_{CC} = \left(g/\sqrt{2}\right)\begin{pmatrix} W_{L\mu}^+ & W_{R\mu}^+ \end{pmatrix}\begin{pmatrix} J_L^{\mu+} \\ J_R^{\mu+} \end{pmatrix} + h.c. \qquad (1.7)$$

where h.c. is the Hermitian conjugate term in the Lagrangian. The complete Lagrangian looks like this

$$\begin{aligned} L_{CC} &= \frac{g}{\sqrt{2}}\begin{pmatrix} W_{L\mu}^+ & W_{R\mu}^+ \end{pmatrix}\begin{pmatrix} J_L^{\mu+} \\ J_R^{\mu+} \end{pmatrix} + \frac{g}{\sqrt{2}}\begin{pmatrix} W_{L\mu}^- & W_{R\mu}^- \end{pmatrix}\begin{pmatrix} J_L^{\mu-} \\ J_R^{\mu-} \end{pmatrix} = \\ &= \frac{g}{\sqrt{2}}\left[W_{L\mu}^+ J_L^{\mu+} + W_{R\mu}^+ J_R^{\mu+}\right] + \frac{g}{\sqrt{2}}\left[W_{L\mu}^- J_L^{\mu-} + W_{R\mu}^- J_R^{\mu-}\right] \end{aligned} \qquad (1.8)$$

The left and right currents are determined by the sum of the currents for quarks $\left(J_q\right)$ and leptons $\left(J_\ell\right)$ as follows:

$$J_L^{\mu+} = \bar{u}_i \gamma_\mu P_L \left(V_L\right)_{ij} d_j + \bar{\nu}_i \gamma_\mu P_L \left(U_L\right)_{ij} \ell_j \qquad (1.9)$$

$$J_L^{\mu-} = \bar{d}_i \gamma_\mu P_L \left(V_L\right)_{ij} u_j + \bar{\ell}_i \gamma_\mu P_L \left(U_L\right)_{ij} \nu_j \qquad (1.10)$$

$$J_R^{\mu+} = \bar{u}_i \gamma_\mu P_R \left(V_R\right)_{ij} d_j + \bar{\nu}_i \gamma_\mu P_R \left(U_R\right)_{ij} \ell_j \quad (1.11)$$

$$J_R^{\mu-} = \bar{d}_i \gamma_\mu P_R \left(V_R\right)_{ij} u_j + \bar{\ell}_i \gamma_\mu P_R \left(U_R\right)_{ij} \nu_j \quad (1.12)$$

where $P_{L,R} = \frac{1}{2}\left(1 \mp \gamma_5\right)$ is the chirality projection operator, $V_{L,R}$ are the right and left CKM matrices, $U_{L,R}$ and is the mixing matrix of light neutrinos $\left(\nu\right)$ and heavy right neutrinos $\left(N\right)$. Considering that:

$$W_L^{\pm} = W_1^{\pm} \cos\zeta + W_2^{\pm} \sin\zeta \quad (1.13)$$

$$W_R^{\pm} = e^{i\omega}\left(-W_1^{\pm} \sin\zeta + W_2^{\pm} \cos\zeta\right)$$

the Lagrangian can be written in the following form:

$$L_{CC} = \frac{g}{\sqrt{2}} \begin{pmatrix} W_{L\mu}^{+} & W_{R\mu}^{+} \end{pmatrix} \begin{pmatrix} \cos\zeta & -e^{i\omega}\sin\zeta \\ \sin\zeta & e^{i\omega}\cos\zeta \end{pmatrix} \begin{pmatrix} J_L^{\mu+} \\ J_R^{\mu+} \end{pmatrix} + \quad (1.14)$$

$$+ \frac{g}{\sqrt{2}} \begin{pmatrix} W_{L\mu}^{-} & W_{R\mu}^{-} \end{pmatrix} \begin{pmatrix} \cos\zeta & -e^{i\omega}\sin\zeta \\ \sin\zeta & e^{i\omega}\cos\zeta \end{pmatrix} \begin{pmatrix} J_L^{\mu-} \\ J_R^{\mu-} \end{pmatrix} =$$

$$= \frac{g}{\sqrt{2}} \left[ W_{1\mu}^{+}\left(\cos\zeta J_L^{\mu+} - e^{i\omega}\sin\zeta J_R^{\mu+}\right) + W_{2\mu}^{+}\left(\sin\zeta J_L^{\mu+} + e^{i\omega}\cos\zeta J_R^{\mu+}\right)\right] +$$

$$+ \frac{g}{\sqrt{2}} \left[ W_{1\mu}^{-}\left(\cos\zeta J_L^{\mu-} - e^{i\omega}\sin\zeta J_R^{\mu-}\right) + W_{2\mu}^{-}\left(\sin\zeta J_L^{\mu-} + e^{i\omega}\cos\zeta J_R^{\mu-}\right)\right]$$

Further discussion should begin with a consideration of the experimental data.

## 2. Precision studies of neutron decay and the V - A variant of the theory

To proceed to the LRSM analysis, it is necessary to first present the experimental data and conduct the analysis first within the framework of the V - A version of the theory.

The kinematics of neutron decay can be described by expression (2.1).

$$\frac{d^3\Gamma}{dE_e d\Omega_e d\Omega_\nu} = \frac{1}{\left(2\pi\right)^5} p_e E_e \left(E_0 - E_e\right)^2 \times \xi \left[ 1 + a\frac{\vec{p}_e \cdot \vec{p}_\nu}{E_e E_\nu} + b\frac{m_e}{E_e} + \frac{\langle\vec{\sigma}_n\rangle}{\left|\sigma_n\right|} \cdot \left( A\frac{\vec{p}_e}{E_e} + B\frac{\vec{p}_\nu}{E_\nu} + D\frac{\vec{p}_e \times \vec{p}_\nu}{E_e E_\nu} \right) \right] \quad (2.1)$$

Within the framework of the V - A variant of the theory of weak interaction, all the neutron decay parameters presented in Table 2.2 can be expressed through a single parameter of the theory $\lambda_n = g_A / g_V$.

$$\tau_n^{V-A} = \frac{K}{G_F^2 \left|V_{ud}\right|^2 g_V^2 \left(1 + 3\lambda_n^2\right)}, \quad a^{V-A} = \frac{1-\lambda_n^2}{1+3\lambda_n^2}, \quad A^{V-A} = -\frac{2\lambda_n\left(\lambda_n + 1\right)}{1+3\lambda_n^2}, \quad B^{V-A} = \frac{2\lambda_n\left(\lambda_n - 1\right)}{1+3\lambda_n^2}, \quad \left(\frac{A}{B}\right)^{V-A} = -\frac{\lambda_n + 1}{\lambda_n - 1}$$

(2.2)

Table 2.1 . Measured values of neutron decay parameters.

| Parameter | Value | Rel. acc. | PDG | Ref. |
|---|---|---|---|---|
| $\tau_n$ | $877.75^{+36}_{-32}$ | $4 * 10^{-4}$ | 878.4(5) | [ 10 ] |
| A | -0.11985(21) | $1.7 * 10^{-3}$ | -0.119 58 (21) | [ 11 ] |
| B | 0.9802(49) | $5 * 10^{-3}$ | 0.9807(30) | [ 12 ] |
| $a_{ev}^{aCORN}$ | -0.1053(18) | $1.7 * 10^{-2}$ | -0.1044(7) | [ 13 ] |
| $a_{ev}^{aSPECT}$ | -0.10402(82) | $8 * 10^{-3}$ | | [ 14 ] |
| A/B | -0.1223(52) | $5.2 * 10^{-3}$ | -0.1219(43) | |
| D | D = ( − 1.2 ± 2.0 ) × 10 $^{-4}$ | | D = ( − 1.2 ± 2.0 ) × 10 $^{-4}$ | [4] |

Table 2.2. Parameter values $\lambda$ extracted from neutron decay parameters within the V - A version of the theory. When calculating the parameter $\lambda$ from the neutron lifetime, the CKM matrix elements obtained from the unitarity condition were used $V_{ud}^{unit}$.

| Parameter | $\lambda_{best}$ | Rel. acc. | $\lambda_{PDG}$ | Rel. acc. |
|---|---|---|---|---|
| $\tau_n \& V_{ud}^{unit}$ | -1.2761(4) | $3 \cdot 10^{-4}$ | -1.2755(5) | $4 \cdot 10^{-4}$ |
| $\tau_n \& V_{ud}^{00}$ | -1.2773(6) | $5 \cdot 10^{-4}$ | -1.2767(7) | $5 \cdot 10^{-4}$ |
| A | -1.2764(6) | $5 \cdot 10^{-4}$ | -1.2757(6) | $5 \cdot 10^{-4}$ |
| B | -1.360(59) | $4 \cdot 10^{-2}$ | -1.354(36) | $3 \cdot 10^{-2}$ |
| $a_{ev}^{aCORN}$ | -1.271(6) | $5 \cdot 10^{-3}$ | -1.2681(23) | $2 \cdot 10^{-3}$ |
| $a_{ev}^{aSPECT}$ | –1.2668(27) | $2 \cdot 10^{-3}$ | | |
| A/B | -1.2786(17) | $1 \cdot 10^{-3}$ | -1.2777(11) | $9 \cdot 10^{-4}$ |

Figure 2.1 shows the distribution of values $\lambda_n$ extracted from the results of the corresponding measurements: $\lambda_{A/B}$, $\lambda_\tau$, $\lambda_A$ and $\lambda_a$.

As can be seen, the description of the experimental results within the SM framework turned out to be unsatisfactory. The largest discrepancy is observed between $\lambda_{A/B}$ and $\lambda_a$, amounting to 3.7 sigma for the experiment $a_{ev}^{aSPECT2023}$.

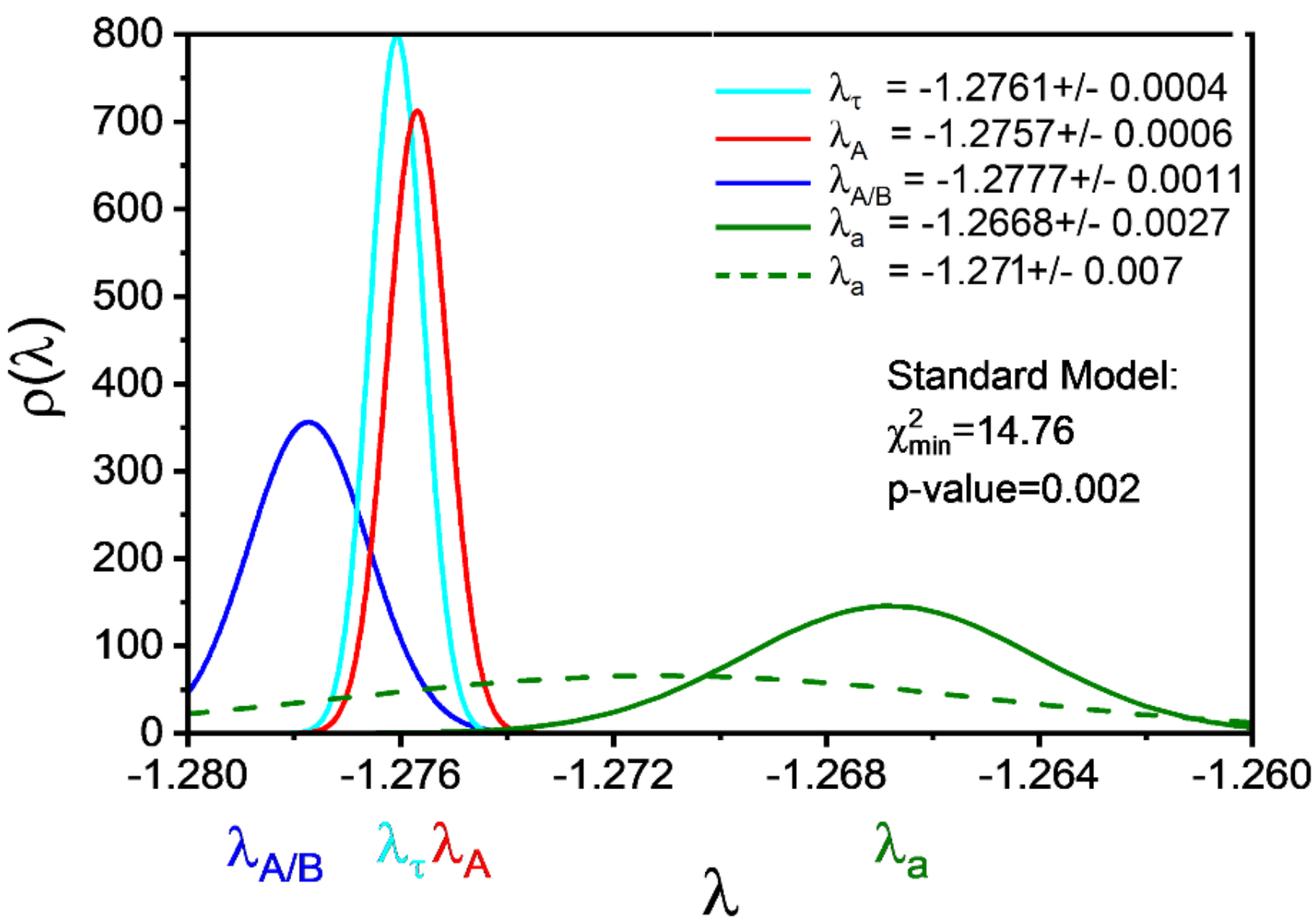

Fig. 2.1. Probability distribution functions for parameters $\lambda$ within the SM framework, extracted from neutron decay parameters and the result of scanning the function profile $\chi^2$

### 3. Left-right manifest (symmetric) weak interaction model (LRSM)

As noted in the introduction, an attempt to reconcile the value $V_{ud}^{n}$ obtained from neutron decay and $V_{ud}^{00}$ from proton decay in nuclei within the simplest left-right manifest model [5-9] failed, so it was necessary to switch to the left-right asymmetry model of weak interaction (LRAM). But before moving on to the analysis of experimental data within the LRAM framework, we will demonstrate what the above-mentioned discrepancy looks like in the definition $V_{ud}^{n}$ from neutron decay and $V_{ud}^{00}$ from proton decay in nuclei within the simplest left-right manifest model. To do this, we will obtain the necessary relations for the neutron lifetime, decay asymmetries, and also for $0^+ - 0^+$ transitions.

The Hamiltonian describing the decay of a nucleon through vector and axial currents has the form [5-9]:

$$H_{V,A}^{(N)} = C_V \langle \bar{e}|\gamma_\mu|\nu\rangle\langle \bar{p}|\gamma_\mu|n\rangle + C_V' \langle \bar{e}|\gamma_\mu\gamma_5|\nu\rangle\langle \bar{p}|\gamma_\mu|n\rangle - C_A \langle \bar{e}|\gamma_\mu\gamma_5|\nu\rangle\langle \bar{p}|\gamma_\mu\gamma_5|n\rangle - C_A' \langle \bar{e}|\gamma_\mu|\nu\rangle\langle \bar{p}|\gamma_\mu\gamma_5|n\rangle + \text{h.c.} \quad (3.1)$$

LRSM is a matrix with the same mixing angle signs for W + and W-, proposed in [7-8].

$$\begin{pmatrix} W_L^- \\ W_R^- \end{pmatrix} = \begin{pmatrix} \cos\zeta & +\sin\zeta \\ -e^{-i\omega}\sin\zeta & e^{-i\omega}\cos\zeta \end{pmatrix} \begin{pmatrix} W_1^- \\ W_2^- \end{pmatrix} \qquad \begin{pmatrix} W_L^+ \\ W_R^+ \end{pmatrix} = \begin{pmatrix} \cos\zeta & +\sin\zeta \\ -e^{i\omega}\sin\zeta & e^{i\omega}\cos\zeta \end{pmatrix} \begin{pmatrix} W_1^+ \\ W_2^+ \end{pmatrix} \tag{3.2}$$

For this vector boson mixing matrix we can obtain:

$$C_V = g_V \frac{G_F V_{ud}}{\sqrt{2}} \left( \cos^2\zeta + e^{2i\omega}\sin^2\zeta + \delta\left(\sin^2\zeta + e^{2i\omega}\cos^2\zeta\right) - \left(1-\delta\right)e^{i\omega}\sin 2\zeta \right) \tag{3.3}$$

$$C_V' = g_V \frac{G_F V_{ud}}{\sqrt{2}} \left( \cos^2\zeta - e^{2i\omega}\sin^2\zeta + \delta\left(\sin^2\zeta - e^{2i\omega}\cos^2\zeta\right) \right) \tag{3.4}$$

$$C_A = g_A \frac{G_F V_{ud}}{\sqrt{2}} \left( \cos^2\zeta + e^{2i\omega}\sin^2\zeta + \delta\left(\sin^2\zeta + e^{2i\omega}\cos^2\zeta\right) + \left(1-\delta\right)e^{i\omega}\sin 2\zeta \right) \tag{3.5}$$

$$C_A' = g_A \frac{G_F V_{ud}}{\sqrt{2}} \left( \cos^2\zeta - e^{2i\omega}\sin^2\zeta + \delta\left(\sin^2\zeta - e^{2i\omega}\cos^2\zeta\right) \right) \tag{3.6}$$

$$|C_V|^2 + |C_V'|^2 = g_V^2 G_F^2 |V_{ud}|^2 \left(1+\delta^2 - \left(1-\delta^2\right)\sin 2\zeta\cos\omega\right),\ |C_A|^2 + |C_A'|^2 = g_A^2 G_F^2 |V_{ud}|^2 \left(1+\delta^2 + \left(1-\delta^2\right)\sin 2\zeta\cos\omega\right)$$

(3.7) For a neutron, the total decay probability is proportional to $\xi$

$$\xi = |C_V|^2 + |C_V'|^2 + 3\left(|C_A|^2 + |C_A'|^2\right) = g_V^2 G_F^2 |V_{ud}|^2 \left(1+3\lambda^2\right)\left[1+\delta^2 - \frac{1-3\lambda^2}{1+3\lambda^2}\left(1-\delta^2\right)\sin 2\zeta\cos\omega\right] \tag{3.8}$$

Calculating the probability of nucleon decay with Hamiltonian (3.1) gives the following result

$$\left(f\tau_n\right)^{-1} \approx \left( |M_F|^2\left(|C_V|^2 + |C_V'|^2\right) + |M_{GT}|^2\left(|C_A|^2 + |C_A'|^2\right) \right) \tag{3.9}$$

where $f$ is the phase volume for a given decay, $\tau_n$ is the lifetime of the particle. In the case of a neutron $M_F = 1$, $M_{GT} = \sqrt{3}$, and the decay occurs through $W^-$

$$\left(f\tau_n\right)^{-1} \approx G_F^2 |V_{ud}|^2 \left(1+3\lambda_n^2\right)\left[1+\delta^2 - \left(1-\delta^2\right)\frac{\left(1-3\lambda_n^2\right)}{\left(1+3\lambda_n^2\right)}\sin 2\zeta\cos\omega\right] \tag{3.10}$$

In the case of $0^+ \to 0^+$ transitions, only the vector current contributes to the decay probability, and decay occurs through $W^+$. The transition of $0^+ \to 0^+$ probability is:

$$\left(f\tau_{00}\right)^{-1} \approx G_F^2 |V_{ud}|^2 \left[1+\delta^2 - \left(1-\delta^2\right)\sin 2\zeta\cos\omega\right] \tag{3.11}$$

The differential distribution of the direction of momentum of electrons and antineutrinos during the decay of a polarized nucleon depending on the electron energy is given in the works [15,16]:

$$\frac{d^3\Gamma}{dE_e d\Omega_e d\Omega_\nu} = \frac{1}{(2\pi)^5} p_e E_e \left(E_0 - E_e\right)^2 \times \xi \left[1 + a\frac{\vec{p}_e\cdot\vec{p}_\nu}{E_e E_\nu} + b\frac{m_e}{E_e} + \frac{\langle\vec{\sigma}_n\rangle}{|\sigma_n|}\cdot\left(A\frac{\vec{p}_e}{E_e} + B\frac{\vec{p}_\nu}{E_\nu} + D\frac{\vec{p}_e\times\vec{p}_\nu}{E_e E_\nu}\right)\right] \tag{3.12}$$

Where $\xi = |M_F|^2\left(|C_V|^2 + |C_V'|^2\right) + |M_{GT}|^2\left(|C_A|^2 + |C_A'|^2\right)$

Formulas for the neutron lifetime $\tau_n$ and asymmetry coefficients $a$, $A$ and $B$ in neutron decay in this work are constructed on the basis of the relationships presented above, without using expansion in small model parameters $\delta$ and $\zeta$ in contrast to those obtained in work [9] and used in work [1]:

$$\tau_n = \frac{\mathrm{K}}{G_F^2 |V_{ud}|^2 g_V^2 \left(1+3\lambda^2\right)\left[1+\delta^2 - \frac{1-3\lambda^2}{1+3\lambda^2}\left(1-\delta^2\right)\sin 2\zeta\cos\omega\right]} \qquad \tau_n(\omega = \pm\pi/2) = \frac{\mathrm{K}}{G_F^2 |V_{ud}|^2 g_V^2 \left(1+3\lambda^2\right)\left[1+\delta^2\right]} \tag{3.13}$$

$$A = \frac{-2\lambda\left(\lambda+1\right)}{1+3\lambda^2} \times \left[\frac{1-\delta^2 + \frac{\lambda}{\lambda+1}\left(1-\delta\right)^2\sin 2\zeta\cos\omega}{1+\delta^2 - \frac{1-3\lambda^2}{1+3\lambda^2}\left(1-\delta^2\right)\sin 2\zeta\cos\omega}\right]\cos 2\zeta \qquad A(\omega = \pm\pi/2) = \frac{-2\lambda\left(\lambda+1\right)}{1+3\lambda^2} \times \left[\frac{1-\delta^2}{1+\delta^2}\right]\cos 2\zeta \tag{3.14}$$

$$a = \frac{1-\lambda^2}{1+3\lambda^2} \times \left[\frac{1+\delta^2 - \frac{1+\lambda^2}{1-\lambda^2}\left(1-\delta^2\right)\sin 2\zeta\cos\omega}{1+\delta^2 - \frac{1-3\lambda^2}{1+3\lambda^2}\left(1-\delta^2\right)\sin 2\zeta\cos\omega}\right] \qquad a(\omega = \pm\pi/2) = \frac{1-\lambda^2}{1+3\lambda^2} \tag{3.15}$$

$$B=\frac{2\lambda(\lambda-1)}{1+3\lambda^2}\times\left[\frac{1-\delta^2+\frac{\lambda}{\lambda-1}(1-\delta)^2\sin 2\zeta\cos\omega}{1+\delta^2-\frac{1-3\lambda^2}{1+3\lambda^2}\left(1-\delta^2\right)\sin 2\zeta\cos\omega}\right]\cos 2\zeta \qquad B(\omega=\pm\pi/2)=\frac{2\lambda(\lambda-1)}{1+3\lambda^2}\times\left[\frac{1-\delta^2}{1+\delta^2}\right]\cos 2\zeta \quad (3.16)$$

$$\frac{A}{B}=-\frac{\lambda+1}{\lambda-1}\times\left[\frac{1-\delta^2+\frac{\lambda}{\lambda+1}(1-\delta)^2\sin 2\zeta\cos\omega}{1-\delta^2+\frac{\lambda}{\lambda-1}(1-\delta)^2\sin 2\zeta\cos\omega}\right] \qquad \frac{A}{B}(\omega=\pm\pi/2)=-\frac{\lambda+1}{\lambda-1} \quad (3.17)$$

$$D=\pm\frac{2\lambda}{1+3\lambda^2}\times\frac{(1-\delta)^2\sin 2\zeta\cos 2\zeta\sin\omega}{1+\delta^2-\frac{1-3\lambda^2}{1+3\lambda^2}\left(1-\delta^2\right)\sin 2\zeta\cos\omega} \qquad D(\omega=\pm\pi/2)=\pm\frac{2\lambda}{1+3\lambda^2}\times\frac{(1-\delta)^2\sin 2\zeta\cos 2\zeta}{1+\delta^2} \quad (3.18)$$

$$D(\omega=0)=0$$

The factor K is given by an expression consisting of fundamental constants: $\mathcal{K}=\frac{2\pi^3\hbar^7c^6}{G_F^2m_e^5c^4}$.

Now we can return to the LRSM analysis using the relationships given above.

Here, a very important experimental circumstance should be immediately noted. The fact is that the coefficient D (which violates T-invariance) has an experimental limit, which imposes a constraint on $\omega$. Its non-zero value would indicate a violation of T-symmetry. The current limits are very small. The PDG reports give an estimate: D = ( −1.2 ± 2.0) × $10^{-4}$. This means that | D | < 3 * $10^{-4}$ at CL 90% or restrictions on $\omega\zeta < 3 * 10^{-4}$. Based on the estimate $\zeta = 3 * 10^{-3}$, we obtain that $\omega < 10^{-1}$. Thus, it is not possible to introduce a complete CP-violation.

On the other hand, if we introduce a complete CP violation, i.e. put $\omega=\pm\pi/2$, then $a=\frac{1-\lambda^2}{1+3\lambda^2}=a^{V-A}$, $\frac{A}{B}=-\frac{\lambda+1}{\lambda-1}=(\frac{A}{B})^{V-A}$. But then there should be no discrepancy between $\lambda_{A/B}$ and $\lambda_a$, which is observed experimentally and is 3.7 sigma. Thus, it is impossible to introduce a complete CP-violation into LRSM for two reasons: 1) due to the experimental limitation on D -asymmetry and 2) due to the experimental discrepancy between $\lambda_{A/B}$ and $\lambda_a$.

Furthermore, there is a third reason: it is impossible to resolve the contradiction between the values of $V_{ud}^{n}$ and $V_{ud}^{00}$. Let us demonstrate what the aforementioned discrepancy looks like in the definition of $V_{ud}^{n}$ from neutron decay and $V_{ud}^{00}$ from proton decay in nuclei within the LRSM (the simplest left-right manifest model) framework for $\omega = 0$.

Let's begin by analyzing the situation with $0^+ \to 0^+$ superallowed transitions. Let's compare proton decay in a nucleus and neutron decay within the left-right symmetric model (LRSM), taking into account that proton decay in a nucleus occurs via W $^+$, while neutron decay occurs via W- . However, the sign of the parameter $\zeta$ for proton decay in a nucleus in the left-right symmetric model (LRSM) does not change to the opposite.

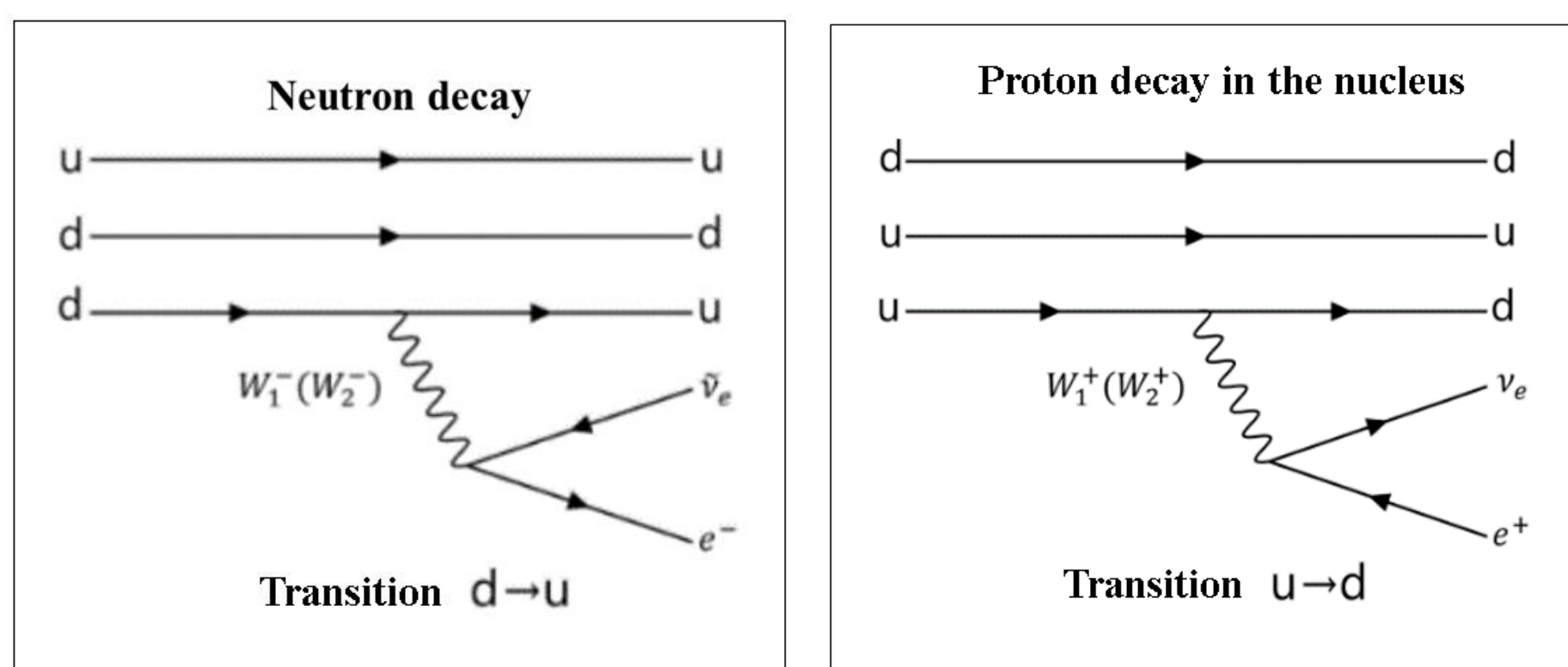

Fig. 3.1. The process of neutron decay and the process of proton decay in the nucleus. The direct transition from the d quark to the u quark and the reverse transition from the u quark to the d quark, but the direct transition is through $W_1^-(W_2^-)$, and the reverse transition is through $W_1^+(W_2^+)$.

$$\left(f\tau_{00}\right)^{-1} \approx G_F^2 \left|V_{ud}\right|^2 \left[1+\delta^2-\left(1-\delta^2\right)\sin 2\zeta \cos\omega\right], \qquad (3.19)$$

$$\left(f\tau_{n}\right)^{-1} \approx G_F^2 \left|V_{ud}\right|^2 \left[1+\delta^2-\left(1-\delta^2\right)\frac{\left(1-3\lambda_n^2\right)}{\left(1+3\lambda_n^2\right)}\sin 2\zeta \cos\omega\right] \qquad (3.20)$$

$\dfrac{\left|\tilde{V}_{ud}^{00}\right|^2-\left|\tilde{V}_{ud}^{n}\right|^2}{\left|\tilde{V}_{ud}^{00}\right|^2+\left|\tilde{V}_{ud}^{n}\right|^2} \approx \dfrac{6\zeta\lambda^2}{3\lambda_n^2+1} \approx 1.66\zeta$ (3.13), where $\tilde{V}_{ud}^{00}=0.97373(32)$ and $\tilde{V}_{ud}^{n}=0.97477(37)$ are experimental values obtained within the framework of the V - A version of the theory.

Note that in the last expression the dependence on $\left|V_{ud}\right|^2$ is reduced.

The discrepancy between the values of $V_{ud}$ for 0+-0+ transitions ( $\tilde{V}_{ud}^{00}$) and $V_{ud}$ for the neutron ( $\tilde{V}_{ud}^{n}$) is: $\left(\tilde{V}_{ud}^{00}\right)^2-\left(\tilde{V}_{ud}^{n}\right)^2 = -(0.00104 \pm 0.00052)\ (2.0\sigma)$. And the discrepancy between the values $V_{ud}$ for 0+-0+ transitions ($\tilde{V}_{ud}^{00}$) and $V_{ud}$ from unitarity ( $\tilde{V}_{ud}^{unit}$) is: $\left(\tilde{V}_{ud}^{00}\right)^2-\left(\tilde{V}_{ud}^{unit}\right)^2 = -\ (0.00079 \pm 0.00036)\ (2.2\sigma)$. The results of determining the model parameter $\zeta$ from the discrepancy between the values $V_{ud}$ for 0+-0+ transitions ($\tilde{V}_{ud}^{00}$) and $V_{ud}$ for the neutron ( $\tilde{V}_{ud}^{n}$) are presented in Fig. 3.2 as a vertical column. The results of determining $\delta$ and $\zeta$ from neutron decay data are also presented here, using formulas for the neutron lifetime $\tau_n$ and the asymmetry coefficients $a$, $A$ and $A/B$. These results are presented in Fig. 3.2 in comparison with the results for 0+-0+ transitions for two cases: $\omega$=0, $\omega=\pi$.

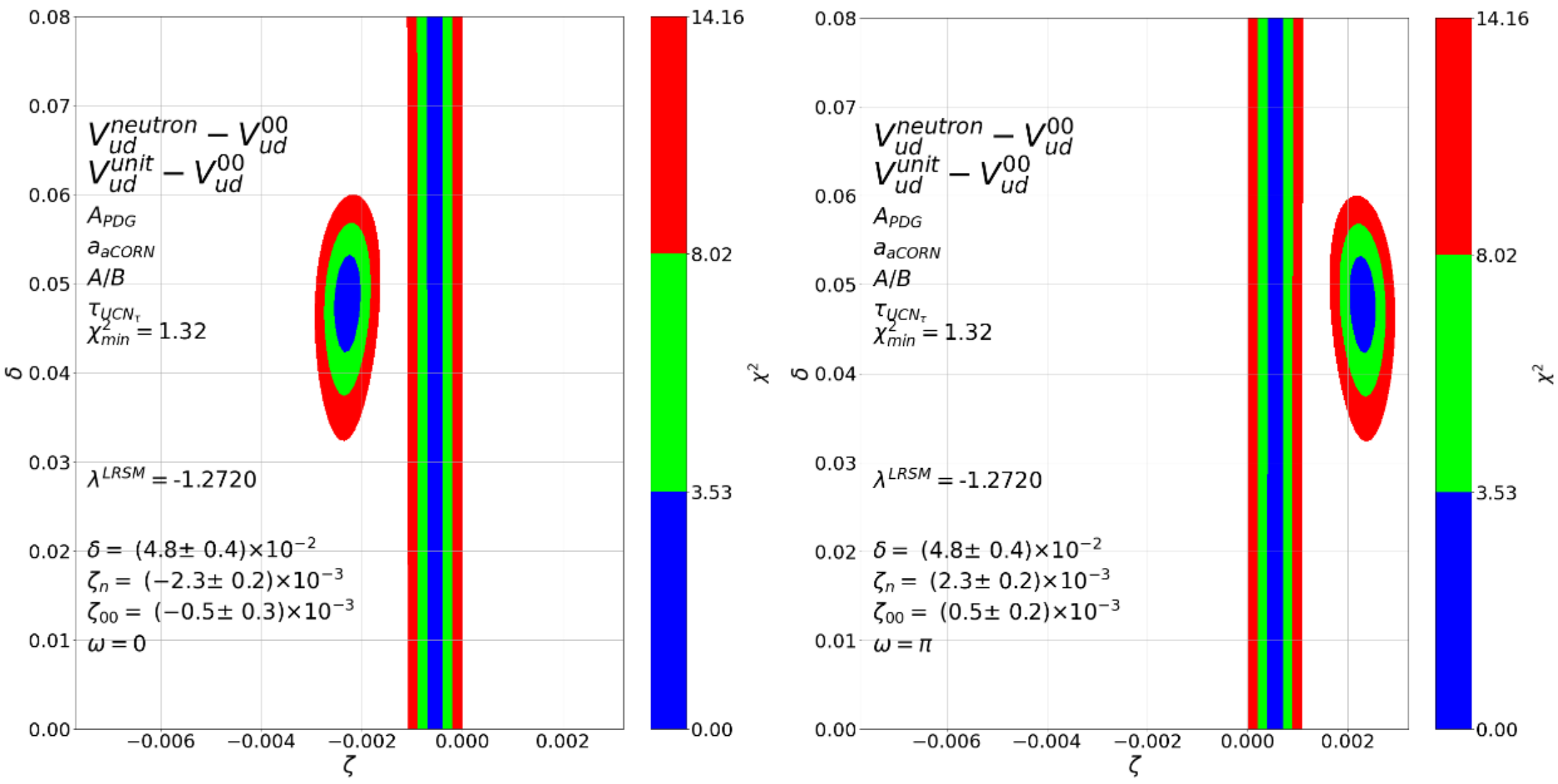

Fig. 3.2 The results of determining the model parameter $\zeta$ from the discrepancy between the values $V_{ud}$ for 0+-0+ transitions ($\tilde{V}_{ud}^{00}$) and $V_{ud}$ for unitarity ( $\tilde{V}_{ud}^{unit}$) and $V_{ud}$ for the neutron ( $\tilde{V}_{ud}^{n}$) in comparison with the results from the decay of the neutron for two cases: $\omega$=0, $\omega=\pi$.

As can be seen from Fig. 3.2, different values of the model parameter were required to describe the experimental data for neutron decay and proton decay in a nucleus $\zeta$. This means that a joint description within this model is impossible. Thus, we have a third indication that the LRSM is incapable of describing the experimental data, so we move on to consider the asymmetry model (LRAM).

## 4. Left-right asymmetry weak interaction model (LRAM)

The Hamiltonian describing the decay of a nucleon through vector and axial currents has the form [5-9]:

$$H_{V,A}^{(N)} = C_V \left\langle \bar{e}\left|\gamma_\mu\right|\nu\right\rangle\left\langle \bar{p}\left|\gamma_\mu\right|n\right\rangle + C_V' \left\langle \bar{e}\left|\gamma_\mu\gamma_5\right|\nu\right\rangle\left\langle \bar{p}\left|\gamma_\mu\right|n\right\rangle - C_A \left\langle \bar{e}\left|\gamma_\mu\gamma_5\right|\nu\right\rangle\left\langle \bar{p}\left|\gamma_\mu\gamma_5\right|n\right\rangle - C_A' \left\langle \bar{e}\left|\gamma_\mu\right|\nu\right\rangle\left\langle \bar{p}\left|\gamma_\mu\gamma_5\right|n\right\rangle + \text{h.c.} \qquad (4.1)$$

$$\begin{pmatrix} W_L^- \\ W_R^- \end{pmatrix} = \begin{pmatrix} \cos\zeta & +\sin\zeta \\ -\sin\zeta & \cos\zeta \end{pmatrix} \begin{pmatrix} W_1^- \\ W_2^- \end{pmatrix} \qquad \begin{pmatrix} W_L^+ \\ W_R^+ \end{pmatrix} = \begin{pmatrix} \cos\zeta & -\sin\zeta \\ +\sin\zeta & \cos\zeta \end{pmatrix} \begin{pmatrix} W_1^+ \\ W_2^+ \end{pmatrix}$$

For a neutron $C_V = g_V \frac{G_F V_{ud}}{\sqrt{2}}\left((1+\delta)-(1-\delta)\sin 2\zeta\right), C'_V = g_V \frac{G_F V_{ud}}{\sqrt{2}}(1-\delta)\cos 2\zeta$ (4.2)

$$C_A = g_A \frac{G_F V_{ud}}{\sqrt{2}}\left((1+\delta)+(1-\delta)\sin 2\zeta\right),\ C'_A = g_A \frac{G_F V_{ud}}{\sqrt{2}}(1-\delta)\cos 2\zeta \tag{4.3}$$

Calculating the probability of nucleon decay with Hamiltonian (4.1) gives the following result

$$\left(f\tau_n\right)^{-1} \approx \left(\left|M_F\right|^2\left(\left|C_V\right|^2+\left|C'_V\right|^2\right)+\left|M_{GT}\right|^2\left(\left|C_A\right|^2+\left|C'_A\right|^2\right)\right) \tag{4.4}$$

where $f$ is the phase volume for a given decay, $\tau_n$ is the lifetime of the particle. In the case of a neutron $M_F = 1$, $M_{GT} = \sqrt{3}$, and the decay occurs in $W^-$

$$\left(f\tau_n\right)^{-1} \approx G_F^2\left|V_{ud}\right|^2\left(1+3\lambda_n^2\right)\left[1+\delta^2-\left(1-\delta^2\right)\frac{\left(1-3\lambda_n^2\right)}{\left(1+3\lambda_n^2\right)}\sin 2\zeta\right] \tag{4.5}$$

In the case of $0^+ \to 0^+$ transitions, only the vector current contributes to the decay probability, and the decay proceeds through $W^+$, which requires a change in sign for $\zeta$ in the coefficients (4.2). The $0^+ \to 0^+$ transition probability is equal to:

$$\left(f\tau_{00}\right)^{-1} \approx G_F^2\left|V_{ud}\right|^2\left[1+\delta^2+\left(1-\delta^2\right)\sin 2\zeta\right] \tag{4.6}$$

The differential distribution of the direction of momentum of electrons and antineutrinos during the decay of a polarized nucleon depending on the electron energy is given in [15,16]:

$$\frac{d^3\Gamma}{dE_e d\Omega_e d\Omega_\nu} = \frac{1}{(2\pi)^5} p_e E_e\left(E_0-E_e\right)^2 \times \xi\left[1+a\frac{\vec{p}_e\cdot\vec{p}_\nu}{E_e E_\nu}+b\frac{m_e}{E_e}+\frac{\langle\vec{\sigma}_n\rangle}{|\sigma_n|}\cdot\left(A\frac{\vec{p}_e}{E_e}+B\frac{\vec{p}_\nu}{E_\nu}+D\frac{\vec{p}_e\times\vec{p}_\nu}{E_e E_\nu}\right)\right] \tag{4.7}$$

Where $\xi = \left|M_F\right|^2\left(\left|C_V\right|^2+\left|C'_V\right|^2\right)+\left|M_{GT}\right|^2\left(\left|C_A\right|^2+\left|C'_A\right|^2\right)$ (4.8).

Formulas for the neutron lifetime $\tau_n$ and asymmetry coefficients $a$, $A$ and $B$ in neutron decay in this work are constructed on the basis of the relationships presented above, without using expansion in small model parameters $\delta$ and $\zeta$ in contrast to those obtained in work [9] and used in work [1]:

$$\tau_n = \tau_n^* \frac{1}{\left[1+\delta^2-(1-\delta^2)\frac{1-3\lambda_n^2}{1+3\lambda_n^2}\sin 2\zeta\right]} \qquad \tau_n^* = \frac{K}{G_F^2\left|V_{ud}\right|^2 g_V^2\left(1+3\lambda_n^2\right)} \tag{4 .9}$$

$$a = a^* \cdot \frac{1+\delta^2-(1-\delta^2)\frac{1+\lambda_n^2}{1-\lambda_n^2}\sin 2\zeta}{1+\delta^2-(1-\delta^2)\frac{1-3\lambda_n^2}{1+3\lambda_n^2}\sin 2\zeta} \qquad a^* = \frac{1-\lambda_n^2}{1+3\lambda_n^2} \tag{4.10}$$

$$A = A^* \cdot \frac{1-\delta^2+\frac{\lambda_n}{\lambda_n+1}(1-\delta)^2\sin 2\zeta}{1+\delta^2-(1-\delta^2)\frac{1-3\lambda_n^2}{1+3\lambda_n^2}\sin 2\zeta}\cdot\cos 2\zeta \qquad A^* = -\frac{2\lambda_n\left(\lambda_n+1\right)}{1+3\lambda_n^2} \tag{4.11}$$

$$B = B^* \cdot \frac{1-\delta^2+\frac{\lambda_n}{\lambda_n-1}(1-\delta)^2\sin 2\zeta}{1+\delta^2-(1-\delta^2)\frac{1-3\lambda_n^2}{1+3\lambda_n^2}\sin 2\zeta}\cdot\cos 2\zeta \qquad B^* = \frac{2\lambda_n\left(\lambda_n-1\right)}{1+3\lambda_n^2} \tag{4.12}$$

$$\frac{A}{B} = \left(\frac{A}{B}\right)^* \cdot \frac{1-\delta^2+\frac{\lambda_n}{\lambda_n+1}(1-\delta)^2\sin 2\zeta}{1-\delta^2+\frac{\lambda_n}{\lambda_n-1}(1-\delta)^2\sin 2\zeta} \qquad \left(\frac{A}{B}\right)^* = -\frac{\lambda_n+1}{\lambda_n-1} \tag{4.13}$$

D = 0

where $\tau_n^*$, $a^*$, $A^*$, $B^*$, $\left(\frac{A}{B}\right)^*$ have the same form of recording as in the V - A version of the theory, but new values of $\lambda$ and $V_{ud}$ must be determined from the analysis of experimental data together with the model parameters $\delta$ and $\zeta$, the factor K is given by the expression: $\mathcal{K}=\frac{2\pi^3\hbar^7c^6}{G_F^2m_e^5c^4}$, where $G_F$ – the Fermi weak interaction coupling constant, determined from μ -decay.

Let's first analyze the situation with superallowed $0^+ \to 0^+$ transitions. Let's compare proton decay in a nucleus and neutron decay within the left-right asymmetry model, taking into account that proton decay in a nucleus occurs via W $^+$, while neutron decay occurs via W- . Therefore, the sign of the parameter $\zeta$ for proton decay in a nucleus in the left-right asymmetry model is reversed.

$$\left(f\tau_{00}\right)^{-1} \approx G_F^2\left|V_{ud}\right|^2\left[1+\delta^2+\left(1-\delta^2\right)\sin 2\zeta\right], \tag{4.14}$$

$$\left(f\tau_n\right)^{-1} \approx G_F^2\left|V_{ud}\right|^2\left[1+\delta^2-\left(1-\delta^2\right)\frac{\left(1-3\lambda_n^2\right)}{\left(1+3\lambda_n^2\right)}\sin 2\zeta\right] \tag{4.15}$$

$\frac{\left|\tilde{V}_{ud}^{00}\right|^2-\left|\tilde{V}_{ud}^{n}\right|^2}{\left|\tilde{V}_{ud}^{00}\right|^2+\left|\tilde{V}_{ud}^{n}\right|^2} \approx \frac{2\zeta}{3\lambda_n^2+1} \approx 0.34\zeta$ (4.16), where $\tilde{V}_{ud}^{00}=0.97373(32)$ and $\tilde{V}_{ud}^{n}=0.97477(37)$ are experimental values obtained within the framework of the V - A version of the theory. Note that in the last expression the dependence on $\left|V_{ud}\right|^2$ is reduced.

From this difference, using formula (4.16) it is possible to make estimates for the model parameter $\zeta$. Figure 4.1a) shows the experimental situation of discrepancies between the values of $V_{ud}^{00}$ and $V_{ud}^{n}$, as well as between the values of $V_{ud}^{00}$ and $V_{ud}^{unt}$. Figure 4.1b) shows the result of calculating the model parameter $\zeta$ from discrepancies in the definition of $V_{ud}$.

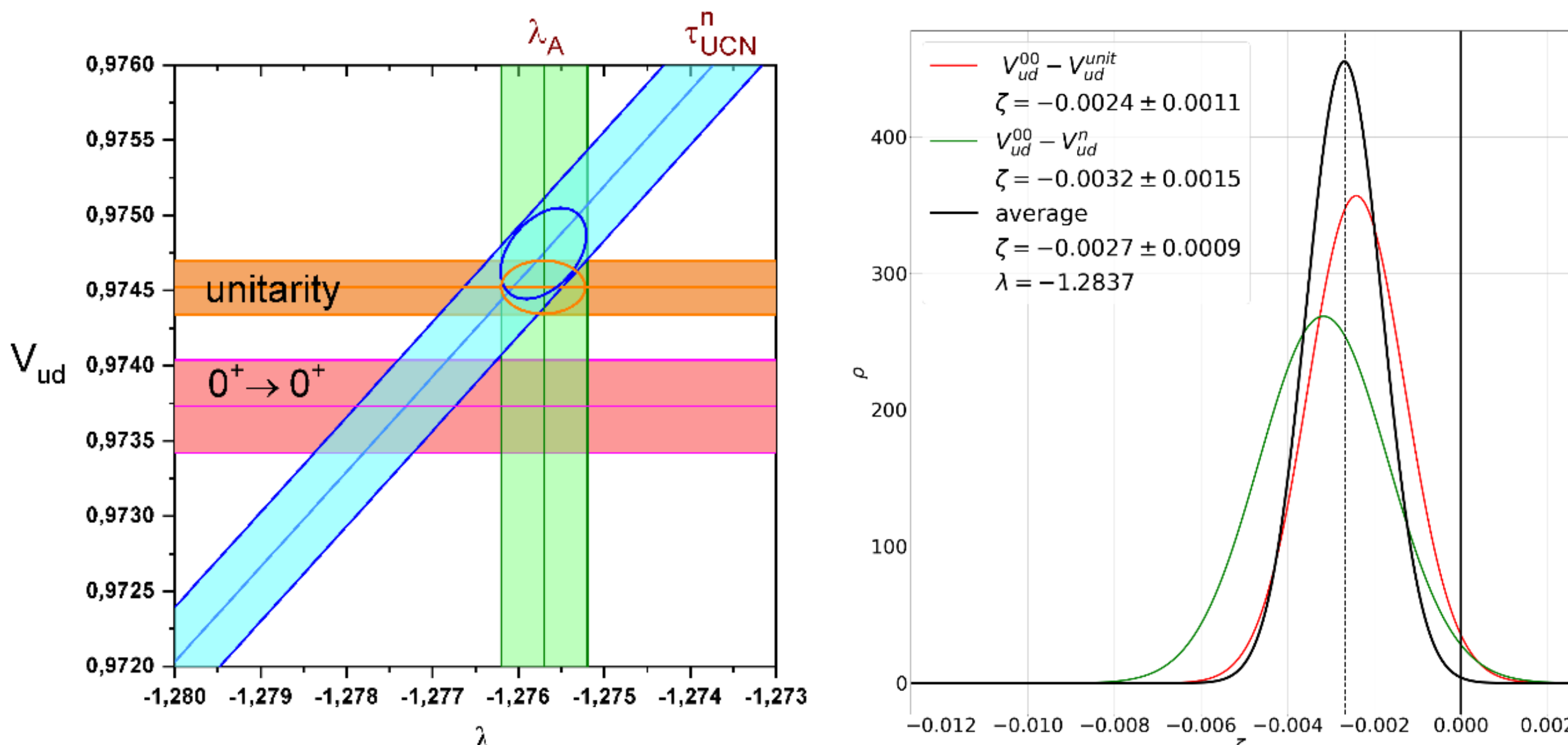

Fig. 4.1a) shows the experimental situation in determining the value $V_{ud}$ from neutron data ( $\tau_n$ and $A$ ), 00 transitions and unitarity. b) the result of calculating the model parameter $\zeta$ with $V_{ud}^{00}$, with $V_{ud}^{n}$, with $V_{ud}^{unt}$ and the weighted average value of the model parameter $\zeta$.

Note that the value $V_{ud}^{unt}$ is more accurate and can be considered as an alternative way of determining $V_{ud}^{n}$, so it is advisable to use both results and provide a weighted average. These calculations use the optimal value of $\lambda$, which will be obtained later by a complete fit of all data (on neutron decay, $0^{+} \rightarrow 0^{+}$ transitions, and unitarity) using the $\chi^{2}$ method.

The optimal values of $\delta$, $\zeta$, $\lambda$, and $V_{ud}$ should also be found by the method $\chi^{2}$ using all experimental data for neutron decay and for $0^{+} \rightarrow 0^{+}$ transitions. In this set of experimental data, the results used for the coefficient "a" were $a_{ev}^{PDG}$= **-0.1053** (18). from aSPECT experiment. The result of such an analysis is presented in Fig. 4.2. On the plane $\delta$, $\zeta$ - Fig. 4.2.a) and on the plane $M_{W_R}$, $\zeta$ - Fig. 4.2.b).

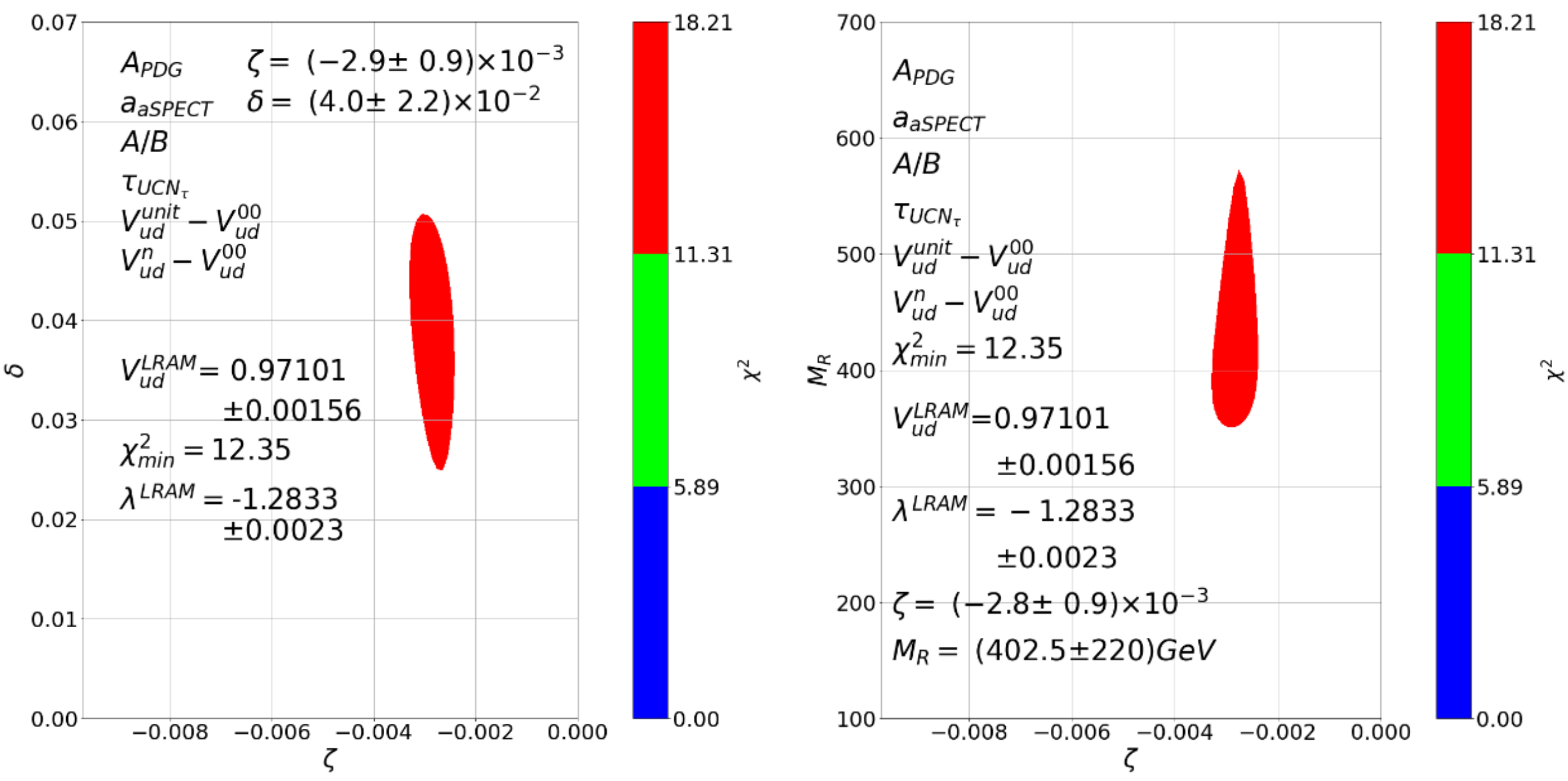

Fig. 4.2. Distribution $\chi^2$ for the model parameters: a) On the plane $\delta$, $\zeta$, b) on the plane $M_{W_R}$, $\zeta$. This analysis used data from the aSPECT experiment.

As can be seen, the contradiction between 0+ -0+ transitions and neutron decay is eliminated.

Below is a verification table of the obtained results for convergence with the experimental data.

Table 4.1.

| Parameter | Exp. value | Calculation | Deviation,σ |
|---|---|---|---|
| $\tau_n$ | $877.75^{+36}_{-32}$ | 877.7 5(4.30) | **0** |
| A | -0.11958(21) | -0.11954 (120) | -0.03 |
| B | 0.9802(49) | 0.98413 (333) | -1.2 |
| $a_{ev}^{aSPECT}$(2023) | -0.10402(82) | -0.10675 (38) | 2.1 |
| C | -0.2377(26) | -0.23762 (93) | 0.03 |
| A/B | -0.1223(52) | -0.12147 (122) | -0.33 |

Now let us consider the analysis results with the experimental data of aCORN experiment (Fig. 4.3).

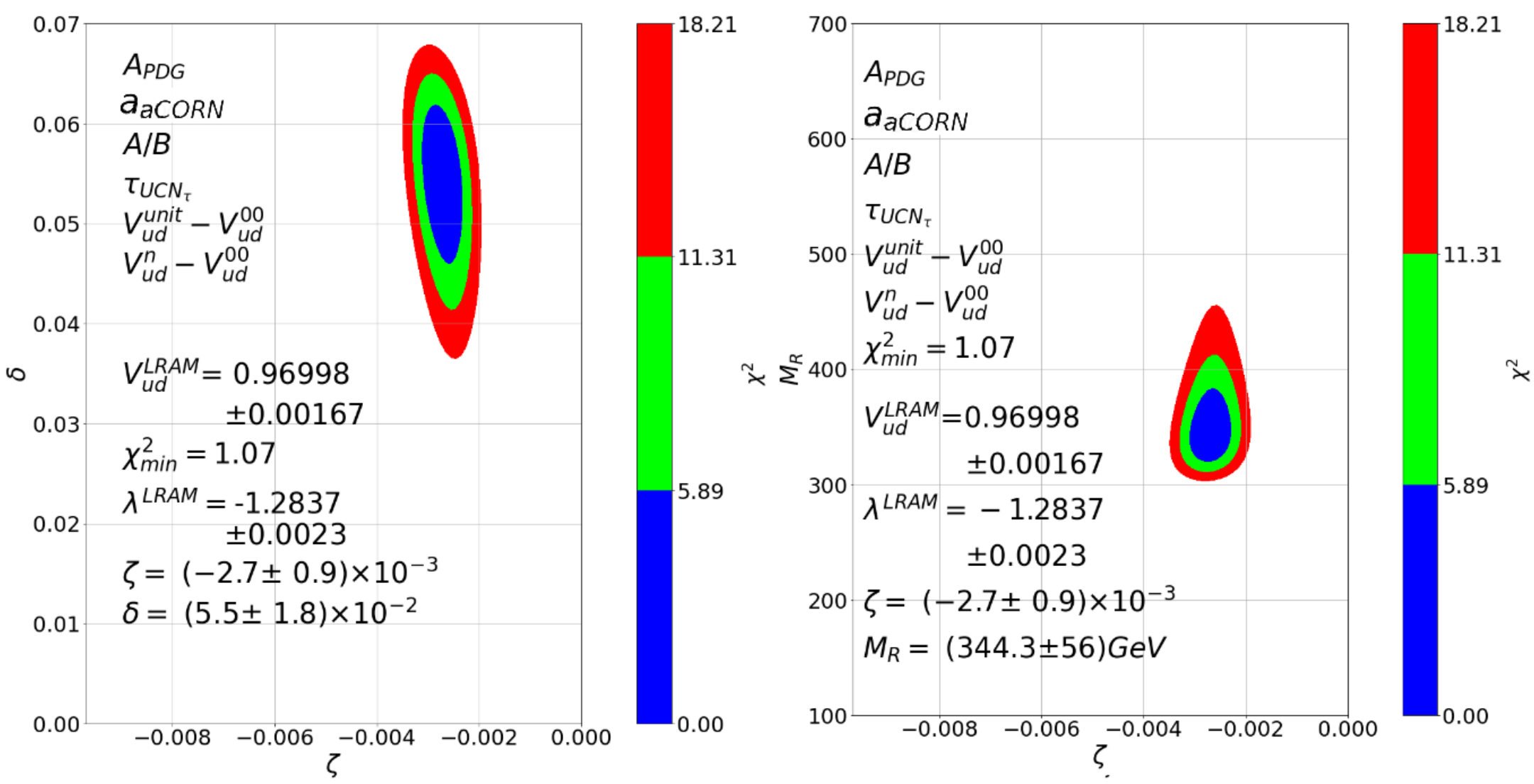

Fig.4.3. Distribution $\chi^2$ for the model parameters: a) on the plane $\delta$, $\zeta$, b) on the plane $M_{W_R}$, $\zeta$ using the data from the aCORN experiment.

To check the convergence of the experimental data (for $a_{ev}^{\mathrm{aCORN}}$) with the calculated data, using the LRAM formulas and the optimal values of the model parameters: $\delta$, $\zeta$, $\lambda$ and $V_{ud}$ (with their errors), Table 4.2 is presented, where the last column shows the discrepancy between the experimental data and the calculated data in the number of standard deviations.

Table 4.2.

| Parameter | Exp. value | Calculation | Deviation,σ |
|---|---|---|---|
| $\tau_n$ | $877.75^{+36}_{-32}$ | 877.7 5(4.46) | 0 |
| A | -0.11958(21) | -0.11957 (118) | 0 |
| B | 0.9802(49) | 0. 9814 (37) | -0.2 |
| $a_{ev}^{aCORN}$ | -0.1053(18) | -0.1070 (10) | 0.83 |
| C | -0.2377(26) | -0.2369 (10) | 0.3 |
| A/B | -0.1223(52) | -0.1218 (12) | -0.05 |

Thus, the results of the analysis can be considered self-consistent. The model parameter, the mixing angle $\zeta = -(2.7 \pm 0.9)\cdot 10^{-3}$, was derived from neutron decay — lifetime and decay asymmetries — as well as from studies of proton decay in nuclei, so-called superallowed $0^+ - 0^+$ transitions. The $W_R$ mass estimate $M_{W_R} = (340 \pm 60)$ GeV is given by the relation: from the parameter $\delta = (5.5 \pm 1.8)\times 10^{-2}$, which is the ratio of the squares of the mass states of vector bosons.

Let's make some clarifications. Figures 4.2 and 4.3 show sections of the distribution $\chi^2$ in the minimum region, and the errors were obtained by analyzing four parameters at once, and, therefore, the magnitude of the error was influenced by the correlation of the parameters.

In conclusion of this analysis, we present the correlation matrix for the model parameters:

| | $\zeta$ | $\delta$ | $\lambda$ | $V_{ud}$ |
|---|---|---|---|---|
| $\zeta$ | 1 | −0.005 | 0.965 | 0.8442 |
| $\delta$ | −0.005 | 1 | −0.024 | −0.096 |
| $\lambda$ | 0.965 | −0.024 | 1 | 0.935 |
| $V_{ud}$ | 0.8442 | −0.096 | 0.935 | 1 |

From this matrix it can be seen that there is a strong relationship between the parameters $\zeta, \lambda, V_{ud}$ within the left-right asymmetry model. Let us present the $\chi^2$ projections onto planes for all parameters of the left-right asymmetric model.

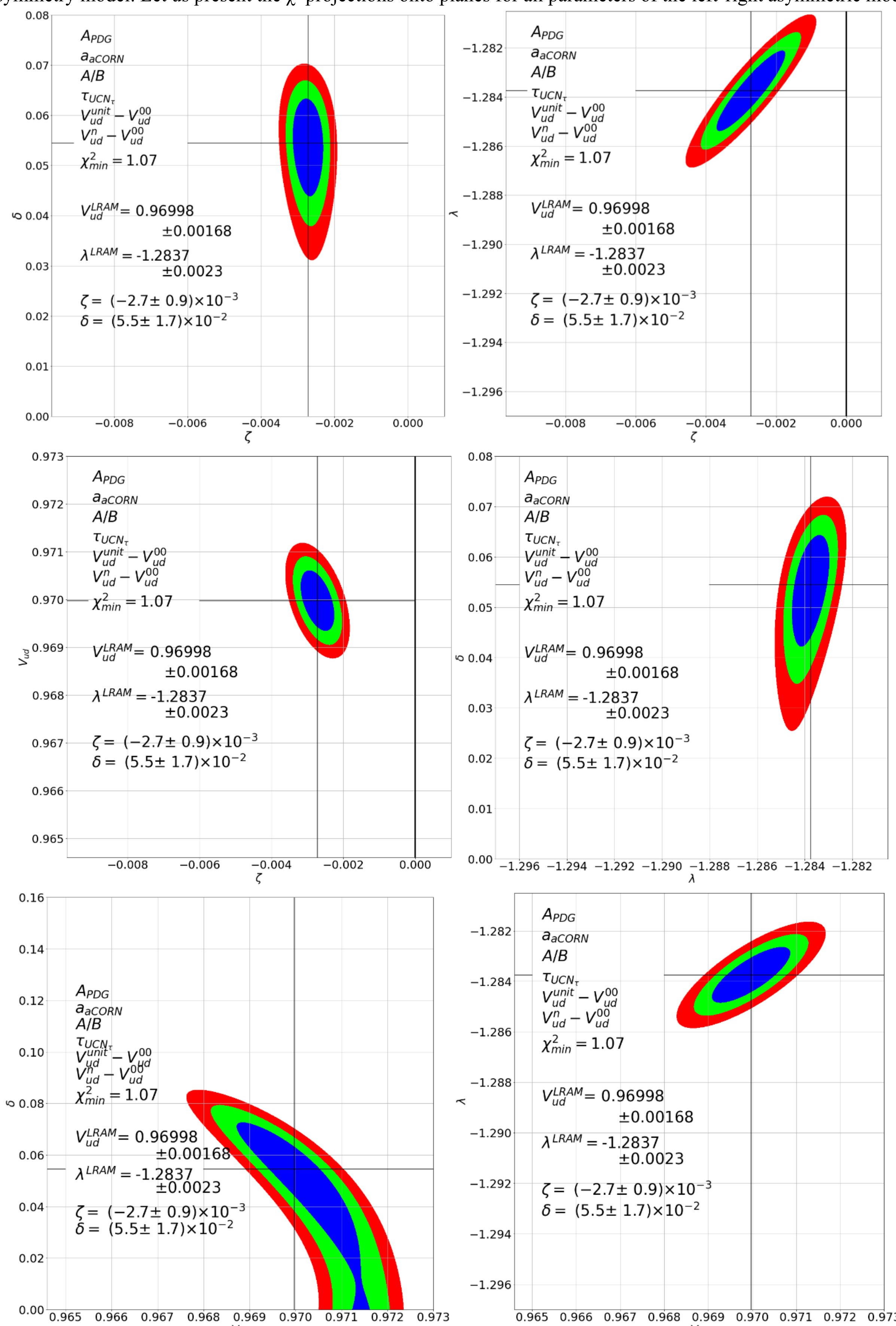

Fig. 4.5. Projections $\chi^2$ onto planes of all parameters of the left-right asymmetry model.

Figure 4.5 shows all projections of $\chi^2$ onto the parameter plane. It is evident that the projections, in pairs, except for $\delta, V_{ud}$, have a deviation of more than $3\sigma$ from the Standard Model parameters, which confirms the need to use the left-right asymmetry model (LRAM) to describe neutron decay, and, as a consequence, accept the need to extend the Standard Model with a right vector boson.
Finally, we need to compare the results of the new analysis of the model parameters $\delta$ , $\zeta$ with their constraints from the TWIST experiment [17] (Fig. 4.4).

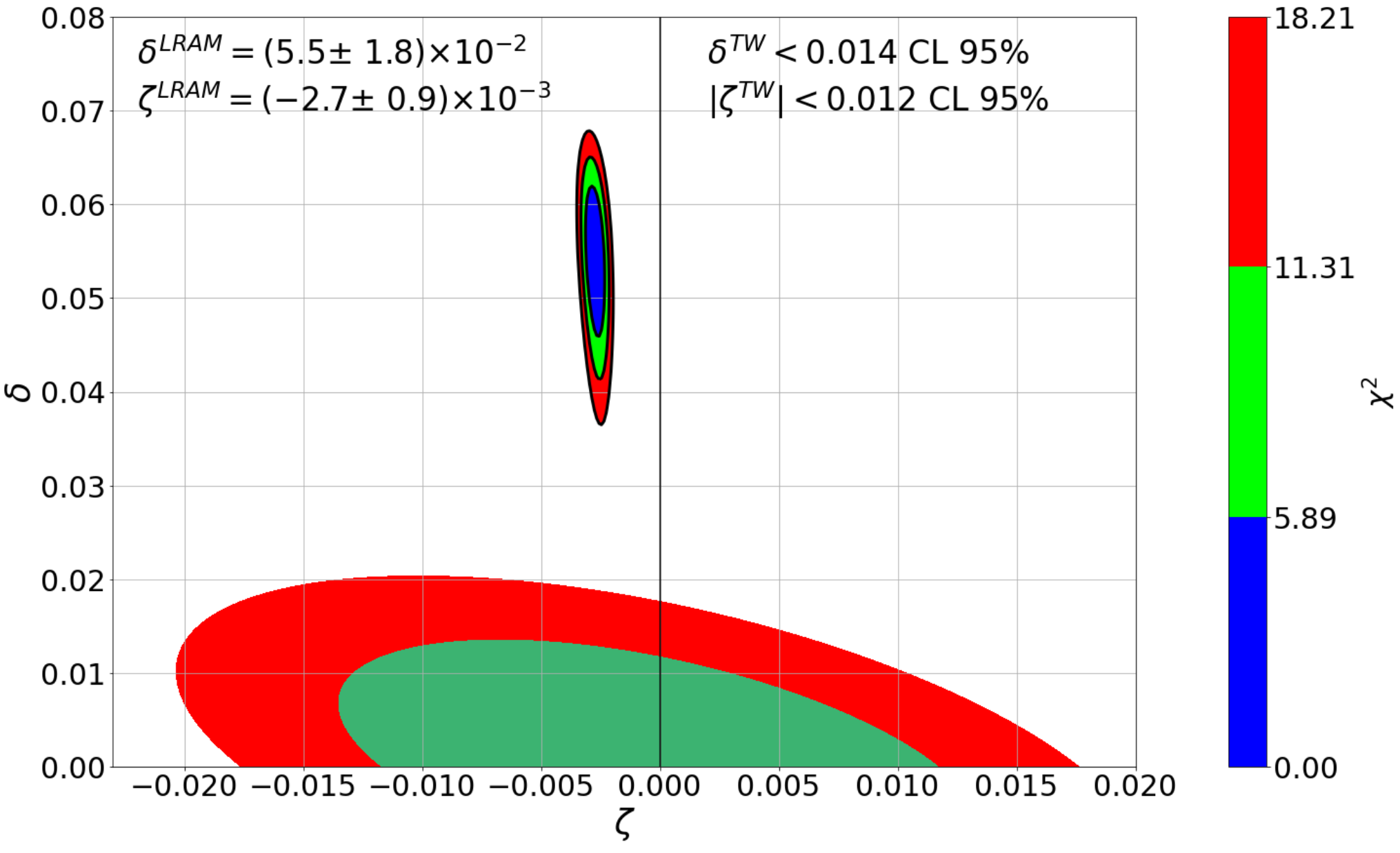

Fig. 4.4. Comparison of the results of the analysis within the framework of the left-right asymmetry model and the result of the TWIST experiment .

The constraints on the model parameters $\delta$ and $\zeta$ for charged and neutral particles do not overlap, as shown in Fig. 4.4. The result of the TWIST experiment [17] does not contradict the result of our analysis of neutron decay, since the TWIST experiment was carried out with charged particles. Within the framework of the left-right asymmetry model, the manifestation of the CP violation effect for decays of charged particles with final-state leptons is impossible due to charge conservation. Therefore, the parameters $\delta$ and $\zeta$ with the values from our analysis could not manifest themselves. From the TWIST experiment, there is only a constraint on $\left|\zeta^{TWi}\right| < 1.2 \times 10^{-2}$ CL 95% and limit on $\delta^{TW} < 1.4 \times 10^{-2}$ CL 95%. The constraints from the TWIST experiment were obtained within the left-right manifest model. Our results, however, were obtained within the left-right asymmetry model, so a direct comparison is invalid. However, it can be concluded that the results of our analysis $\delta = (5.5 \pm 1.8) \times 10^{-2}$ , $\zeta = -(2.7 \pm 0.9) \cdot 10^{-3}$ do not contradict the TWIST experiment, since it should be noted that there is no reason to expect CP violation in muon decay.
In conclusion of the analysis, we can draw conclusions from the comparison of LRSM and LRAM using the obtained formulas. The value of the coefficient D is an important criterion determining T-invariance. At D = 0, the probabilities of direct and inverse processes are equal for both particles and antiparticles. This is the case for LRAM and for LRSM at $\omega = 0, \pi$ . For LRAM, with the equality of direct and inverse processes, there is a difference in the probability for particles and antiparticles, which we are forced to interpret as CP violation while preserving T-invariance, and thus with CPT-violation. As a result, a difference in the lifetimes of particles and antiparticles arises and the possibility of explaining the asymmetry of the Universe. For LRSM at $\omega = 0$, it is impossible to introduce CP violation, and for LRSM at $\omega = \pm\pi/2$ CP violation has a different sign for particles and antiparticles, which does not allow us to explain the asymmetry of the Universe and experimental asymmetries in neutron decay (a and A/ B).

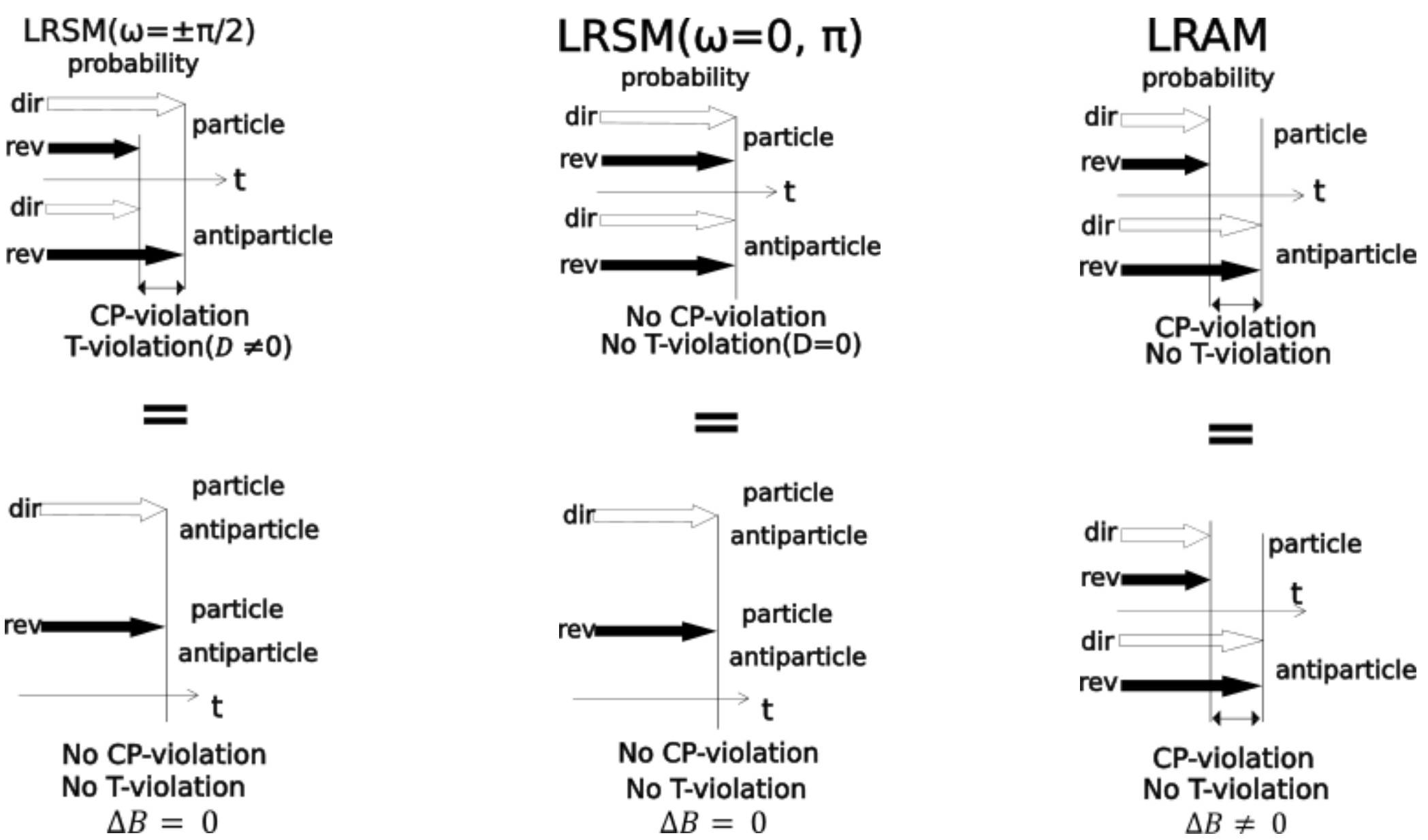

Fig. 4.6. Diagram showing the differences in the representation of CP -violation in LRSM at $\omega = 0, \pi/2$ and in LRAM. ΔB = 0 implies the absence of baryon asymmetry in the Universe.

Now let's consider the question of what CP violation means while maintaining T-invariance, and thus CPT violation. First, let's recall that CPT invariance must be strictly satisfied for truly elementary and charged particles—such as the electron, positron, muon, and antimuon—which is a consequence of their elementary nature and the law of charge conservation, which prohibits mutual transitions. The situation is entirely different for neutral and composite particles—such as the neutron, antineutron, and meson (antimeson). Below is a picture of what happens to the K-meson during charge and spatial conjugation (CP conjugation), which transforms particles into antiparticles. (Fig. 4.7). As is well known, this violates the strangeness by two units ( $\Delta S = 2$ ). But strangeness is not preserved, as has been shown experimentally. Oscillations $K^0 \to \tilde{K}^0$ are possible, and the probability $K^0 \to \tilde{K}^0$ is equal to the probability $\tilde{K}^0 \to K^0$ (due to the fulfillment of T-invariance), while simultaneously violating CP-invariance and strangeness. And there is no CPT-violation process; it is merely a consequence, the cause of CP-violation. There is no reason to insist on the conservation of CPT. The quark structure is simply being restructured while the total charge is conserved. Crucially, the sign of CP-violation changes as we move from $K^0 \to \tilde{K}^0$ process to $\tilde{K}^0 \to K^0$ .

Concluding the question of what CP violation means when T-invariance is preserved, and thus CPT violation arises, it should be said that the assumption of CPT violation in a neutral composite system (like neutral mesons and neutrons) does not in itself contradict Lorentz invariance [20]. Moreover, if CPT violation is possible for neutral composite systems, then unitarity is satisfied [21]. Finally, the briefest conclusion is that there is no CPT violation process as such, it is just a consequence, and the cause is CP violation while maintaining T-invariance. Fig. 4.7 illustrates the application of the C and P conjugation operation, which transforms a particle into an antiparticle, after which the application of the operation $t \to -t$ transforms the antiparticle into a particle while preserving T-invariance. There is a notion that CP violation is compensated by T-invariance violation to preserve CPT invariance. However, it is possible that the decay rates of particles and antiparticles differ, but the probability of transition from a particle to an antiparticle and vice versa is the same. Then, with sufficiently frequent oscillations, the CP violation effect is compensated, but if the decay rate is much smaller than the oscillation period, the CP violation effect will be clearly manifested. This is precisely the situation that arises for neutral mesons. This will be shown in Sections 5 and 6. The question of T-invariance conservation will be demonstrated by relation 5.28, which shows the equality of the probability of transitions from a particle to an antiparticle and vice versa.

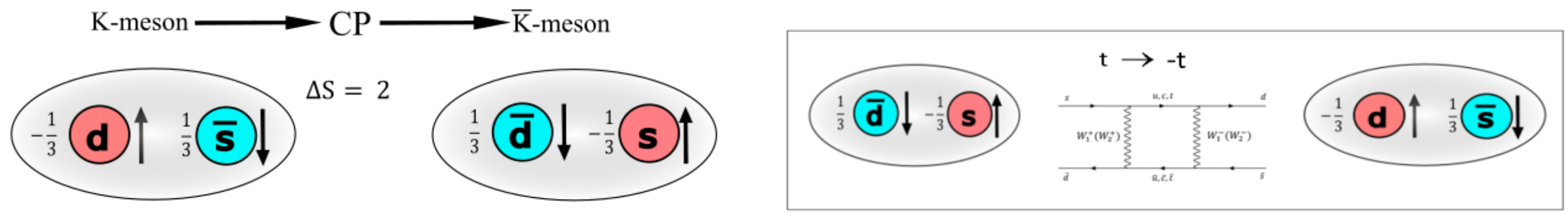

Fig. 4.7. Schemes illustrating the CP transition for kaon-antikaon systems and the transition t ⟶ - t with preservation of T-invariance in the process of kaon-antikaon oscillations.

A similar situation is possible for the neutron and antineutron. In this case, neutron-antineutron oscillations and a difference in lifetime for neutrons and antineutrons in LRAM are possible, and the absence of experimental observation [18] of neutron-antineutron oscillations does not mean the absence of a difference in lifetime between the neutron and antineutron. If the oscillation period is much longer than the lifetime, then it is impossible to observe oscillations. For example, for $D^0$ meson, the oscillation period is 250 times longer than the lifetime, and CP violation is observed due to the difference in lifetime for $D^0$ and $\tilde{D}^0$. The experimental value of the CP violation effect is A = (6.8±0.65)·10 $^{-3}$ [19], and the calculated value based on LRAM is: $A_{CP}^{D^0}$ =5.3 × $10^{-3}$ (See Section 6.) Thus, we have an example of the same situation for $D^0$ meson that we assume for a neutron.

In conclusion of the presented analysis, we must give important clarifications concerning the radiative corrections that must be taken into account when obtaining experimental results for the neutron lifetime $\tau_n$ and asymmetry coefficients $a$, $A$ and $B$ in neutron decay, as well as corrections in the calculations $f\tau_{00}$ for different $0^+ - 0^+$ transitions. The experimental results presented in Table 4.2 already include radiative corrections. Corrections in the calculations $f\tau_{00}$ for different $0^+ - 0^+$transitions were carefully analyzed in [2, 3], and the value $\tilde{V}_{ud}^{00}$ obtained for a whole set (about 20) of nuclei takes into account the corrections in these works.

Below we comment on the fact that taking into account radiative corrections [22-24] plays an important role in precision experiments to measure the neutron lifetime - $\tau_n$ and correlation coefficients of neutron decay asymmetry - $A$, $B$ and $a$ to search for possible deviations from the predictions of the Standard Model [25-27]. In addition to radiative corrections, corrections for weak magnetism and the finite mass of the nucleon (non-zero momentum of the proton) are also taken into account [28]. The neutron lifetime $\tau_n, \lambda_n$ and the square of the modulus of the matrix element $V_{ud}$ of the CKM matrices are related by the following expression [26]:

$$\frac{1}{\tau_n} = \frac{G_F^2 {V_{ud}}^2}{2\pi^3} m_e^5 (1+3\lambda_n^2)(1+\Delta_\mathrm{R}) f \qquad (4.17)$$

where $f$ is the coefficient associated with the phase space [28] and taking into account the Fermi function, as well as the finite size and mass of the nucleus $f = 1.6887(1)$, $\Delta_\mathrm{R}$ is of the order of the fine structure constant $\alpha/\pi$ [23,24]. The so-called master formula is given in [25]:

$${V_{ud}}^2 = \frac{4905.7(1.7)s}{\tau_n(1+3\lambda_n^2)} \qquad (4.18),$$

which is obtained from (4.17) taking into account the radiative corrections $\Delta_\mathrm{R}$ equal to 0.03947(32). There is also a new calculation result equal to 0.03983(27) from [29]. But this does not change the results of our calculations much, so we will use the old value for the formula (4.18).

Radiative corrections in the form of a coefficient $(1+\Delta_\mathrm{R})$ can be represented as a product $(1+\Delta_\mathrm{R}) = (1+\delta_R)(1+\Delta_R^\mathrm{V})$, where the contribution of the "external" radiative correction $\delta_R$ was obtained by calculating one-loop QED diagrams and bremsstrahlung diagrams under the assumption that nucleons are point particles and the contribution of the "internal" radiative correction $\Delta_R^V$ depends on the details of the electroweak theory of the Standard Model and allows one to study the hadronic structure of the nucleon [30]. In the work [25], the main formula (4.18) was obtained from (4.17) with $\Delta_R^V$= 0.02426(32). New calculations $\Delta_R^V$ based on the data-driven dispersion relation can be found in [31–33], and based on lattice QCD in [34]. The mean value is $\Delta_R^V$= 0.02467(27)

When measuring the correlation coefficient $A$ in the work [11], when fitting the experimental value of asymmetry, the necessary corrections are made for the final mass of nucleons (recoil protons) and weak magnetism, $g_V - g_A$ interference. In addition, the experimental significance of asymmetry $A$ takes into account radiation corrections, the value of which is of the order of magnitude $10^{-3}$ and errors $10^{-4}$.

When measuring the correlation coefficient $a$ in [35], the proton energy spectrum was fitted to the formulas from [36], which take into account radiative corrections of the order of the fine structure constant. Corrections for weak magnetism and Coulomb interaction were also taken into account.

In the work [30] expressions are given for the dependence of the coefficients $a$, $A$ and $B$ on the electron energy with an accuracy of corrections of the order of $10^{-4}$. It turns out that the influence of radiative corrections on the value of the coefficient $B$ less [37,25] than the neutron lifetime. In the above analysis, the accuracy of the radiation corrections was already taken into account, with a correction with an error of 1.7 s in formula (4.18). Table 4.3 presents a comparison of the experimental accuracy and the accuracy of the corresponding calculated corrections, indicating their errors.

Table 4.3

| Value | Experimental error, % | Correction, % | Correction error, % | Ref. |
|---|---|---|---|---|
| $\tau_n$ | 0 .040 | 3.947 | 0.032 | [25] |
| $A$ | 0.176 | - 0.100 | 0.01 | [11] |
| $a$ | 0.788 | 0.005 | 0.005 | [13] |
| $B$ | 0.306 | <0.1 | <0.1 | [38] |

As can be seen from Table 3.2, the most significant corrections are those for the neutron lifetime; the accuracy of their inclusion is exceptionally high—comparable to the accuracy of experimental measurements. The next most important correction is for the A -asymmetry. This correction is taken into account and is almost half the experimental error. The corrections for the a-asymmetry and B-asymmetry are exceptionally small because they are related to neutrinos.

In concluding this section, it is necessary to make some important clarifications. The fact is that the mixing angle $\zeta$ in this analysis is an order of magnitude smaller compared to our analysis in [1], and it is necessary to clarify what is going on here. The fact is that the new analysis presents a complete set of formulas for the neutron lifetime $\tau_n$ and asymmetry coefficients $a$ , $A$ and $B$ in neutron decay, which was obtained by successive application of the above-mentioned relations (4.1) - (4.8) without an approximate expansion in small model parameters, in contrast to the work [9]. Here it should be noted that in the work [9] expressions (C16-C24) are given for the coefficients a, A and B, which do not contain linear terms in small model parameters $\delta$ and $\zeta$ . In this regard, a detailed analysis of the derivation of expressions (C16-C24) for the coefficients a, A and B was carried out in the work [9], and it turned out that in these expressions there is no linear term in the parameter $\zeta$ because, even when writing formulas (36) - (39) for $C_V$ and $C_A$, only a linear approximation was used, but when passing to $|C_V|^2+|C'_V|^2$ and $|C_A|^2+|C'_A|^2$ it was also necessary to take into account the quadratic terms. When constructing the asymmetry (as a relative difference to the sum), the linear terms in the parameter $\zeta$ in the numerator and denominator of this ratio were factored out and canceled. Therefore, only quadratic terms in the parameter $\zeta$ remained. Such manipulations with approximate quantities led to erroneous formulas for asymmetries. In the work [9], instead of the exact formulas (4.2) and (4.3) (on the left in the table below), the following approximations were used (on the right in the table below (formulas (36-39) from the work [9]), where the quadratic terms were omitted.

Table 4.4.

| This work | Reference [9] formulas (36-39) |
|---|---|
| $C_V = g_V a_{LL}\big((1+\delta)-(1-\delta)\sin 2\zeta\big)$ | $C_V = g_V a_{LL}(1-2\zeta+\delta)$ |
| $C'_V = g_V a_{LL}(1-\delta)\cos 2\zeta$ | $C'_V = g_V a_{LL}(1-\delta)$ |
| $C_A = g_A a_{LL}\big((1+\delta)+(1-\delta)\sin 2\zeta\big)$ | $C_A = g_A a_{LL}(1+2\zeta+\delta)$ |
| $C'_A = g_A a_{LL}(1-\delta)\cos 2\zeta$ | $C'_A = g_A a_{LL}(1-\delta)$ |

Thus, it was revealed that our work [1] presented an analysis based on erroneous formulas from [9]. This error has been corrected in the present work, as precise formulas are used. However, the estimate of the CP-violation effects from a linear but small parameter $\zeta$ turned out to be practically the same. The overestimated previous estimates of CP-violation in quadratic form practically coincided with the estimates with the correct small parameter $\zeta$, now presented in linear form.

## 5. Analysis of CP-violation processes in neutral meson oscillations within the framework of the left-right asymmetry model

However, let's begin our consideration with the decay of charged $\pi$-mesons to demonstrate the fundamental difference with processes involving neutral mesons. While oscillations occur between neutral mesons and antimesons, with CP violation, charged $\pi$-mesons are strictly CPT-constrained on the difference in lifetimes $\pi^+$ and $\pi^-$. Neutral mesons undergo oscillations with CP violation and no CPT violation, even though these are transitions between particles and antiparticles. It is generally accepted that the oscillations involve a rearrangement of the flavor states of quarks, and it is the mixing of quarks that causes CP violation. However, in the left-right asymmetry model, CP violation arises due to the presence of a right vector boson admixture, and the sign of CP violation is opposite for particles and antiparticles, i.e. the CP violation phase is $\pi/2$, which corresponds to 100% CP violation. Thus, the nature of CP violation is related to vector bosons.

Neutral mesons decay via a vector current similarly to the case $0^+ \to 0^+$. The difference is that kaons have negative intrinsic parity rather than positive. Therefore, we must change the sign before $\zeta$ relative to the sign in $0^+ \to 0^+$ transitions.

For decay through $W^+$ we write the probability for decay $\bar{K}^0$ in the final state $e^+\pi^-\nu$ as follows:

$$\Gamma^{W^+} \propto \left|V_{us}^{LRAM}\right|^2 \left[\left(1+\delta^2\right)-\left(1-\delta^2\right)\sin 2\zeta\right] \qquad (5.1)$$

For the decay $K^0$ when in the final state $e^-\pi^+\bar{\nu}$ (decay through $W^-$), we obtain:

$$\Gamma^{W^-} \propto \left|V_{us}^{LRAM}\right|^2 \left[\left(1+\delta^2\right)+\left(1-\delta^2\right)\sin 2\zeta\right] \qquad (5.2)$$

Using the formulas we obtain an expression for $T$ - violating asymmetry

$$A_T = \frac{\Gamma\left(\bar{K}^0 \to e^+\pi^-\nu\right)-\Gamma\left(K^0 \to e^-\pi^+\bar{\nu}\right)}{\Gamma\left(\bar{K}^0 \to e^+\pi^-\nu\right)+\Gamma\left(K^0 \to e^-\pi^+\bar{\nu}\right)} = -\frac{\left(1-\delta^2\right)}{\left(1+\delta^2\right)}\sin 2\zeta \qquad (5.3)$$

First, we estimate the integral estimate of the CP violation effect for the system $K^0$ $\bar{K}^0$. During the oscillations of the system $K^0$ $\bar{K}^0$ decay into a state may occur $e^-\pi^+\bar{\nu}$ or in a state $e^+\pi^-\nu$.

$$A_T = -\frac{\left(1-\delta^2\right)}{\left(1+\delta^2\right)}\sin 2\zeta \approx -\left(1-2\delta^2\right)\times 2\zeta \approx (5.4\pm 1.8)\times 10^{-3} \quad (5.4)$$

This value is within the available accuracy and in agreement with the experimentally measured asymmetry [39] $A_T^{\exp} = \left(6.6\pm 1.3\pm 1.0\right)\times 10^{-3}(4\sigma)$.

Now let us consider in detail the process of CP violation in the oscillations of neutral mesons. The process of oscillation of neutral mesons involves the transition of particles into antiparticles and back from antiparticles into particles. Transitions between neutral particles do not violate CPT invariance, but these transitions may violate CP invariance. The assumption of CPT violation in a neutral composite system (like mesons) does not in itself contradict Lorentz invariance [20].

Moreover, if CPT violation for neutral composite systems is possible, then unitarity is satisfied [21].

The manifestation of CP violation with mixing of right- and left W bosons is possible for neutral mesons, such as $K^0-\bar{K}^0$, $D^0-\bar{D}^0$, $B^0-\bar{B}^0$, $B_s^0-\bar{B}_s^0$ mesons, as well as for $n-\bar{n}$ oscillations. The specific structure of transitions between quark states is presented below.

$$K^0(d\bar{s})-\bar{K}^0(\bar{d}s),\ D^0(c\bar{u})-\bar{D}^0(\bar{c}u),\quad B^0(d\bar{b})-\bar{B}^0(\bar{d}b),\ B_s^0(s\bar{b})-\bar{B}_s^0(\bar{s}b),\ n(udd)-\bar{n}(\bar{u}\bar{d}\bar{d}) \qquad (5.5)$$

There are experimental constraints on the violation of CPT invariance. These can be divided into direct constraints from experiments with $\pi^+$and $\pi^-$ mesons and indirect constraints from experiments with $K^0-\bar{K}^0$oscillations. In experiments with $\pi^+$and $\pi^-$ mesons, the lifetimes of the particle and antiparticle can be measured directly and compared. CPT violation in decays of $\pi^+$and $\pi^-$[40] has already been experimentally constrained to a precision level of $10^{-3}$.

$\left(\tau_{\pi^+}-\tau_{\pi^-}\right)/\tau = (5.5\pm 7.1)\times 10^{-4}$, i.e. $\Delta\tau_{\pi^\pm}/\tau = (2.25\pm 3.35)\times 10^{-4}$ or $/\Delta\tau_{\pi^\pm}/\tau/ < 0{,}9\times 10^{-3}$ with CL 95%. (5.6)

Although due to the law of conservation of charge the difference $\tau_{\pi^+}-\tau_{\pi^-}$ must be strictly equal to zero.

The same experimental constraints proving $CPT$ conservation exist from decays $K^0, \bar{K}^0$[41]

$\left(\Gamma_{K^0} - \Gamma_{\bar{K}^0}\right) / \Gamma = (5.4 \pm 5.4) \times 10^{-4}$, i.e. $\Delta\Gamma_{K^0} / \Gamma = (2.7 \pm 2.7) \times 10^{-4}$ or $/\Delta\Gamma_{K^0} / \Gamma / < 0.8 \times 10^{-3}$ with CL 95%. 5.7)

However, this is an indirect limitation, since it is not possible to measure the lifetime separately for $K^0$ and separately for $\bar{K}^0$, since $K^0, \bar{K}^0$ are in the mode of mutual transformations, and the lifetime of each of them cannot be determined, as demonstrated by the experimental results in Fig. 5.1.

Estimate (5.7) was obtained in [41], where a mixing matrix with different decay probabilities for $K^0$ and $\bar{K}^0$ on the diagonal of the matrix was considered to account for CPT violation, but CP violation was taken into account in the off-diagonal matrix elements. This result should be interpreted as meaning that after taking into account the CP violation in the off-diagonal matrix elements within the SM framework, there is nothing left for the CPT violation in the diagonal matrix elements with the specified accuracy. Therefore, one should rely on this experimental result and adopt a position based on the results of the experiment [41]. Namely, in the left-right asymmetry model for neutral mesons, CP invariance is violated, but CPT invariance is not violated. The CPT prohibition is directly related to $\pi^+$and $\pi^-$mesons.

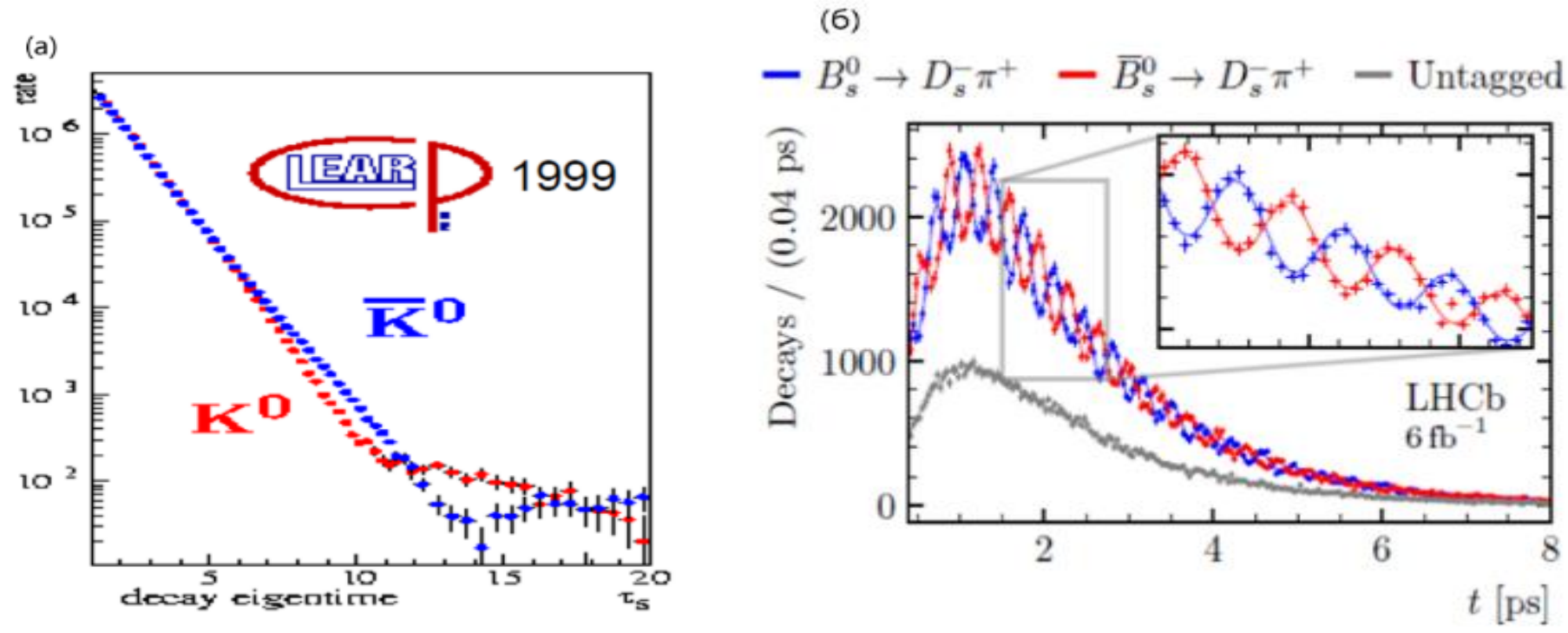

Fig. 5.1. Experimental results of flavor oscillations: a) for oscillations of K $^0$ mesons [42], b) for oscillations $B_s^0 \bar{B}_s^0$[43].

Thus, from equation (5.4) it follows that $A_T = \Delta\Gamma_{K^0} / \Gamma = -\left(1 - 2\delta^2\right) \times 2\zeta \approx (5.4 \pm 1.8) \times 10^{-3}$, and the CPT limitation on $/\Delta\Gamma_{K^0} / \Gamma / < 0.8 \times 10^{-3}$ with CL 95%. Thus, the asymmetry estimate $A_T \approx (5.4 \pm 1.8) \times 10^{-3}$ from equation (5.4) exceeds the experimental CPT limitation by almost 5 times and is nothing other than the effect of CP violation in $K^0 \bar{K}^0$ oscillations, calculated within the framework of the left-right asymmetry model.

We now turn to a description of the CP violation process in neutral meson oscillations. The scheme for analyzing neutral meson oscillations in accelerator experiments is represented by the matrix (5.8). This is the scheme of the Standard Model. The scheme for analyzing neutral meson oscillations within the left-right asymmetry model is represented by the matrix (5.9).

$$\begin{pmatrix} M - i\dfrac{\Gamma_0}{2} & \Delta m - i\dfrac{\Delta\Gamma}{2} \\ \Delta m - i\dfrac{\Delta\Gamma}{2} & M - i\dfrac{\Gamma_0}{2} \end{pmatrix} \quad (5.8) \qquad \begin{pmatrix} M - i\dfrac{\Gamma_0 - \delta\Gamma}{2} & \Delta m - i\dfrac{\Delta\Gamma}{2} \\ \Delta m - i\dfrac{\Delta\Gamma}{2} & M - i\dfrac{\Gamma_0 + \delta\Gamma}{2} \end{pmatrix} \quad (5.9)$$

The mixing matrix for neutral mesons in the Standard Model takes into account the effects of flavor oscillations and CP violation in off-diagonal matrix elements, with CP violation effects being accounted for through the complex phase in the CKM matrix as a phenomenological parameter. The physical causes of CP violation are unknown. With such a SM matrix, flavor oscillations and CP violation effects in off-diagonal matrix elements interfere, making their separation significantly difficult. It is generally accepted that the oscillation process involves a rearrangement of quark flavor states, and it is quark mixing that triggers CP violation.

The mixing matrix for neutral mesons in the left-right asymmetry model is structured differently. It takes into account the effects of flavor oscillations in the off-diagonal matrix elements and the effects of CP violation in the diagonal matrix elements through the parameter $\delta\Gamma$. However, in the left-right asymmetry model, CP violation arises due to the presence of a right vector boson admixture, and the sign of CP violation is opposite for particles and antiparticles, i.e., the CP violation phase is $\pi / 2$, which corresponds to 100% CP violation. Thus, the nature of CP violation is associated with vector bosons, with $\delta\Gamma \approx -\left(1 - 2\delta^2\right) \times 2\zeta \times \Gamma \approx (5.4 \pm 1.8) \times 10^{-3} \times \Gamma$.

Let's compare two schemes for describing the oscillation process of neutral mesons. The effective Hamiltonian is written as follows:

$$H = \begin{pmatrix} m_0 - i\frac{\Gamma_0}{2} & \Delta m - i\frac{\Delta\Gamma}{2} \\ \Delta m - i\frac{\Delta\Gamma}{2} & m_0 - i\frac{\Gamma_0}{2} \end{pmatrix} \quad (5.10) \qquad \begin{pmatrix} m_0 - i\frac{\Gamma_0}{2} & \Delta m - i\frac{\Delta\Gamma}{2} \\ \Delta m - i\frac{\Delta\Gamma}{2} & m_0 - i\frac{\Gamma_0}{2} \end{pmatrix} \begin{pmatrix} a \\ b \end{pmatrix} = \omega \begin{pmatrix} a \\ b \end{pmatrix} \quad (5.11)$$

Solution of the Schrödinger equation for the Hamiltonian $H\Psi(t) = i\hbar \frac{\partial \Psi(t)}{\partial t}$ we will search for in the form of a two-dimensional vector $\Psi(t) = \begin{pmatrix} a \\ b \end{pmatrix} e^{-i\omega t}$. The solution to the Schrödinger equation is reduced to solving the problem of eigenvalues – $\omega$ and eigenvectors - $\begin{pmatrix} a \\ b \end{pmatrix}$. The equation for the eigenfrequencies – $\omega$ is determined from condition (5.11).

$$\det \begin{Vmatrix} m_0 - i\frac{\Gamma_0}{2} - \omega & \Delta m - i\frac{\Delta\Gamma}{2} \\ \Delta m - i\frac{\Delta\Gamma}{2} & m_0 - i\frac{\Gamma_0}{2} - \omega \end{Vmatrix} = 0 \qquad (5.12)$$

Frequency difference: $\omega_+ - \omega_- = \Delta\omega = 2\Delta m - i\Delta\Gamma$. Natural frequencies: $\omega_\pm = m_0 \pm \Delta m - i\frac{\Gamma_0 \pm \Delta\Gamma}{2}$.

For $\omega_+ = m_0 + \Delta m - i\frac{\Gamma_0 + \Delta\Gamma}{2}$ the normalized eigenvector is equal to: $\Psi_L(t) = \frac{1}{\sqrt{2}} \begin{pmatrix} 1 \\ 1 \end{pmatrix} e^{-i\,\omega_+ t}$

For $\omega_- = m_0 - \Delta m - i\frac{\Gamma_0 - \Delta\Gamma}{2}$ the normalized eigenvector is equal to: $\Psi_S(t) = \frac{1}{\sqrt{2}} \begin{pmatrix} 1 \\ -1 \end{pmatrix} e^{-i\omega_- t}$

If at a moment in time $t = 0$ we have a particle state, then $\psi(t) = \frac{1}{\sqrt{2}}\left(\Psi_L(t) + \Psi_S(t)\right)$ (5.13).

If at the initial moment of time $t = 0$ the state of the antiparticle, then $\bar{\psi}(t) = \frac{1}{\sqrt{2}}\left(\Psi_L(t) - \Psi_S(t)\right)$ (5.14).

When the particle is in the initial state, the probability of detecting the particle will be represented by the expression (5.15), and the probability of detecting the antiparticle will be represented by the expression (5.16)

$$\psi^*(t)\psi(t) = \frac{1}{4} e^{-\Gamma_0 t}\left[e^{-\Delta\Gamma t} + e^{\Delta\Gamma t} + 2\cos(2\Delta m t)\right] \quad (5.15),$$

$$\bar{\psi}^*(t)\bar{\psi}(t) = \frac{1}{4} e^{-\Gamma_0 t}\left[e^{-\Delta\Gamma t} + e^{\Delta\Gamma t} - 2\cos(2\Delta m t)\right] \quad (5.16),$$

When the initial state is an antiparticle, the probability of detecting the antiparticle will be represented by the expression (5.17), and the probability of detecting the particle will be represented by the expression (5.18)

$$\bar{\psi}^*(t)\bar{\psi}(t) = \frac{1}{4} e^{-\Gamma_0 t}\left[e^{-\Delta\Gamma t} + e^{\Delta\Gamma t} + 2\cos(2\Delta m t)\right] \quad (5.17),$$

$$\psi^*(t)\psi(t) = \frac{1}{4} e^{-\Gamma_0 t}\left[e^{-\Delta\Gamma t} + e^{\Delta\Gamma t} - 2\cos(2\Delta m t)\right] \quad (5.18)$$

It's clear that particles and antiparticles behave symmetrically: when the initial conditions change, it's as if the particle and antiparticle swap places. We'll use this fact to estimate the magnitude of the asymmetry between the decay probabilities of particles and antiparticles, taking mixing into account.

The next step is to introduce CP violation. In the Standard Model, CP violation is introduced through the complex phase in the CKM matrix. However, we propose introducing CP violation into the diagonal elements of the mixing matrix.

Let's analyze the results of experiments with neutral mesons using the mixing matrix representation (5.9), in which the parameters responsible for flavor oscillations ( $\Delta\Gamma$ and $\Delta$m) and the parameters responsible for CP violation effects ($\delta\Gamma$) are separated. Parameters $\Delta\Gamma$ and $\Delta$m are taken from the experimental data, and the parameter $\delta\Gamma \approx -\left(1-2\delta^2\right)\times 2\zeta \times \Gamma \approx (5.4 \pm 1.8)\times 10^{-3} \times \Gamma$. It is important to emphasize that it is here that the parameters of the left-right asymmetry model are used, and in a unified manner to describe the CP violation effect in the oscillations of all neutral mesons.

Thus, if we introduce in the diagonal matrix elements of the Hamiltonian $\pm\,\delta\Gamma$ for a particle and an antiparticle, we can write:

$$H = \begin{pmatrix} m_0 - i\frac{\Gamma_0 - \delta\Gamma}{2} & \Delta m - i\frac{\Delta\Gamma}{2} \\ \Delta m - i\frac{\Delta\Gamma}{2} & m_0 - i\frac{\Gamma_0 + \delta\Gamma}{2} \end{pmatrix} \qquad (5.19).$$

Different sign of $\delta\Gamma$ means that the sign of the CP violation effect changes when moving from particle to antiparticle. Thus, in the presence of sufficiently frequent oscillations, the average lifetime for particles and antiparticles is the same, and the CP violation effect averages to zero. However, when the lifetime is significantly shorter than the oscillation period, CP violation occurs. Thus, the criterion for the validity of the CP violation scheme within the extended left-right model is the following:

No. 1. The integral (average value over several decay periods) effect of CP violation is practically zero when the oscillation period is significantly shorter than the decay time.

No. 2. The integral effect of CP violation appears when the oscillation period is significantly longer than the decay time.

No. 3. The differential effect of CP violation (dependence on the decay time) changes the sign of the effect to the opposite when passing from a particle to an antiparticle, when the oscillation period is significantly shorter than the decay time.

So, oscillations are possible, but not obligatory, but CP violation is always present.

These conclusions follow from the fact that the transition from particle to antiparticle occurs through a positive or negative vector boson, which has a different sign of the mixing angle ζ with an impurity right vector boson, which will change the sign of the CP violation effect. Our task is to calculate the behavior of the integral and differential effects of CP violation depending on the above conditions and compare with experimental results.

Let's move on to solving the problem of CP-violated oscillations within the extended left-right model. Modifying the Hamiltonian with corrections $\delta\Gamma$ will change both the eigenvalues and eigenvectors in the previous problem:

$$\omega_+ = m_0 - i\frac{\Gamma_0}{2} + \frac{\Delta\omega}{2},$$

$$\omega_- = m_0 - i\frac{\Gamma_0}{2} - \frac{\Delta\omega}{2}, \text{ where } \Delta\omega = \sqrt{(2\Delta m)^2 - [(\Delta\Gamma)^2 + (\delta\Gamma)^2] - 2i(\Delta\Gamma)(2\Delta m)} \quad (5.20)$$

The presence of corrections from $\delta\Gamma$ leads to the fact that in the first order in $\delta\Gamma$ the probability of detecting a particle, calculated in the absence of $\delta\Gamma$ in the Hamiltonian, receives an additive $\varepsilon_{pp}(t)$. When in the initial state the particle (p), then the probability of detecting the particle (p) will be (5.21), and the probability of detecting the antiparticle (a) will be equal to (5.22). Accordingly, the total probability of the process is equal to (5.23).

$$|\psi(p,p)|^2 = \psi^*(t)\psi(t) = \frac{1}{4}e^{-\Gamma_0 t}[e^{-\Delta\Gamma t} + e^{\Delta\Gamma t} + 2\cos(2\Delta m t)] + \varepsilon_{pp}(t) \quad (5.21)$$

$$\left|\bar{\psi}(p,a)\right|^2 = \bar{\psi}^*(t)\bar{\psi}(t) = \frac{1}{4}e^{-\Gamma_0 t}[e^{-\Delta\Gamma t} + e^{\Delta\Gamma t} - 2\cos(2\Delta m t)] \quad (5.22)$$

$$|\psi(p,p)|^2 + \left|\bar{\psi}(p,a)\right|^2 = \frac{1}{2}e^{-\Gamma_0 t}[e^{-\Delta\Gamma t} + e^{\Delta\Gamma t}] + \varepsilon_{pp}(t) \quad (5.23)$$

When the initial state is an antiparticle (a), the probability of detecting the antiparticle (a) will be (5.24), and the probability of detecting the particle (p) (5.25). Accordingly, the total probability of the process is equal to (5.26).

$$\left|\bar{\psi}(a,a)\right|^2 = \bar{\psi}^*(t)\bar{\psi}(t) = \frac{1}{4}e^{-\Gamma_0 t}[e^{-\Delta\Gamma t} + e^{\Delta\Gamma t} + 2\cos(2\Delta m t)] - \varepsilon_{pp}(t) \quad (5.24)$$

$$|\psi(a,p)|^2 = \psi^*(t)\psi(t) = \frac{1}{4}e^{-\Gamma_0 t}[e^{-\Delta\Gamma t} + e^{\Delta\Gamma t} - 2\cos(2\Delta m t)] \quad (5.25)$$

$$\left|\bar{\psi}(a,a)\right|^2 + |\bar{\psi}(a,a)|^2 = \frac{1}{2}e^{-\Gamma_0 t}[e^{-\Delta\Gamma t} + e^{\Delta\Gamma t}] - \varepsilon_{pp}(t) \quad (5.26)$$

Thus, the CP asymmetry of the process will be:

$$A_{CP}(t) = \frac{[\psi(p,p)]^2 + [\psi(p,a)]^2 - [\psi(a,a)]^2 - [\psi(a,p)]^2}{[\psi(p,p)]^2 + [\psi(p,a)]^2 + [\psi(a,a)]^2 + [\psi(a,p)]^2} = \frac{2\delta\Gamma}{\sqrt{(\Delta\Gamma)^2 + (2\Delta m)^2}} \times \frac{(\Delta\Gamma)\mathrm{sh}(\Delta\Gamma t) + (2\Delta m)\sin(2\Delta m t)}{\sqrt{(\Delta\Gamma)^2 + (2\Delta m)^2}\,\mathrm{ch}(\Delta\Gamma t)} \quad (5.27)$$

Note that $\left|\bar{\psi}(p,a)\right|^2 = |\psi(a,p)|^2$, (5.28)

which indicates the fulfillment of T-invariance in the process of meson-anti-meson oscillations.

For approximate calculations of asymmetry in the expansion by the smallness parameter, $x = \delta\Gamma/((2\Delta m)^2 + (\Delta\Gamma)^2)^{1/2}$ we use the formula:

$$\varepsilon_{pp}(t) = \frac{\delta\Gamma}{\sqrt{(2\Delta m)^2 + (\Delta\Gamma)^2}} \times \frac{e^{-\Gamma_0 t}((\Delta\Gamma)sh(\Delta\Gamma t) + (2\Delta m)sin(2\Delta m t))}{\sqrt{(2\Delta m)^2 + (\Delta\Gamma)^2}} \quad (5.29)$$

The calculated results for the probability of detecting a particle or antiparticle for the two schemes are presented in Fig. 5.2. As can be seen, the calculated results with $\delta\Gamma = 0$ and $\delta\Gamma / \Gamma \approx (5.4 \pm 1.8) \times 10^{-3}$ do not differ within the graphical accuracy for all mesons. This is because the flavor oscillation process is successfully described by the off-diagonal parameters of the mixing matrices, while the diagonal parameter $\delta\Gamma$, responsible for CP violation, has little effect on the oscillation process. However, as will be shown below, the diagonal parameter $\delta\Gamma$, plays a decisive role in calculating the CP-violating decay asymmetry. Thus, in the mixing matrix (5.9) in the left-right asymmetry model, flavor oscillations and CP-violation effects are separated.

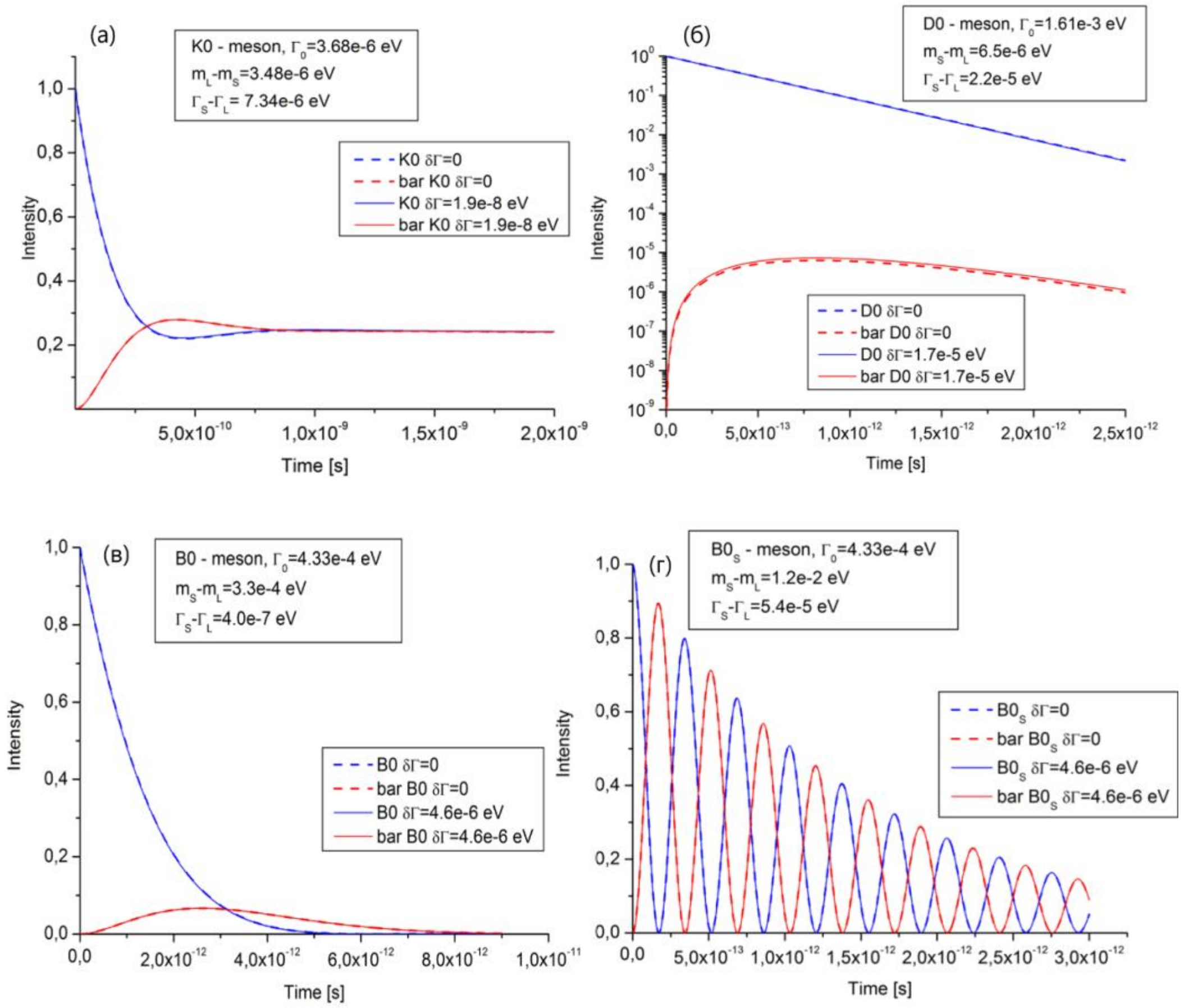

Fig. 5.2. The process of oscillations of neutral mesons. The blue curves correspond to the state of the particle, the red ones to the state of the antiparticle, the solid curves correspond to the calculations carried out with the matrix (5.8) (SM), the dashed curves correspond to the calculations carried out with the matrix (5.9) (LRAM). In the calculations, the off-diagonal matrix elements are defined with $2\Delta m = (m_S - m_L),\ 2\,\Delta\Gamma = (\Gamma_S - \Gamma_L)$, a) - results of calculations for $K^0$ -meson, b) - for $D^0$ -meson, c) - for $B^0$ -meson, d) for $B_S^0$ -meson.

## 6. Behavior of integral and differential effects of CP violation and comparison with experimental results

Now let's move on to calculating the integral asymmetry. For this, we take the ratio of the integral $\varepsilon_{pp}$ to the integral $\rho_{pp}$.

$$A_{CP}(t) = \frac{2\delta\Gamma}{\sqrt{(\Delta\Gamma)^2 + (2\Delta m)^2}} \times \frac{(\Delta\Gamma)\operatorname{sh}(\Delta\Gamma t) + (2\Delta m)\sin(2\Delta m t)}{\sqrt{(\Delta\Gamma)^2 + (2\Delta m)^2}\,\operatorname{ch}(\Delta\Gamma t)}\,, \qquad (6.1)$$

Table 6.1 lists the integral asymmetries $\tilde{A}_{p\bar{p}}$ for $K^0,\ D^0,\ B^0$ and $B_S^0$-mesons, calculated using wave functions (Exact) and approximate formulas (Approx). As can be seen, the approximate formulas agree reasonably well with the exact computer calculations. However, it is important to note that the formulas allow for an analytical analysis of various dependencies and relationships, which are extremely useful.

The absolute values of the calculated and experimental asymmetries are quite difficult to compare. The point is that the experiment uses a specific decay mode, while the calculation considers transitions from particles to antiparticles, which cover all decay channels. Therefore, there is a problem with normalizing the asymmetry value.

$$A_{CP}(t) = \frac{2\delta\Gamma}{\sqrt{(\Delta\Gamma)^2 + (2\Delta m)^2}} \times \frac{(\Delta\Gamma)\operatorname{sh}(\Delta\Gamma t) + (2\Delta m)\sin(2\Delta m t)}{\sqrt{(\Delta\Gamma)^2 + (2\Delta m)^2}\,[\operatorname{ch}(\Delta\Gamma t) + 1]} =$$

$$= -\left(1 - 2\delta^2\right) \times 2\zeta \frac{2\Gamma}{\sqrt{(\Delta\Gamma)^2 + (2\Delta m)^2}} \times \frac{(\Delta\Gamma)\operatorname{sh}(\Delta\Gamma t) + (2\Delta m)\sin(2\Delta m t)}{\sqrt{(\Delta\Gamma)^2 + (2\Delta m)^2}\,[\operatorname{ch}(\Delta\Gamma t) + 1]} = A^{LRAM} \times F \times A_{asym}(t) \qquad (6.2)$$

where $A^{LRAM} \approx -\left(1 - 2\delta^2\right) \times 2\zeta$ is a CP-violating asymmetry of the left-right weak interaction model

$F = \dfrac{2\Gamma}{\sqrt{(\Delta\Gamma)^2 + (2\Delta m)^2}}$ - a time-independent factor characterizing a specific meson (6.3),

$A_{asym}(t) = \dfrac{(\Delta\Gamma)\,\mathrm{sh}(\Delta\Gamma t) + (2\Delta m)\sin(2\Delta m t)}{\sqrt{(\Delta\Gamma)^2 + (2\Delta m)^2}\,[\mathrm{ch}(\Delta\Gamma t) + 1]}$ - time-dependent asymmetry characterizing a specific meson (6.4).

The integral value of CP-violating asymmetry is: $A_{CP} = \int A_{CP}(t) e^{-t/\tau} dt$ (6.5)

Table 6.1 presents the oscillation parameters: $2\Delta m = m_S - m_L$, $2\Delta\Gamma = \Gamma_S - \Gamma_L$ and $\Gamma$, as well as the CP-violating parameter $\delta\Gamma$, equal to $\delta\Gamma \approx -(1 - 2\delta^2) \times 2\zeta \times \Gamma \approx (5.4 \pm 1.8) \times 10^{-3} \times \Gamma$.

Time-dependent asymmetry functions $A(t)$ for all mesons will be presented later.

Table 6 .1.

| Meson | $2\Delta m[eV]$ | $2\Delta\Gamma[eV]$ | $\Gamma$ , $\delta\Gamma[eV]$ | Exact $A_{CP}$ | Approx $A_{CP}$ | $\Gamma / 2\Delta m = T_{osc} / \tau_{dec}$ |
|---|---|---|---|---|---|---|
| $K^0$ | $3.48 \times 10^{-6}$ | $7.3 \times 10^{-6}$ | $3.68 \times 10^{-6}$ $(1.9 \times 10^{-8})$ | $5.5 \times 10^{-3}$ | $5.6 \times 10^{-3}$ | 2.13 |
| $D^0$ | $6.5 \times 10^{-6}$ | $2.2 \times 10^{-5}$ | $1.61 \times 10^{-3}$ $(8.5 \times 10^{-6})$ | $5.3 \times 10^{-3}$ | $5.3 \times 10^{-3}$ | 250 |
| $B^0$ | $3.3 \times 10^{-4}$ | $4.0 \times 10^{-7}$ | $4.33 \times 10^{-4}$ $(2.3 \times 10^{-6})$ | $3.3 \times 10^{-3}$ | $3.4 \times 10^{-3}$ | 1.32 |
| $B_S^0$ | $1.2 \times 10^{-2}$ | $5.4 \times 10^{-5}$ | $4.33 \times 10^{-4}$ $(2.3 \times 10^{-6})$ | $\mathbf{6.9 \times 10^{-6}}$ | $\mathbf{6.9 \times 10^{-6}}$ | 0.036 |

In these calculations, the values from PDG [4] were used, presented in Tables 6.2 and 6.3.

Table 6.2.

| | Values in calculations | | Data from PDG | |
|---|---|---|---|---|
| Meson | $m[eV]$ | $2\Delta m[eV]$ | $m$ | $2\Delta m$ |
| $K^0$ | $497.611 \times 10^6$ | $3.48 \times 10^{-6}$ | $K^0$ MASS VALUE (MeV) 497.611±0.013 | $m_{K_L^0} - m_{K_S^0}$ VALUE ($10^{10}\ \hbar\ s^{-1}$) 0.5293 ±0.0009 $\Delta m = (3.48 \pm 0.01) \times 10^{-6} eV$ |
| $D^0$ | $1864.84 \times 10^6$ | $6.5 \times 10^{-6}$ | $D^0$ MASS VALUE (MeV) 1864.84 ±0.05 | $\lvert m_{D_1^0} - m_{D_2^0} \rvert$ VALUE ($10^{10}\ \hbar\ s^{-1}$) 0.997±0.116 $\Delta m = (6.5 \pm 0.8) \times 10^{-6} eV$ |
| $B^0$ | $5279 \times 10^6$ | $3.3 \times 10^{-4}$ | $B^0$ MASS VALUE (MeV) 5279.72±0.08 | $\Delta m_{B^0} = m_{B_H^0} - m_{B_L^0}$ VALUE ($10^{12}\ \hbar\ s^{-1}$) 0.5069±0.0019 $\Delta m = (3.34 \pm 0.01) \times 10^{-4} eV$ |
| $B_S^0$ | $5366 \times 10^6$ | $1.2 \times 10^{-2}$ | $B_s^0$ MASS VALUE (MeV) 5366.93± 0.10 | $\Delta m_{B_s^0} = m_{B_{sH}^0} - m_{B_{sL}^0}$ VALUE ($10^{12}\ \hbar\ s^{-1}$) 17.765 ±0.006 $\Delta m = (1.17 \pm 0.04) \times 10^{-2} eV$ |

Table 6.3.

| | Values in calculations | | Data from PDG | | |
|---|---|---|---|---|---|
| Meson | $\Gamma[eV]$ | $2\Delta\Gamma[eV]$ | $\Gamma$ | | $2\Delta\Gamma$ |
| $K_s^0$ | $7.35 \times 10^{-6}$ | $7.35 \times 10^{-6}$ | $K_L^0$ MEAN LIFE VALUE ($10^{-8}$ s) 5.116±0.021 $\Gamma = (1.29 \pm 0.03) \times 10^{-8} eV$ | $K_S^0$ MEAN LIFE VALUE ($10^{-10}$ s) 0.8954 ±0.0004 $\Gamma = (7.35 \pm 0.01) \times 10^{-6} eV$ | $\Gamma_S \gg \Gamma_L \to \Delta\Gamma \approx \Gamma_S$ $\Gamma = (7.35 \pm 0.01) \times 10^{-6}\ eV$ |
| $K_L^0$ | $1.29 \times 10^{-8}$ | | | | |

| | | | | |
|---|---|---|---|---|
| $D^0$ | 1.60 <br>$\times 10^{-6}$ | 2.2 <br>$\times 10^{-5}$ | $D^0$ MEAN LIFE VALUE ($10^{-15}$ s) 410.3± 1.0 <br>$\boldsymbol{\Gamma = (1.60 \pm 0.01) \times 10^{-6} eV}$ | $(\Gamma_{D_1^0} - \Gamma_{D_2^0})/\Gamma$ VALUE (units $10^{-2}$) 1.394± 0.056 <br>$\boldsymbol{\Delta\Gamma = (2.2 \pm 0.01) \times 10^{-5}\ eV}$ |
| $B^0$ | 4.34 <br>$\times 10^{-4}$ | 4.0 <br>$\times 10^{-7}$ | $B^0$ MEAN LIFE VALUE ($10^{-12}$ s) 1.517±0.004 <br>$\boldsymbol{\Gamma = (4.34 \pm 0.01) \times 10^{-4} eV}$ | $\Delta\Gamma_{B_d^0} / \Gamma_{B_d^0}$ VALUE (units $10^{-2}$) 0.1 ±1.0 <br>$\boldsymbol{\Delta\Delta\Gamma = (4 \pm 40) \times 10^{-7}\ eV}$ |
| $B_S^0$ | 4.34 <br>$\times 10^{-4}$ | 5.4 <br>$\times 10^{-5}$ | $B_s^0$ MEAN LIFE VALUE ($10^{-12}$ s) 1.516±0.006 <br>$\boldsymbol{\Gamma = (4.34 \pm 0.01) \times 10^{-4} eV}$ | $\Delta\Gamma_{B_s^0}/\Gamma_{B_s^0}$ VALUE 0.124±0.007 <br>$\boldsymbol{\Delta\Gamma = (5.4 \pm 0.3) \times 10^{-5}\ eV}$ |

Let us discuss the results of Table 6.1. It can be seen that for $K^0$ -meson, $D^0$–meson and $B^0$–meson we have integral CP asymmetries that are practically determined by the value of the asymmetry: $A^{LRAM} \approx -\left(1-2\delta^2\right)\times 2\zeta = 5.4\times 10^{-3}$.

For $K^0$ -meson, the calculation of the integral asymmetry yielded the result $A_{CP}^{K^0} = 5.4 \times 10^{-3}$. The calculation of integral asymmetries is somewhat simplified. The fact is that for $K^0$ -meson, the CP violation effect can be represented as a mixture of two effects: the mixing effect for $K_S^0$ -meson, when the lifetime is much shorter than the oscillation period of $K_L^0$-meson and the effect in the final state for $K_L^0$ -meson, when the lifetime practically coincides with the oscillation period. Indeed, there is an experimental result for mixing: $A_T^{exp} = (6.6 \pm 1.3 \pm 1.0) \times 10^{-3}$[44], in addition, there is an experimental result for the decay in the final state: $A_{\mathrm{CP}}^{exp} = (3.32 \pm 0.06) \times 10^{-3}$[4], when the initial asymmetry is partially compensated by the oscillation process.

Thus, criterion #2 is satisfied for $K^0$ -meson (the integral effect of CP violation appears when the oscillation period is significantly longer than the decay time). In this case, $\Gamma / 2\Delta m = T_{osc} / \tau_{dec}$ =2.13. Moreover, for $K^0$ -meson, in the same work [39], there is an experimental dependence of the CP-violating asymmetry effect on the decay time - $A_{CP}(t)$. It is shown in Fig. 4.1 in the interval of 20 $K_S$ meson lifetimes. The average value of the asymmetry is $(6.6 \pm 1.3) \times 10^{-3}$[39]. Our calculated dependence of the CP-violating asymmetry effect on the decay time - $A_{CP}(t)$ is shown in the same figure by the blue line. It can be concluded that the experimental results for $K^0$ -meson, within the limits of the available accuracy, can be described in the left-right asymmetry model of weak interaction.

But for $B_S^0$ -meson the CP violation effect is equal to $6.9 \times 10^{-6}$ , since in the process of oscillations of $B_S^0$ meson the CP asymmetry is averaged out. However, it should be noted that the CP-violating process existed throughout the entire oscillation process. This is because the sign of the CP-violating effect reversed during each oscillation period, according to the formula $\varepsilon_{p\bar{p}} \approx e^{-\Gamma_0 t} \sin(2\Delta m t)\ (\frac{\delta\Gamma}{2\Delta m})$, and was therefore compensated. Experimental observations with more statistical data in 2021 confirmed the absence of asymmetry [43] with an accuracy of $1.1 \cdot 10^{-3}$. (Fig. 6.2).

For $D^0$ -meson, the ratio of the decay rate to the oscillation frequency is $\Gamma / 2\Delta m = T_{осц} / \tau_{расп} = 2.5 \times 10^2$, so the oscillation process has time to decay before it even begins. In this case, criterion #2 is satisfied with good accuracy for $\mathrm{D}^0$ -meson (The integral effect of CP violation manifests itself when the oscillation period is significantly longer than the decay time). The experimental value of the CP violation effect is A = (6.8±0.65) $10^{-3}$ [44], and the calculated value $A_{CP}^{D^0} = 5.3 \times 10^{-3}$. The calculated dependence of the CP violation effect on the decay time, which was used for integration, is shown in Fig. 6.3. As can be seen, the asymmetry at large times ($2 \cdot 10^{-10}$ s) reaches 68%, but the decay time of $D^0$-meson is $\tau(D^0) = 4.1\times 10^{-13}\,s$ , therefore, only the initial section of the differential asymmetry at $\tau(D^0) = 4.1\times 10^{-13}\,s$ contributes to the integral value of the asymmetry. Thus, $A_{CP}^{D^0} = \int A_{CP}(t) e^{-t/\tau} dt = 5.3 \times 10^{-3}$.

The calculated integral asymmetry taking into account the error $A_T^{LRAM}$ gives the value $A_{CP}^{D^0}$ =( $5.4 \pm 1.8) \times 10^{-3}$. It can be concluded that the experimental results for $D^0$ -meson, within the limits of available accuracy, can be described in the left-right model of weak interaction.

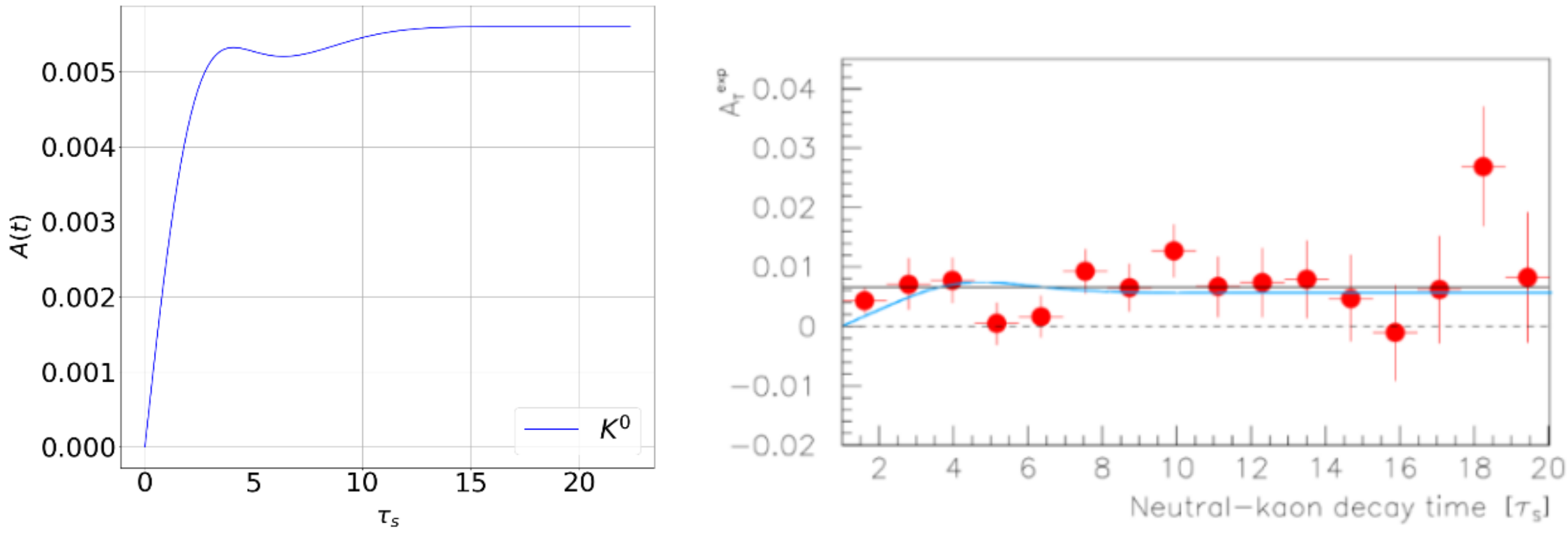

Fig. 6.1. Calculated dependence of the CP-violating asymmetry effect on the decay moment $A_{CP}(t)$ for $K^0$ -meson.

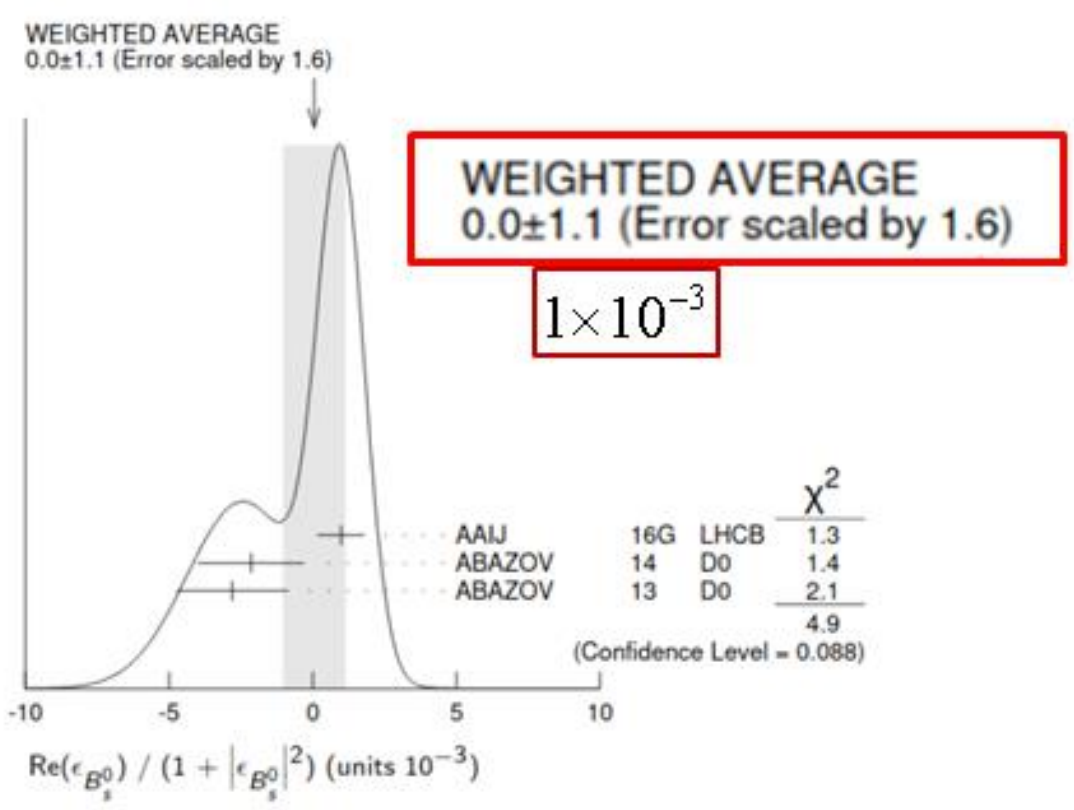

Fig. 6.2. Experimental data from PDG for the integral effect of CP violation for $B_S^0$ -meson.

Fig. 6.3. Calculated dependence of the CP violation effect on the decay time for $D^0$ -meson, which was used for integration.

For $B^0$ -meson, the ratio of the decay rate to the oscillation frequency is $\Gamma / 2\Delta m = T_{osc} / \tau_{dec} = 1.32$, so the oscillation process is on the verge of satisfying criterion #2 (The integral effect of CP violation manifests itself when the oscillation period is significantly longer than the decay time), while criterion #3 is quite clearly satisfied (when the differential effect of CP violation changes the sign of the effect to the opposite upon transition from particle to antiparticle). The experimental effect of CP violation as a function of the decay time is shown in Fig. 6.4, and the calculated time-dependent component of the asymmetry:

$$A_{asym}(t) = \frac{(\Delta\Gamma)\mathrm{sh}(\Delta\Gamma t) + (2\Delta m)\sin(2\Delta m t)}{\sqrt{(\Delta\Gamma)^2 + (2\Delta m)^2}[\mathrm{ch}(\Delta\Gamma t) + 1]}$$

, characterizing $B^0$ meson is shown in the lower Fig. 6.4. As can be seen from Fig. 6.4, the experimental time-dependent asymmetry reaches a fairly large level - much larger than the average value of the CP-violating asymmetry $A^{LR}$ for all decay modes during the oscillation process $B^0 - \bar{B}^0$. However, it should be noted that the experimental result is presented for a decay mode that accounts for only $5 \cdot 10^{-5}$ of the total decay probability of $B^0$ meson. Therefore, the contribution to the average integral value of the CP-violating asymmetry $A_T^{LR}$ for all decay modes during the oscillation process $B^0 - \bar{B}^0$ will be significantly suppressed. We additionally note that the average integral value of the CP-violating asymmetry for $B^0$ -meson in the PDG data is 0.005 ±0.012±0.014 [4]. Thus, for $B^0$ -meson we have fairly clear experimental confirmation that criterion No. 3 is met, i.e., a change in the sign of CP violation occurs upon the transition from $B^0$ to $\bar{B}^0$.

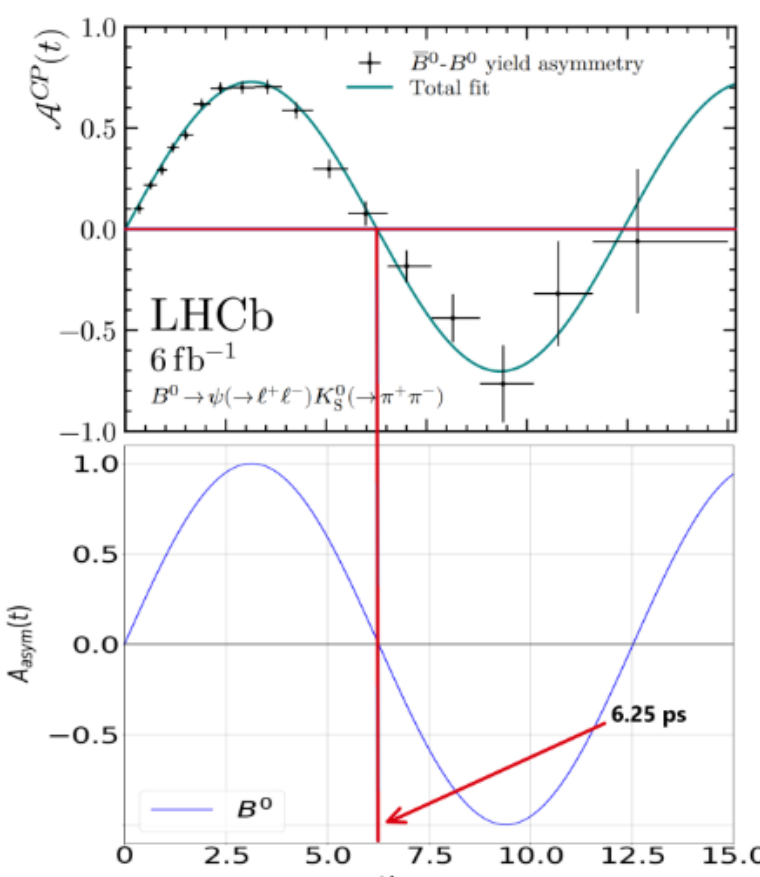

Fig. 6.4. Comparison of the results of measuring the time dependence for the decay asymmetry $B^0 \to \psi K_S^0$ from work [45] and the calculated time-dependent component of the asymmetry $A_{asym}(t)$ for $B^0$ meson within the left-right model.

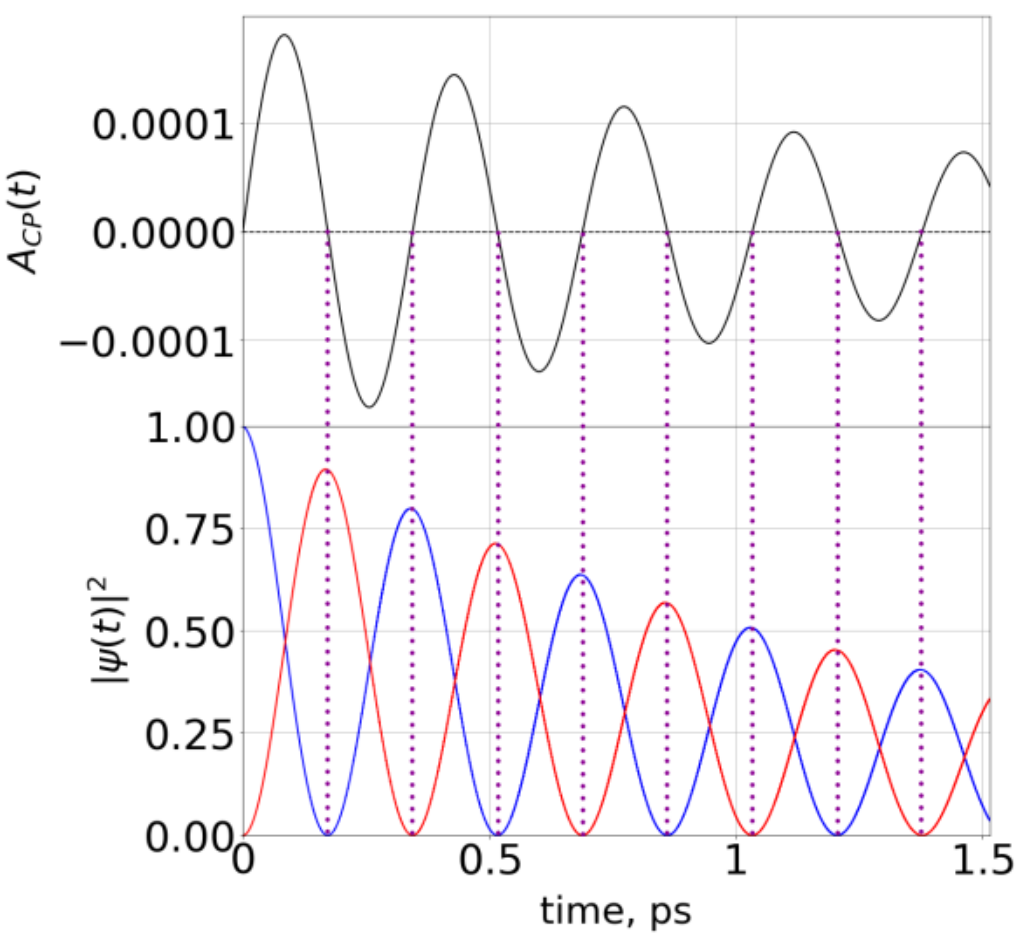

Fig. 6.5. CP-violating asymmetry for $B_S^0$ -meson and the probabilities of states during oscillations for the particle (blue curve) and antiparticle (red curve). Clear synchronization in the change of sign of CP violation for $B_S^0$ -meson (shown by the dotted line).

Finally, for $B_S^0$ -meson, the ratio of the decay rate and the oscillation frequency is $\Gamma / 2\Delta m = T_{osc} / \tau_{dec} = 3.6 \times 10^{-2}$, i.e., much less than unity, and therefore the oscillation process is represented by a large number of periods. However, due to the high frequency of oscillations, it is difficult to obtain good statistical significance experimentally for $B^0$ -meson. A clear synchronization of the change in the sign of CP violation is demonstrated in calculations for $B_S^0$ -meson in Fig. 6.5.

As can be seen, the positive sign of CP violation changes to a negative sign at 0.17 ps, when the process of transitions from particles to antiparticles changes to a process of transitions from antiparticles to particles. Moreover, the new sign of the CP violation process is determined by the direction of the transition: from plus to minus during the transition from particles to antiparticles (at 0.17 ps) and from minus to plus during the transition from antiparticles to particles (at 2 $\times$ 0.17 ps). This constitutes the essence of the CP-violation process, since the sign of the CP violation depends on the direction of the process: a transition from particles to antiparticles or vice versa.

Thus, it can be concluded that the calculation results within the left-right asymmetry model are confirmed by experimental results. In the left-right asymmetry model with CP violation, a change in the sign of mixing with the right vector boson was incorporated in the transition from $W^-$(particle) to $W^+$(antiparticle). This leads to the fact that when mixing with quarks through $W^-$ the CP violation process with a positive sign occurs, and when mixing with quarks through $W^+$ the CP violation process with a negative sign occurs. The nature of CP violation is associated with the presence of an admixture of the right vector boson, with different signs of the mixing angle. It is shown that within the framework of this model, the effects of CP violation in the decays of $K^0$ -mesons, $D^0$ -mesons, $B^0$ -mesons, and $B_s^0$ -mesons can be successfully described. Moreover, the sign of CP violation is opposite for particles and antiparticles, i.e. the phase of CP violation is equal to $\pm \pi / 2$, which corresponds to 100% CP violation.

This situation with the mixing phase $\pi / 2$ in neutral mesons is reminiscent of the situation with the maximum mixing phase $\pi / 2$ in neutrino oscillations - in the T2K [46] and NOvA [47] experiments. Judging by the idea of the nature of CP violation in the left-right asymmetry model, such a generalization is possible.

In the left-right asymmetry model, CP -violation has a special structure. It is associated with the mixing of left- and right currents (parameter ζ). This leads to a phase of $\pm \pi/2$ in the effective Hamiltonians (as we saw in the matrices (5.9)). Such a phase can manifest itself in both meson oscillations and neutrino oscillations, hence the analogy. The key idea is that if in the left-right asymmetry model CP violation is arranged so that it yields a phase of $\pm \pi /2$, then this model can simultaneously explain both CP asymmetry in mesons and CP violation in neutrino oscillations. T2K and NOvA [46,47] are experiments studying neutrino oscillations. They measure the transition probabilities: $\nu_\mu \to \nu_e$ and $\bar{\nu}_\mu \to \bar{\nu}_e$, and compare them to find the CP -asymmetry:

$$A_{CP}^{\nu} = \frac{P(\nu_\mu \to \nu_e) - P(\bar{\nu}_\mu \to \bar{\nu}_e)}{P(\nu_\mu \to \nu_e) + P(\bar{\nu}_\mu \to \bar{\nu}_e)}, \tag{6.6}$$

which is analogous to the asymmetry measured in mesons:

$$A_{CP}^{M} = \frac{\mathrm{P}(M^0 \to f) - \mathrm{P}(\bar{M}^0 \to f)}{\mathrm{P}(M^0 \to f) + \mathrm{P}(\bar{M}^0 \to f)} \quad (6.7)$$

The left-right model provides a single mechanism of CP-violation, which works for both quarks (mesons, nucleons) and leptons (neutrinos). This is no mere coincidence: in both systems, CP-violation arises from the same structure—the mixing of left- and right currents. Therefore, if the model contains a parameter ζ, which gives a phase of $\pm\pi/2$, it can simultaneously explain both effects.

In addition, there is what is called the gallium anomaly [48]. The gallium anomaly is the observed deficit in the number of electron neutrinos detected in gallium experiments compared to the theoretical prediction. Recently, the BEST experiment [49] confirmed a deficit of $\sim 20\text{–}24\ \%$. However, such a large deficit is not observed in experiments with reactor antineutrinos (DANSS experiment) [50]. In particular, the KATRIN experiment [51] with tritium antineutrinos does not confirm the possible mixing angles for neutrinos in the gallium anomaly. Thus, we have an asymmetry in the processes between neutrinos and antineutrinos, which can be explained by the CP violation effect, which has different signs within the left-right asymmetry model for particles and antiparticles, and in this case between neutrinos and antiteutrinos. Figure 4.6 in Section 4 showed that CP-violation effects have different probabilities in the LRAM, including between neutrinos and antineutrinos. To test this hypothesis, an experiment with an artificial antineutrino source should be conducted at the gallium setup.

Finally, this analogy can be continued for the case of neutron-antineutron asymmetry, i.e. for CP violation in nucleons:

$$A_{CP}^{n} = \frac{P(n \to p, e^-, \bar{\nu}_e) - P(\bar{n} \to \bar{p}, e^+, \nu_e)}{P(n \to p, e^-, \bar{\nu}_e) + P(\bar{n} \to \bar{p}, e^+, \nu_e)} \quad (6.8)$$

In this case, there is an asymmetry in the decay of neutrons and antineutrons, meaning neutron-antineutron oscillations and a difference in lifetimes for neutrons and antineutrons are possible. As we have already noted earlier, the absence of experimental observation [18] of neutron-antineutron oscillations does not mean the absence of a difference in the lifetimes of a neutron and an antineutron. Indeed, we have an example of such a situation for $D^0$ meson, which we assume for a neutron. In this section, it was shown that for $D^0$ meson, the oscillation period is 250 times longer than the lifetime, and CP violation is observed due to the difference in the lifetimes of $D^0$ and $\tilde{D}^0$.

The difference in lifetime between the neutron and antineutron is of great importance for the formation of the baryon asymmetry of the Universe, so we will consider this issue in detail in the next section.

## 7. Baryon asymmetry of the Universe

The baryon asymmetry of the Universe is a fundamental question in physics—the main mystery of particle physics. Baryon asymmetry is the observed predominance of matter (baryons) over antimatter (antibaryons) in the Universe. When a particle and antiparticle meet, annihilation occurs—they mutually annihilate with the release of energy. If matter and antimatter had been equal in the early Universe, they would have completely annihilated, and today the Universe would be filled only with photons and neutrinos. But we observe a world composed of matter, meaning that initially there was an excess of matter. The standard model of particle physics in its classical form cannot explain the observed asymmetry. It assumes symmetry between matter and antimatter, which contradicts reality.

The main idea that offers an explanation for the baryon asymmetry of the Universe belongs to A.D. Sakharov [52,53]. More precisely, he formulated three necessary conditions for the emergence of baryon asymmetry in the famous article of A.D. Sakharov [52,53]. In this article, the following are formulated: (quote)

"Three main prerequisites for the cosmological formation of baryon asymmetry (BA).

**I.** Absence of the law of conservation of baryon charge.

**II.** The difference between particles and antiparticles, manifested in the violation of CP invariance.

**III.** Non-stationarity. The formation of BA is possible only under non-stationary conditions in the absence of local thermodynamic equilibrium."

In an equilibrium system, processes and their inverses proceed at the same rate, preventing the accumulation of excess matter. The early Universe must have gone through a phase of nonequilibrium expansion.

Now, assuming that the nature of CP violation is related to the difference between left- and right weak interactions, we can propose the following scenario for the formation of the baryon asymmetry of the Universe within the framework of the extended left-right model. At temperatures of the order of $10^4$ GeV , the degree of symmetry between left- and right processes (processes with opposite CP-parity) was quite high. However, as the temperature decreased below the mass of $W_R$ , an advantage arose in processes involving $W_L$ , but with an admixture of $W_R$ with an opposite sign of the mixing angle for $W^+$ and $W^-$ , leading to CP-violation.

Thus, within the left-right asymmetry model, condition " **II.** Distinction of particles from antiparticles, manifested in CP violation" is satisfied due to processes involving $W_R$ with different sign of mixing angle for $W^+$ and $W^-$ . For baryon number violation within the left-right asymmetry model, we have a difference in the neutron and antineutron lifetimes, which can be calculated from the relations:

$$\left(f\tau_n\right)^{-1} \approx G_F^2 \left|V_{ud}^{LRAM}\right|^2 \left(1+3\lambda_n^2\right)\left[1+\delta^2-\left(1-\delta^2\right)\frac{\left(1-3\lambda_n^2\right)}{\left(1+3\lambda_n^2\right)}\sin 2\zeta\right] \qquad (7.1)$$

$$\left(f\tilde{\tau}_n\right)^{-1} \approx G_F^2 \left|V_{ud}^{LRAM}\right|^2 \left(1+3\lambda_n^2\right)\left[1+\delta^2+\left(1-\delta^2\right)\frac{\left(1-3\lambda_n^2\right)}{\left(1+3\lambda_n^2\right)}\sin 2\zeta\right] \qquad (7.2)$$

$$(\tau_n-\tilde{\tau}_n)/\tau_n \approx 2\times\left[\left(1-\delta^2\right)\frac{\left(1-3\lambda_n^2\right)}{\left(1+3\lambda_n^2\right)}\sin 2\zeta\right] \approx (7.1\pm 2.3)\times 10^{-3} \qquad (7.3)$$

Condition " **III.** Non-stationarity. The formation of BA is possible only under non-stationary conditions in the absence of local thermodynamic equilibrium." can also be satisfied, since the decay rate of neutrons and antineutrons is much smaller than the expansion rate of the Universe during the hadronization of quark-gluon plasma. Indeed, the decay rate of neutrons and antineutrons is of the order of $10^{-3}$ $s^{-1}$ , and the expansion rate of the Universe during the hadronization of quark-gluon plasma is $10^{3}$ $s^{-1}$ .
Thus, it is necessary to find a solution to the balance equation:

$$\frac{dn(t)}{dt}+3H(t)n(t)=-\frac{n(t)}{\tau} \quad (7.4) \qquad H(t)=\frac{0.53}{t[s]} \quad (7.5) \qquad \frac{dn(t)}{dt}+1.6\frac{n(t)}{t}=-\frac{n(t)}{\tau} \quad (7.6)$$

$\tau$ - particle lifetime, $n(t)$ - particle density. We use the Hubble value - $H(t)$ in the radiation-dominated stage. Let the particle density be $n_0$ at the initial moment $t_0$, then the solution yields the following relationships for the particle density and the asymmetry between neutrons and antineutrons.

$$n(t)=n_0\left(\frac{t}{t_0}\right)^{-1.6} e^{-\left(\frac{t-t_0}{\tau}\right)} \quad (7.7) \qquad A(t)=\frac{n(t)-\bar{n}(t)}{n(t)+\bar{n}(t)}=\frac{1-e^{-\left(t-t_0\right)\left(\frac{1}{\tau_n}-\frac{1}{\tilde{\tau}_n}\right)}}{1+e^{-\left(t-t_0\right)\left(\frac{1}{\tau_n}-\frac{1}{\tilde{\tau}_n}\right)}} \quad (7.8)$$

Since in (5.7) $t>t_0$, taking into account the expansion of space leads to the particle density decreasing faster than exponentially. For two different lifetimes of a particle and antiparticle— $\tau_n$ and $\tilde{\tau}_n$ ( $\tau_n>\tilde{\tau}_n$ )—and since the exponent is much less than unity, it turns out that the asymmetry is equal to:

$$A(t)=\frac{\left(t-t_0\right)}{2\tau_n}\frac{\tau_n-\tilde{\tau}_n}{\tau_n} \quad (7.9) \qquad A\left(t=10^{-3}\text{ s}\right)=-\frac{t}{2\tau}\frac{\tau_n-\tilde{\tau}_n}{\tau_n}=(3.55\pm 1.15)\times 10^{-9} \quad (7.10)$$

Let's make an order-of-magnitude estimate. The lifetime of a neutron and antineutron $\tau \sim 10^3$ , $t_0$- the moment of synthesis of neutrons and antineutrons is the process of hadronization of quarks at a temperature of 150 MeV; $t_0 \sim 10^{-5}$ s; $t$ is the current time. We choose the current time of $10^{-3}$ s, in accordance with the assumption that by this moment the process of hadron annihilation is almost complete. Thus, the asymmetry at the current time of $10^{-3}$ s is equal to $(3.55\pm 1.15)\times 10^{-9}$ . The current moment in time is determined by the completion of the hadronization process at approximately $10^{-3}$ s. And the beginning of the hadronization process is determined by the fact that during the expansion of space, free quarks combine into hadrons: mesons and baryons. This occurs approximately at $t_0 \sim 10^{-5}$ s. Thus,

$$A\left(t=10^{-3}s\right)=(3.55\pm 1.15)\times 10^{-9}=\frac{n-\bar{n}}{n+\bar{n}} \qquad (7.11)$$

Now we need to compare the obtained result with the observed asymmetry of the Universe based on the microwave background radiation. According to the most accurate modern data, the magnitude of the baryon asymmetry of the Universe is $\eta = \frac{n_B - n_{\bar{B}}}{n_\gamma} = 6.1 \times 10^{-10}$. This means that there are about $10^9$ relic photons per 1 baryon. The result we obtained $A\left(t = 10^{-3}s\right) = (3.55 \pm 1.15) \times 10^{-9}$ means that there were approximately $3 * 10^8$ (in total) neutrons and antineutrons, or (taking into account protons and antiprotons) $6 * 10^8$nucleons or $1.8 * 10^9$quarks. In principle, all quarks can annihilate, ultimately, into gamma quanta and neutrinos. Thus, there will be three times as many gamma quanta per surviving neutron as there are neutrons and antineutrons. Thus, the relationship between the observed asymmetry of the Universe in the microwave background radiation $\eta = \frac{n_B - n_{\bar{B}}}{n_\gamma}$ and the neutron-antineutron asymmetry at the end of the quark-gluon plasma hadronization process is given by the following relationship: $\eta = \frac{1}{3} A\left(t = 10^{-3}s\right) = (1.2 \pm 0.4) \times 10^{-9}$, and according to the relic radiation $\eta = \frac{n_B - n_{\bar{B}}}{n_\gamma} = 6.1 \times 10^{-10}$.

Of course, comparisons can only be made on an order-of-magnitude basis. The estimate of the number of gamma quanta is highly arbitrary, but examining the entire sequential process of formation of the cosmic microwave background radiation density is impossible. It should also be noted that the estimate of the completion of the hadronization process and the annihilation moment is approximate. However, in this analysis, a fundamental approach is important, explaining the origin of the baryon asymmetry of the Universe in the extended left-right weak interaction model.

As for the condition " **I.** Absence of the law of conservation of baryon charge." it is fulfilled due to the difference in the decay times of neutrons and antineutrons in the short time interval between the beginning and the end of the process of hadronization of quark-gluon plasma in the conditions of a rapidly expanding Universe.

In addition, in the article of A.D. Sakharov is discussed the violation of both baryon and lepton asymmetry, and notes the conservation of the B - L difference . (quote)

"If at temperatures exceeding the low-temperature region $T = 10^{2} - 10^{4}$ GeV, a baryon-lepton asymmetry with $B \neq L$ arises— then in the low-temperature region, with high accuracy, a state will be established that corresponds to the maximum value of entropy for a given constant value $B — L$ = const (and when the condition of electroneutrality is met).

Note that the right neutrino's mass $W_R$ lies in the specified range of $T = 10^2 - 10^4$ GeV , so the appearance of sterile neutrinos in this scenario is easy to predict. Indeed, the existence of a right $W_R$ presupposes the presence of right (so-called sterile) antineutrinos, which have a significantly larger mass than active neutrinos.

In the left-right asymmetry model, non-conservation of baryon and lepton numbers can occur due to the fact that heavy (1 – 100 keV) sterile neutrinos do not thermalize and leave the cosmic plasma, carrying with them a lepton asymmetry with a sign opposite to the baryon asymmetry. However, strongly interacting baryons remain in the cosmic plasma. Thus, the ratio between the number of baryons and leptons in the cosmic plasma is disrupted. Baryon-lepton asymmetry arises, with B - L conservation .

In the left-right asymmetry model, deviation from thermodynamic equilibrium occurs again due to the escape of sterile neutrinos from the cosmic plasma. This process occurs under nonequilibrium conditions as the universe cools, preventing the reverse reaction and perpetuating the asymmetry.

Thus, in the left-right asymmetry model, all three conditions of A.D. Sakharov's theorem are met. Sterile neutrinos play an important role. In this regard, let us consider the process of dark matter formation due to sterile neutrinos.

## 8. The process of dark matter formation by sterile neutrinos

There must be the same number of sterile neutrinos as active neutrinos (Fig. 8.1). They can mix with active neutrinos, primarily with chirality preserved, but the lepton number changes by two units $\Delta L = 2$. This is important for fulfilling A.D. Sakharov's lepton number violation condition.

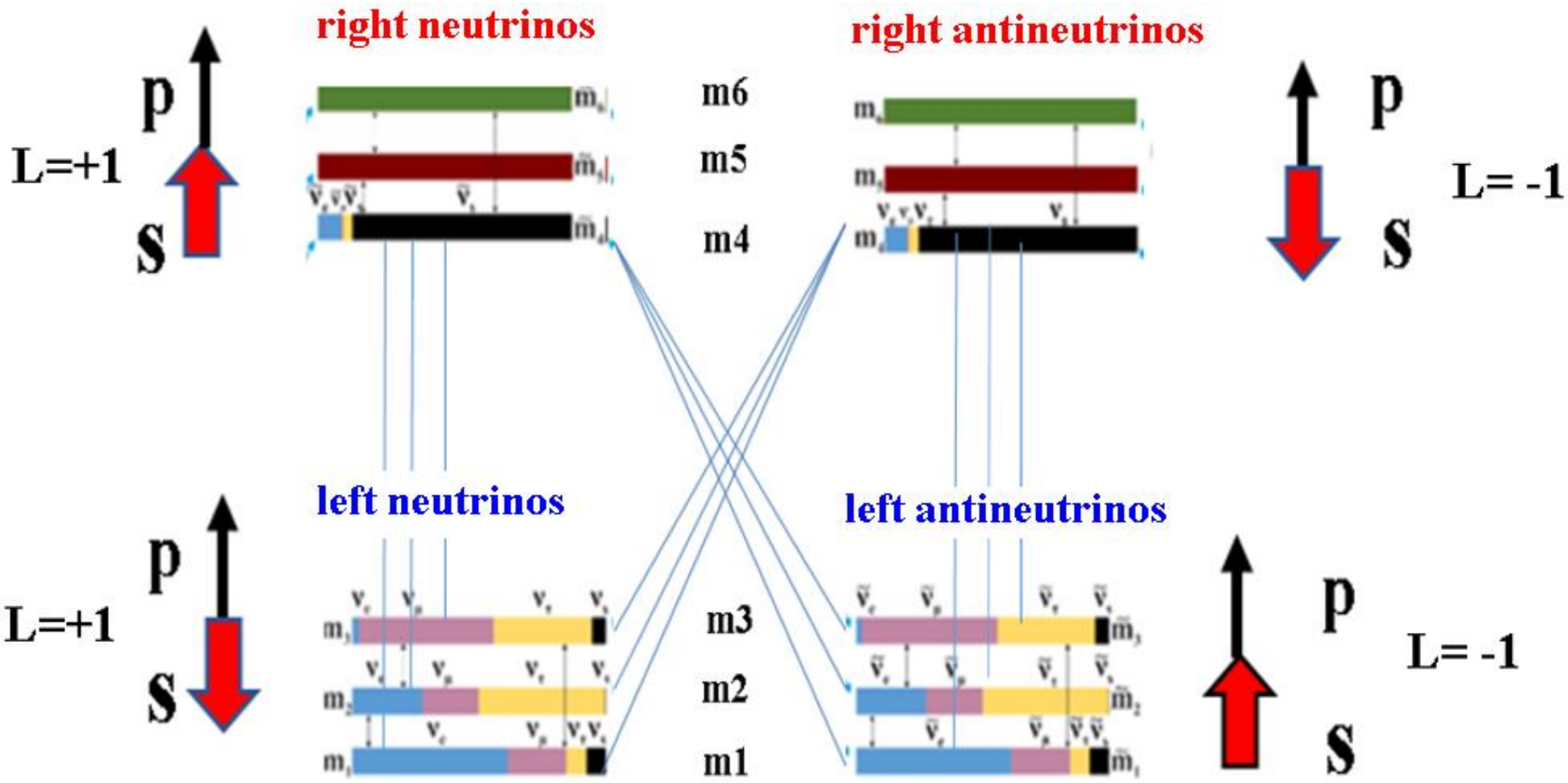

Fig. 8.1. Scheme of active and sterile neutrinos.

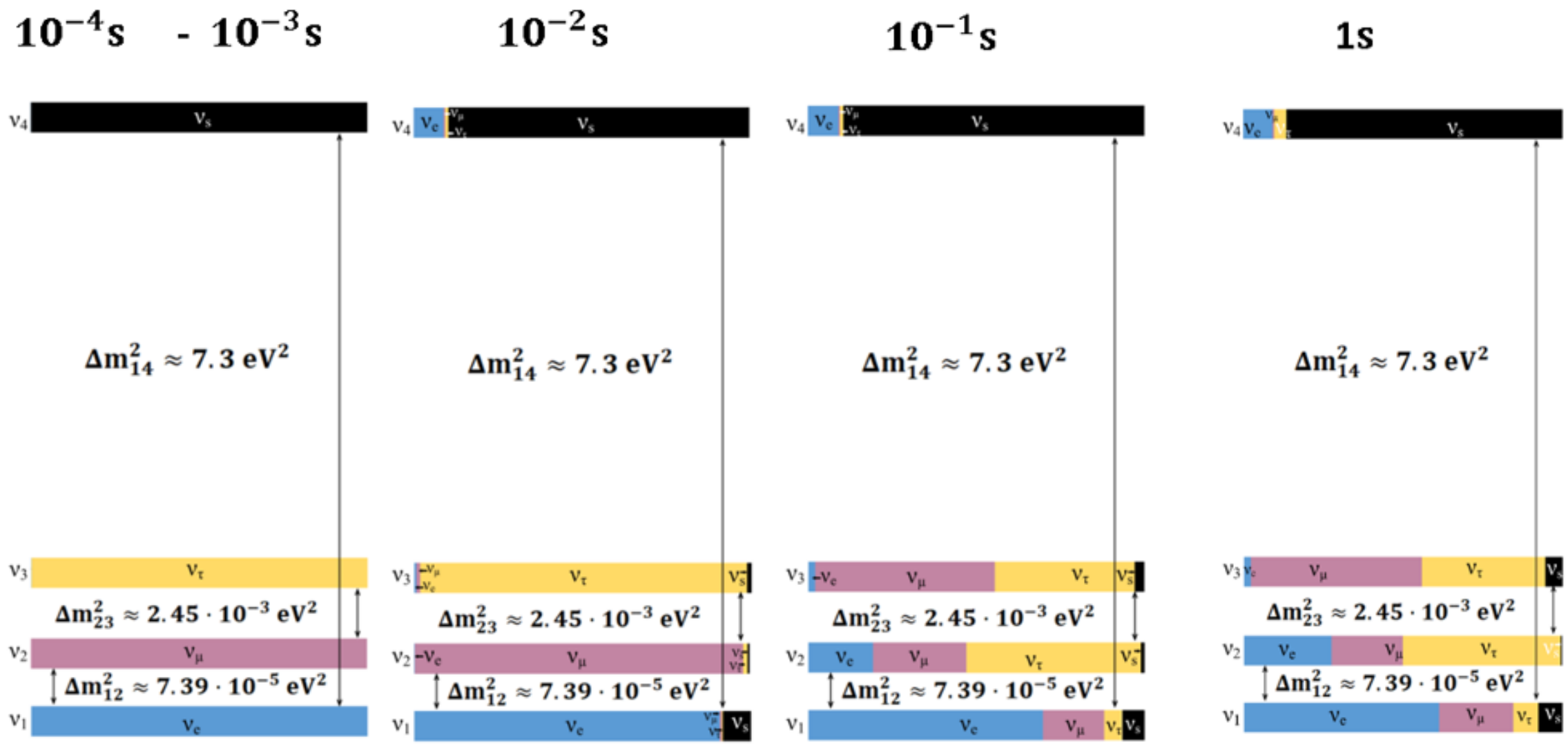

Fig. 8.2. Scheme of mixing of active and sterile neutrinos.

The scheme of mixing the flavors of active neutrinos and sterile neutrinos is illustrated in Fig. 8.1. Mass states $m_1$, $m_2$, $m_3$ are a mixture of electron, muon, and tau flavors with a small fraction of the sterile state. Therefore, the mass states $m_1$, $m_2$,$m_3$ have weak interactions, while mass states $m_4$, $m_5$, $m_6$are mostly sterile and exhibit weak interactions only due to the small contribution of electron, muon, and tau flavors. As a result, sterile neutrinos $m_4$, $m_5$, $m_6$ arising from active neutrino oscillations will propagate in the cosmic plasma for a considerable time before interacting and transforming back into active neutrinos. This circumstance creates the possibility of accumulation of sterile neutrinos and the possibility of their separation from the cosmic plasma at an earlier stage than active neutrinos.

In what follows, we will consider a simplified scheme for mixing three active neutrinos with one sterile neutrino and apply this scheme three times for all sterile neutrinos. This significantly simplifies the problem.

In our work [54], we considered the process of interaction of sterile and active neutrinos in cosmic plasma in this 3+1 model. In this case, the mixing matrix was supplemented with one column and row for one sterile neutrino.

As is well known, the process of neutrino oscillations in matter is altered by the interaction of neutrinos with matter (the Mikheev-Smirnov-Wolfenstein effect). This process is particularly pronounced in cosmic plasma. Here, we present only a pictorial representation of how the mixing process in the plasma potential changes due to the weak interaction between neutrinos (Fig. 8.2). Detailed calculations can be found in our paper [54].

As can be seen from Fig. 8.1 the plasma potential strongly suppresses the oscillations of both active and sterile neutrinos. The dynamics of the sterile neutrino density are influenced by three processes: 1) the expansion of the universe, 2) transitions of active neutrinos into sterile ones as a result of collisions and 3) reverse transitions of a sterile neutrino to an active state. The reverse transition of a sterile neutrino is considered as an oscillation of the sterile state into an active state, followed by interaction of the active component.

Below is an equation that takes into account the generation of sterile neutrinos and their "sink." Equation (8.1) includes the effective interaction of the sterile neutrino with the plasma due to oscillations:

$$\frac{dn_{\nu_s}}{dt} + 3Hn_{\nu_s} = \frac{1}{2}\left(\frac{\sin^2 2\theta_{m\,14}\, n_{\nu_e}}{\tau_{\nu_e}} + \frac{\sin^2 2\theta_{m\,24}\, n_{\nu_\mu}}{\tau_{\nu_\mu}} + \frac{sin^2 2\theta_{m\,34}\, n_{\nu_\tau}}{\tau_{\nu_\tau}}\right) - \frac{1}{2}\left(\frac{\sin^2 2\theta_{m\,14}}{\tau_{\nu_e}} + \frac{\sin^2 2\theta_{m\,24}}{\tau_{\nu_\mu}} + \frac{\sin^2 2\theta_{m\,34}}{\tau_{\nu_\tau}}\right) n_{\nu_s} \quad (8.1)$$

where $n_{\nu_s}$, $n_{\nu_e}$, $n_{\nu_\mu}$and $n_{\nu_\tau}$— the densities of sterile, electron, muon, and tau neutrinos corresponding to the Fermi-Dirac distribution with zero chemical potential [55]. We used the following values for the squares of the sines of the double angle $\sin^2 2\theta_{14} = 0.36$, $\sin^2 2\theta_{24} = 0.024$ and $\sin^2 2\theta_{34} = 0.043$. These are estimates from the Neutrino-4 and IceCube experiments [54, 56].

Equation (8.1) is a simplification that ignores the effect of transitions of active neutrinos to sterile ones on the density of active neutrinos themselves. It considers only the transition to sterile neutrinos followed by the transition of sterile neutrinos to active ones. The initial conditions are set to zero, indicating the density of sterile neutrinos at the initial moment of time.

This equation can be applied up to the neutrino quenching temperature, that is, up to the temperature at which the density decreases so much that the interaction of neutrinos with plasma matter can be neglected. At this point, the interaction of neutrinos with matter ceases, and only the expansion of space influences the subsequent density dynamics.

The rate of increase in the sterile neutrino density is determined by the difference in the probabilities of sterile neutrino production and annihilation. Both processes are proportional to the amplitude of electron-neutrino-to-sterile neutrino oscillations (and vice versa) with a factor of 1/2. The process of sterile neutrino production is proportional to the electron-neutrino density $n_{\nu_e}$ and the electron-neutrino interaction frequency $\frac{1}{\tau_{\nu_e}}$. The process of sterile neutrino-to-electron neutrino conversion is proportional to the sterile neutrino density $n_{\nu_s}$ and the electron-neutrino interaction frequency $\frac{1}{\tau_{\nu_e}}$. Therefore, the factor $\frac{1}{2}\frac{\sin^2 2\theta_{m\,14}}{\tau_{\nu_e}}$ is included in both the generation and the "sink" of sterile neutrinos.

The electron neutrino density $n_{\nu_e}(T)$ and $H(T)$ depend on temperature, and temperature depends on time. The oscillation time depends on density, density on temperature, and temperature on time. All these dependencies are taken into account to solve the differential equation over time.

$$n_{\nu_e}(T) = \frac{3}{4}\frac{\zeta(3)}{\pi^2}T^3 \qquad H(T) = \frac{T^2}{M_{Pl}^*} \qquad \tau_{osc} = \tau_0 \frac{sin2\theta_m}{sin2\theta_0} \quad (8.2)$$

The equation also includes the Hubble parameter H, which depends on the number of relativistic degrees of freedom. We use the value 43/4 given in the PDG review for temperatures below the muon mass. In the ultrarelativistic case, the Hubble constant is related to the temperature by the following expression (8.2), where $M_{Pl}^*$ is the reduced Planck mass (formula (3.32) from [55]), $\tau_0$ is the period of oscillations in vacuum $\tau_0 = 4\pi E/\Delta m^2$, $\Delta m^2$ – the difference between the squares of the neutrino masses, $\sin 2\theta_m$ is the sine of the double mixing angle of two neutrinos in the plasma, $\sin 2\theta_0$ - sine of twice the mixing angle of two neutrinos in a vacuum.

We are interested in the question of what mixing angles for heavy neutrinos account for the sterile neutrino contribution to dark matter below the 25% limit. The calculation result is shown in Figure 8.3a). This result indicates that heavy sterile neutrinos must have small mixing angles to comply with cosmological constraints on the total contribution of dark matter to the energy density of the Universe. This dependence of the angle on mass can be explained simply.

Figure 8.3b) shows that light sterile neutrinos at sufficiently large mixing angles reach thermodynamic equilibrium with the plasma, and their density becomes comparable to the density of active neutrinos. However, for large masses, thermodynamic equilibrium is unacceptable, as it would lead to exceeding the 25% threshold. Therefore, to remain within the limit, the mixing angle must decrease as the neutrino mass increases.

A decrease in the mixing angle results in the sterile neutrino not having time to reach equilibrium with the electron neutrino before the neutrino separates from the plasma, i.e., the ratio $n_{\nu_s}/n_{\nu_e}$remains less than 1. In Figure 8.3a), 7 points are highlighted on the plane ( $\sin^2 2\theta_{14}$ , $\Delta m^2_{14}$), which correspond to the points in Figure 8.3b), where the curves of the ratio of the number of sterile neutrinos to the number of electron neutrinos are plotted. With a decrease in the mixing angle and an increase in mass, this ratio decreases.

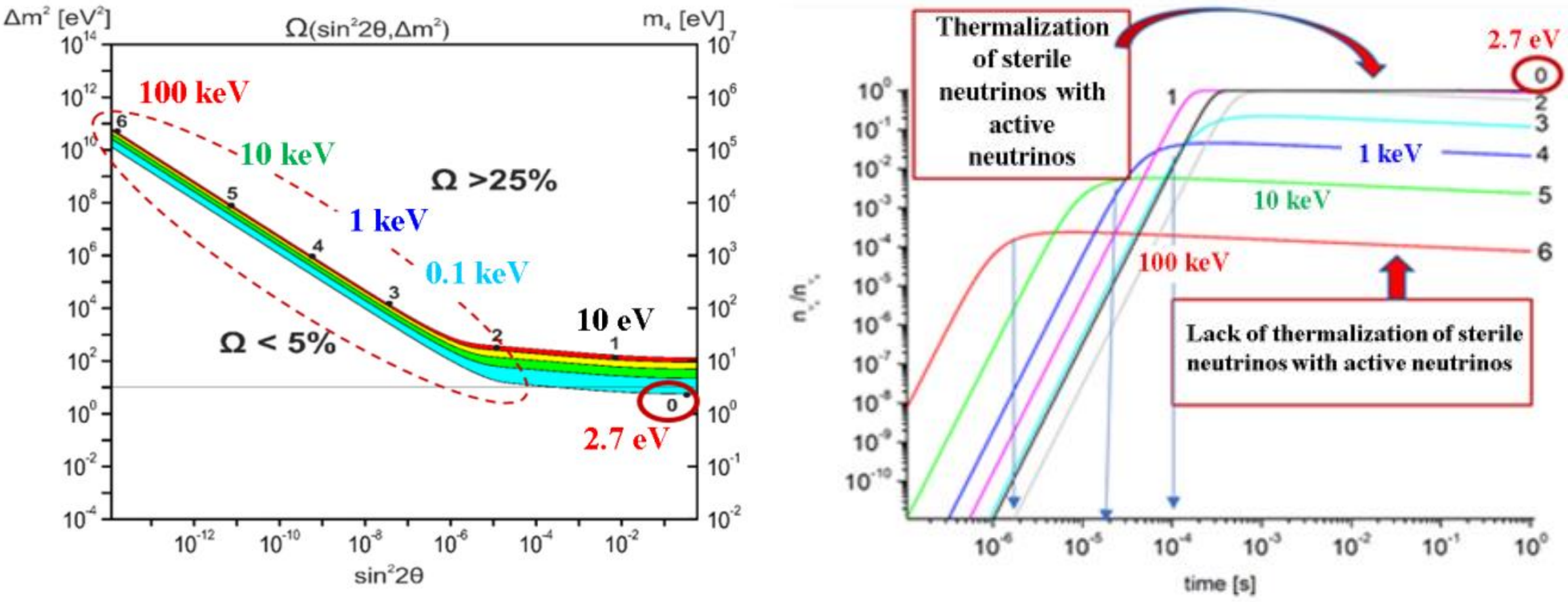

Fig. 8.3. a) The region $\Delta m^2_{14}$ and $\sin^2 2\theta_{14}$ leading to acceptable values of the contribution to dark matter. b) Dynamics of generation of dark matter consisting of right neutrinos. The ratio of the number of sterile neutrinos to the number of electron neutrinos for several values of the parameters on the plane ( $\sin^2 2\theta_{14}$ , $\Delta m^2_{14}$). Point "0" corresponds to $\Delta m^2 = 7.3$ eV$^2$, $\sin^2 2\theta = 0.36$.

The following conclusions can also be drawn from this analysis.

1. A sterile neutrino with parameters $\Delta m^2_{14} = 7.3$ eV$^2$, $\sin^2 2\theta_{14} = 0.36$ contributes approximately 5% to dark matter, but is relativistic and does not explain the structure of the Universe.
2. To explain the structure of the Universe, heavy sterile neutrinos with very small mixing angles are needed.
3. Expanding the neutrino model by introducing two more heavy sterile neutrinos will allow us to explain the structure of the Universe and bring the contribution of sterile neutrinos to the dark matter of the Universe to a level of 27%.

The next important issue to be discussed is the lifetime of sterile neutrinos. It was shown in [56–59] that right neutrinos can decay via two-particle and three-particle channels. Figure 8.4b) shows the decay time as a function of the neutrino mass. The vertical axis shows the ratio of the lifetime to the age of the Universe—13.8 billion years. The decay time of sterile neutrinos with a mass of 1 MeV is equal to the age of the Universe. Thus, right neutrinos with a mass of 1 MeV or greater are no longer suitable for dark matter.

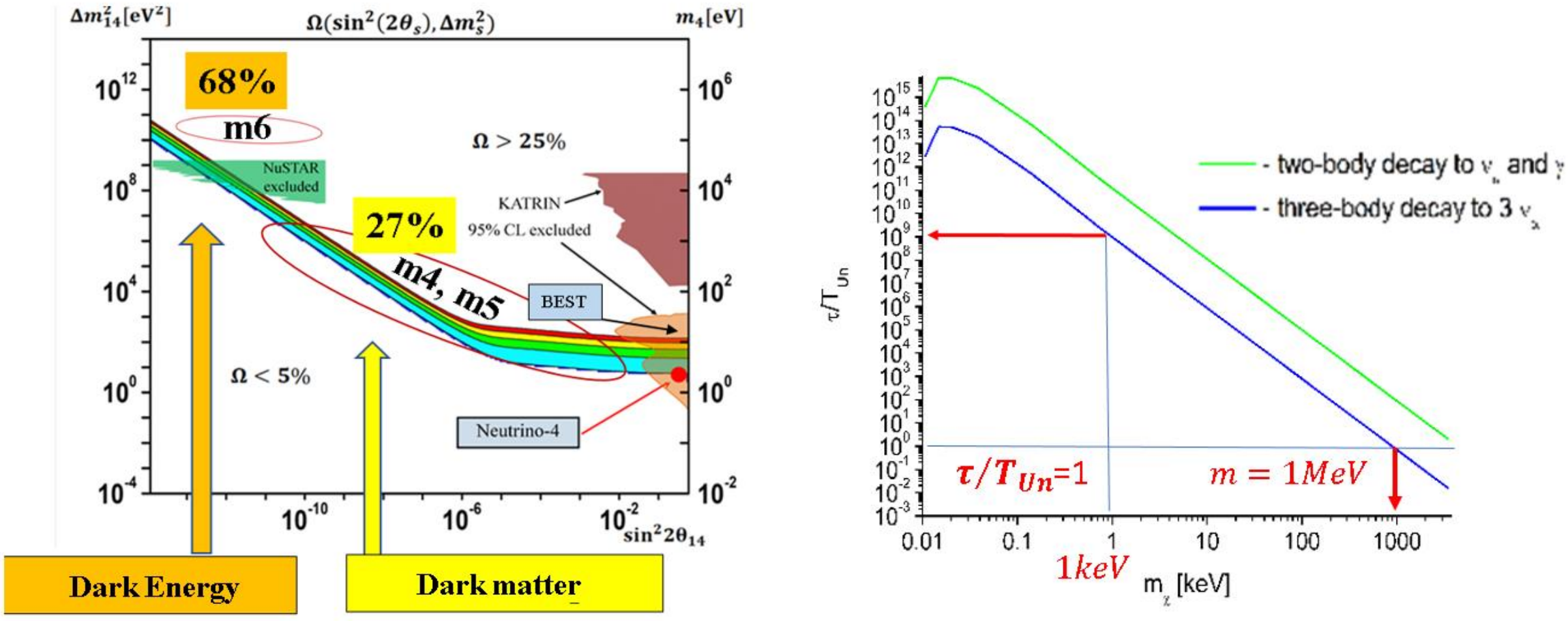

Fig. 8.4. a) Laboratory and astrophysical constraints on the parameters of sterile neutrinos. Red dots – result of Neutrino-4 experiment and possible mass of heavy right neutrinos; green area – constraints of NuSTAR experiment [62]; orange area – KATRIN, excluded with 95% confidence interval – constraints of KATRIN experiment for sterile neutrinos [51]; red area – 95% confidence constraints of experiments on measuring the mass of electron neutrinos from [51]; b) Lifetime of a heavy neutrino as a function of its mass. Lifetime is normalized to the lifetime of the Universe. Decay time of right neutrinos in the channel of two-particle (green line) and three-particle (blue line) decay. Lifetime is normalized to the age of the Universe.

Three-particle decay is the primary decay mode, and the probability of two-particle decay is two orders of magnitude lower. However, two-particle decay is significant because it produces monochromatic gamma rays with energies equal to half the mass of a heavy neutrino $E_{\gamma} = m_{\nu_R} / 2$. Since the mass of an active neutrino is extremely small, the energies of a gamma ray and an active neutrino are virtually equal. The presence of such monochromatic radiation would confirm the existence of heavy neutrinos with corresponding masses. It is precisely because of their absence that the experimental limits end at a mass of 6 keV. Thus, dark matter with right neutrino masses $m_{\nu_R}$ < 6 keV is quite stable, since the decay time is seven orders of magnitude longer than the lifetime of the Universe.
In the region of small sterile neutrino masses, there are laboratory constraints from the KATRIN experiment [60,61]. They reject the results of the Neutrino-4 experiment and the BEST experiment [49]. Thus, the region of sterile neutrino masses in the keV range is most suitable for the formation of so-called warm dark matter.

In the left-right asymmetry model, there are three types of sterile neutrinos: electron, muon, and tau sterile neutrinos, corresponding to the three types of leptons. Sterile neutrinos of different types arise from the decays of neutral mesons: K -mesons, B -mesons, D -mesons, and apparently tau mesons, which are the heaviest and shortest-lived. It should be expected that the mass hierarchy of sterile neutrinos will also be arranged in ascending order, like the mass hierarchy of leptons and mesons, as well as the ascending mass hierarchy of quarks. Moreover, the mass states $m_4$, $m_5$, $m_6$ (electron, muon and tau) have weak interactions only due to the small contribution of electron, muon and tau flavors.

The heaviest sterile neutrinos are the first to separate from the plasma and have the smallest mixing angle with left active neutrinos. The lighter sterile neutrinos separate from the plasma later, at a larger mixing angle, which is also small, so thermalization does not occur. The separated heavy sterile neutrinos already begin to form gravitational regions, since at times of $10^{-6}$ – $10^{-5}$ s, the density of dark matter particles is sufficiently high, and the law of gravitational attraction $1/r^2$ operates quite effectively. Thus, dark matter forms, creating gravitational wells for the subsequent concentration of baryonic matter and the formation of galaxies.
Extension of the neutrino model by introducing two sufficiently heavy sterile neutrino mass states $m_4$, $m_5$(electron, muon) allows us to explain the structure of the Universe and bring the contribution of sterile neutrinos to the dark matter of the Universe to the level of 27%. Therefore, in Figure 8.4, the red ellipse indicates the mass region $m_4$, $m_5$ and mixing angles at which dark matter is formed, accounting for 27% of the energy density of the Universe. Figure 8.5 shows a schematic diagram of the dynamic process of generating dark energy and dark matter based on three sterile neutrinos. $m_4$, $m_5$, $m_6$(electron, muon and tau).

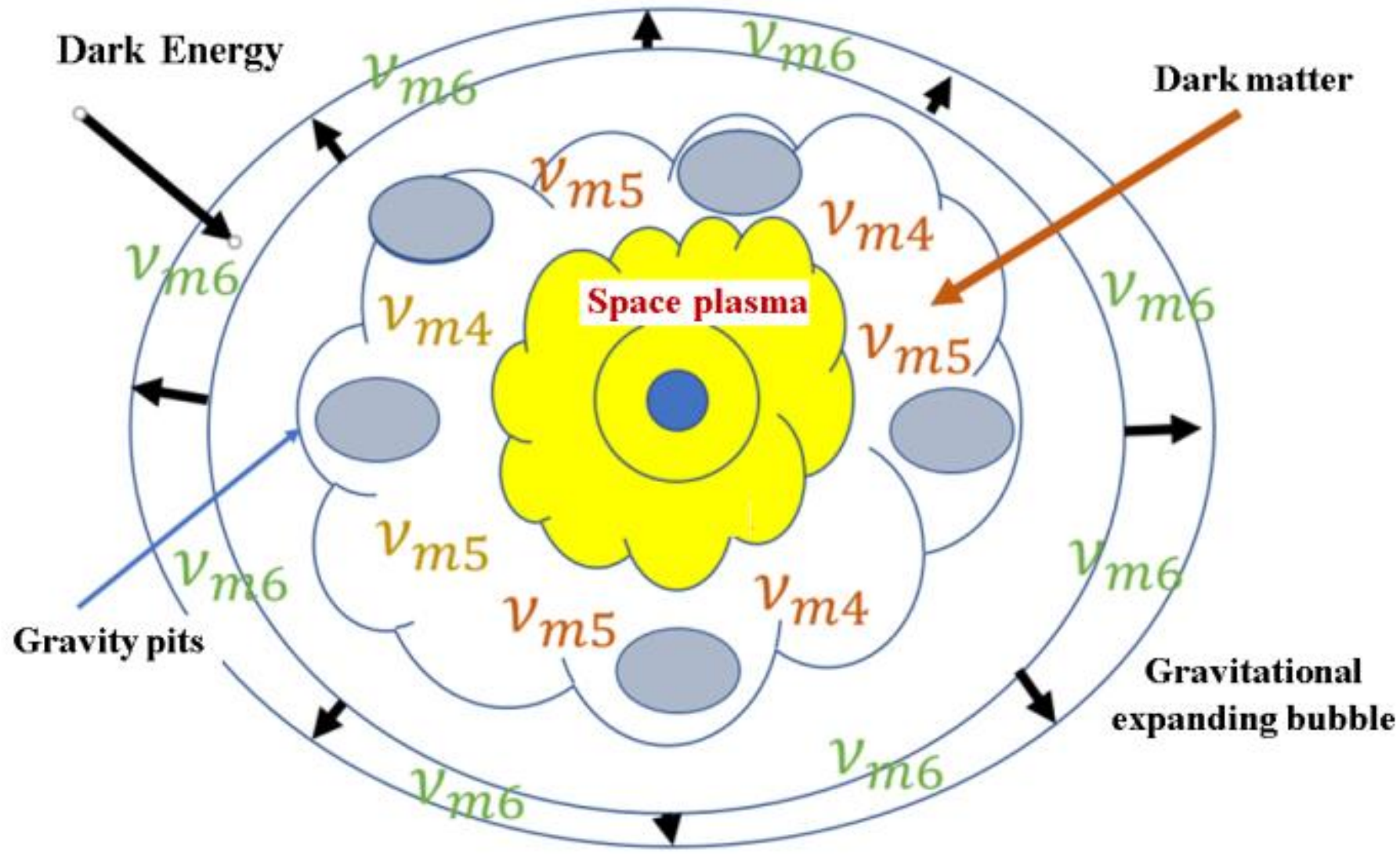

Fig. 8.5. A schematic and illustrative diagram of the dynamic process of generation of dark energy and dark matter based on three sterile neutrinos $m_4$, $m_5$, $m_6$(electron, muon and tau).

Clearly, it's tempting to interpret 68% of the universe's dark energy density as the heaviest sterile neutrinos $m_6$(tau), which are the first to separate from the plasma with the smallest mixing angle and form an expanding gravitational bubble. Essentially, the expansion of the gravitational bubble caused the gravitational stretching of space. However, this issue requires a separate discussion and is beyond the scope of this article.

## 9. Generation of lepton asymmetry of the Universe

Lepton asymmetry is generated during the decay of neutral mesons, also during the hadronization of quark-gluon plasma. This may occur somewhat ahead of nucleon formation, as neutral mesons consist of two quarks. However, the decay rate of neutral mesons is quite high compared to the expansion rate of space, so thermodynamic equilibrium is not violated in the same way as with neutrons. However, a deviation from thermodynamic equilibrium occurs due to the escape of sterile neutrinos from the cosmic plasma. This process leads to nonequilibrium conditions during the cooling of the universe, which prevents the reverse reaction and perpetuates the asymmetry.

Thus, the basic idea behind the mechanism for the emergence of lepton asymmetry is that the lepton asymmetry that existed at that point in time is frozen due to the escape of sterile neutrinos from the cosmic plasma. Moreover, the lepton asymmetry is frozen to the extent that the number of sterile neutrinos that escaped from the plasma is proportional to the number of active neutrinos that continue to participate in the thermodynamic process. Thus, the lepton asymmetry is equal to:

$L_\nu = \frac{n_\nu - n_{\bar{\nu}}}{n_\nu + n_{\bar{\nu}}} = \frac{n_{\nu_S}}{n_\nu} \times A^{LRAM}$ where $n_{\nu_s}$ is the density of sterile neutrinos that have left the plasma, (9.1),

$n_\nu$ - the density of active neutrinos participating in the thermodynamic process, $A_{LR}$ - the asymmetry of CP violation in the decays of neutral mesons equal to $A^{LRAM} \approx -\left(1-2\delta^2\right)\times 2\zeta = (5.4 \pm 1.8)\times 10^{-3}$.

Using the values of $\delta = (5.5 \pm 1.8)\times 10^{-2}$ and $\zeta = -(2.7 \pm 0.9)\cdot 10^{-3}$ obtained within the left-right model from neutron decay, we have: $A^{LRAM} \approx -\left(1-2\delta^2\right)\times 2\zeta = (5.4 \pm 1.8)\times 10^{-3}$. For the ratio, $n_{\nu_s} / n_\nu$ we must use the result of calculations from the balance equation (8.1), which takes into account the generation of sterile neutrinos and their "sink".

As can be seen from Fig. 8. 3b) the ratio $n_{\nu_s} / n_\nu$ depends on the mass of the sterile neutrino, which is unknown. Therefore, we have a problem of choice, which is easier to solve in the presence of different possibilities. The hierarchy of sterile neutrino masses is unknown, and we must consider different options. In the case of light sterile neutrino masses in the eV range, the ratio $n_{\nu_s} / n_\nu$ is almost unity. But this does not mean that we should use this value in formula (9.1). Quite the contrary, this means that Boltzmann equilibrium will arise and the original CP asymmetry will be compensated. Such light sterile neutrinos do not create lepton asymmetry, and they, as already noted, do not explain the structure of the Universe. To explain the structure of the Universe due to dark matter, we need to move to the keV mass range. These can be masses $m_4$ and $m_5$(electron and muon sterile neutrinos), or, in fact, only $m_5$.

For the keV range of sterile neutrino masses, the ratio $n_{\nu_s} / n_\nu$ is $10^{-2}$, and for a 100 keV sterile neutrino mass, it is $10^{-4}$. Thus, it turns out that the lepton asymmetry depends on the type of sterile neutrino. However, it should be noted that equation (8.1) was written without taking into account the mixing of sterile neutrinos, so this splitting is the effect of an approximate consideration of the independent generation of sterile neutrinos. Sterile neutrinos, like active neutrinos, apparently mix, so a unified ratio apparently exists, but for now we can propose a lepton asymmetry in the range of two or even three orders of magnitude, if we take into account the sterile tau neutrino:

$L_\nu = \frac{n_{\nu_S}}{n_\nu} \times A^{LRAM} = (5.4 \pm 1.8)\times(10^{-5} \div 10^{-8})$ (9.2),

In conclusion of this part of the analysis, we recall that the indicated range of values is the result of an approximate consideration of the independent generation of sterile neutrinos without mixing them with each other.

However, we have not yet taken into account the decay of charged pi -mesons ( $\pi^\pm$), which produce neutrinos and antineutrinos. This is the most common decay type. The main channel is $\pi^+ \to \mu^+ + \nu_\mu$, $\pi^- \to \mu^- + \nu_\mu$. When $\pi^+$ decays, it produces a positive muon ( $\mu^+$) and a muon neutrino ( $\nu_\mu$), and $\pi^-$ decays into a negative muon ( $\mu^-$) and a muon antineutrino ( $\bar{\nu}_\mu$). In these decays, due to CPT conservation, there is no asymmetry in the production of neutrinos and antineutrinos, since this is the decay of charged particles.

Thus, to obtain the correct lepton asymmetry, we must introduce a correction that is related to $\pi^+\pi^-$ and $\mu^+\mu^-$, as well as to the charged K-mesons, D -mesons, and B -mesons. In general, this correction factor is at least a factor of 5. We can rewrite the lepton asymmetry estimate (9.2) as follows:

$L_\nu^{cor} = \frac{n_{\nu_S}}{n_\nu^{cor}} \times A_{LR} = (1.08 \pm 0.36)\times(10^{-5} \div 10^{-8})$ (9.3),

Now it is necessary to make some refinements to the estimation of the seed CP-violating asymmetry $A_{LR}$ in a space plasma environment. The fact is that, in such an environment, oscillations are suppressed, yet the process of CP violation continues to operate effectively. Most importantly, the $B_S^0$ meson ceases to oscillate and, consequently, stops compensating for the CP-violation process. The same applies to the $B^0$ and $K^0$ mesons; only for the $D^0$ meson is this space-plasma effect less significant.

To do this, we'll examine the manifestation of CP-violation effects in cosmic plasma and consider the role of the weak interaction potential in the plasma. To do this, we need to use the following matrix:

$$\begin{pmatrix} M+\Delta U-i\dfrac{\Gamma-\delta\Gamma}{2} & \Delta m-i\dfrac{\Delta\Gamma}{2} \\ \Delta m-i\dfrac{\Delta\Gamma}{2} & M-\Delta U-i\dfrac{\Gamma+\delta\Gamma}{2} \end{pmatrix} \quad (9.4)$$

Indeed, in the early Universe, there was no asymmetry in the number of particles and antiparticles, and the cause of this asymmetry is unknown. However, in the left-right CP-violated model, an asymmetry arises in the interaction potential for particles and antiparticles, which was previously estimated using the model parameters $\delta$ and $\varsigma$ by the following relation $A^{LRAM} \approx -\left(1-2\delta^2\right)\times 2\zeta = (5.4\pm 1.8)\times 10^{-3}$.

The weak interaction potential is discussed in detail in [63] and was also used in [54] and can be represented by the following equation:

$$U=\eta\frac{11\zeta(3)}{\pi^2\sqrt{2}}G_F T^3-\frac{14}{45}\frac{\pi\left(3-\sin^2\theta_W\right)\sin^2\theta_W}{\alpha}G_F^2 T^4 E \qquad U=\eta\times 1.1\times 10^{-23}\left[\frac{1}{\mathrm{eV}^2}\right]T^3-1.1\times 10^{-44}\left[\frac{1}{\mathrm{eV}^4}\right]T^4 E \quad (9.5)$$

or in numerical terms: where $G_F$ is the Fermi constant, $T$ is the plasma temperature, $E$ is the particle energy (neutrino, antineutrino)

$\eta=\dfrac{N_f-N_{\bar f}}{N_f+N_{\bar f}}$ - lepton asymmetry, $\dfrac{11\zeta(3)}{\pi^2\sqrt{2}}T^3$ - density of cosmic plasma.

We will assume that in the early Universe, the lepton asymmetry $\eta$ was zero. But neutral mesons, being in the particle-antiparticle state, will have different interaction potentials with the medium even in the case where there is no asymmetry in the number of particles and antiparticles in the medium, i.e. $\eta=0$. Then the potential difference for neutral mesons and antimesons will be equal to: $\Delta U\approx \pm A^{LR}\dfrac{11\zeta(3)}{\pi^2\sqrt{2}}G_F T^3$ or, in numerical terms, $\Delta U\approx \pm 6\cdot 10^{-23}\left[\mathrm{eV}^2\right]\cdot T^3\left[\mathrm{eV}^3\right]$. Using this dependence of the potential difference on the temperature of the space plasma, it is possible to calculate the influence of the potential difference on the process of oscillation suppression, which is shown in Fig. 9.1. As can be seen, with an increase in the distance between the levels for particles and antiparticles, the oscillation process is suppressed.

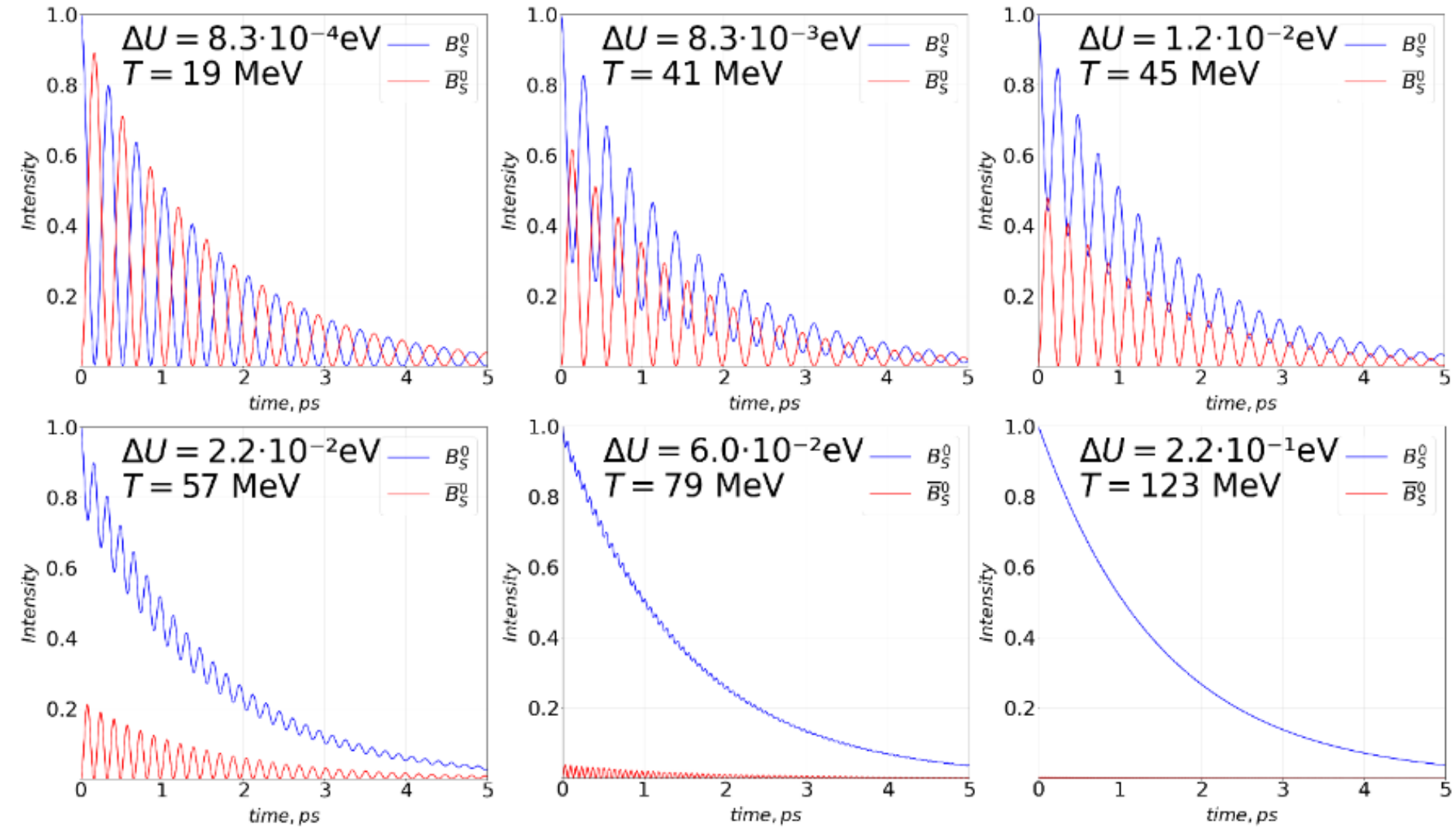

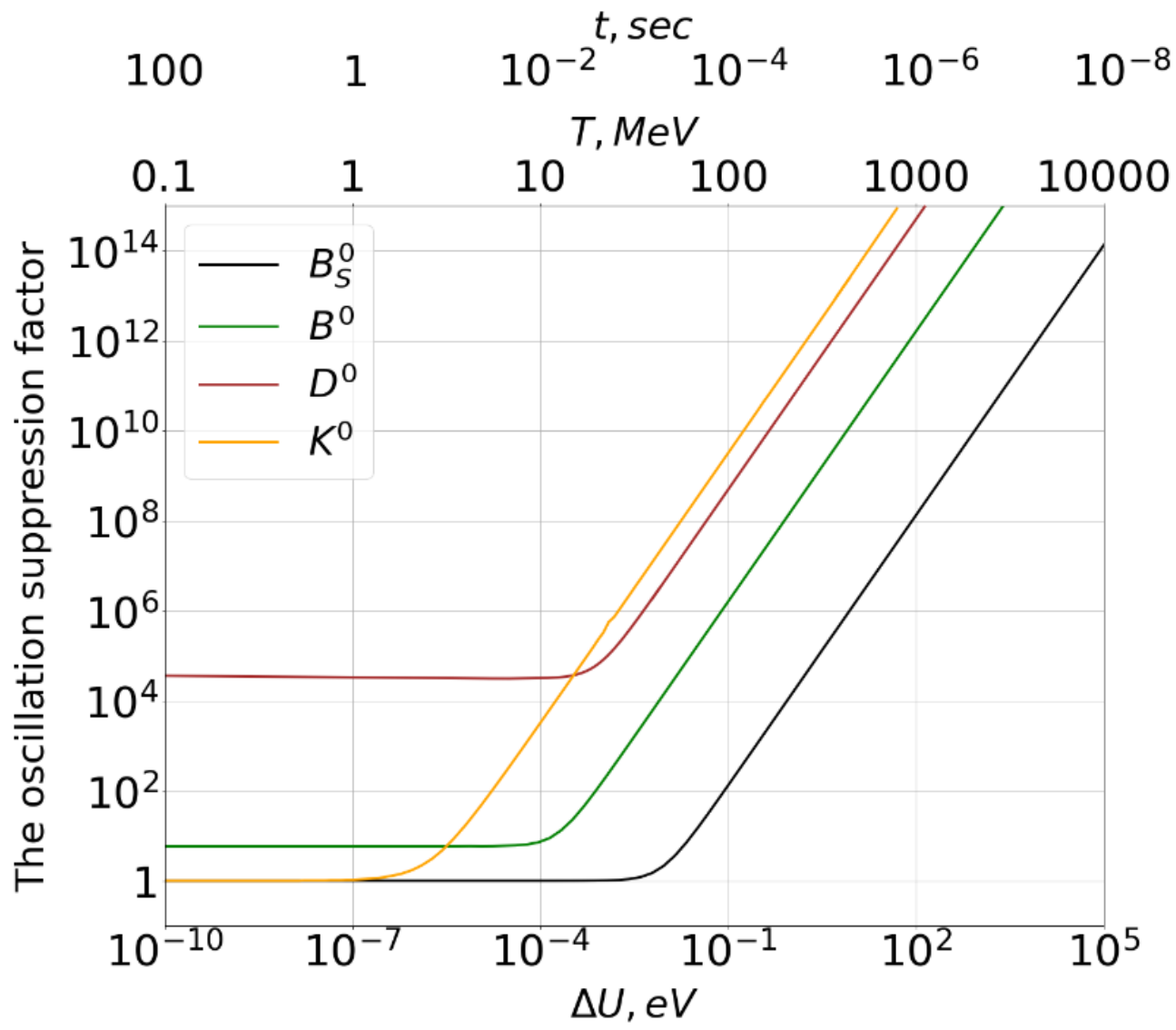

Fig. 9.1. a) The process of suppression of oscillations with an increase in the distance between the levels for particles and antiparticles $\Delta U$ , which depends on the plasma density, which depends on the plasma temperature. b) The oscillation suppression factor for $B_S^0$ -meson, $B^0$-meson, $D^0$-meson, $K^0$-meson is presented.

As shown by the calculations in Fig. 9.1b), at cosmic plasma temperatures above 100 MeV, the process of neutral meson oscillations is suppressed by the potential difference of the weak interaction. At the same time, the process of their mutual annihilation is also suppressed; it is at this stage that lepton asymmetry arises, since the predominant decay of antimesons occurs due to the CP violation process. Thus, the CP asymmetry for $B^0$ -meson and $B_S^0$ − meson is again $A^{LR} \approx -\left(1-2\delta^2\right)\times 2\zeta = (5.4\pm1.8)\times10^{-3}$ , since oscillations are suppressed during its formation upon the escape of sterile neutrinos from the plasma. Thus, a single estimate of the CP asymmetry can be used for all neutral mesons $A^{LRAM} \approx -\left(1-2\delta^2\right)\times 2\zeta = (5.4\pm1.8)\times10^{-3}$ . Therefore, the previously made estimate of the lepton asymmetry (9.3) remains valid, that is, the lepton asymmetry is: $L_\nu^{cor} = \frac{n_{\nu_S}}{n_\nu^{cor}}\times A^{LRAM} = (1.08\pm0.36)\times(10^{-5}\div10^{-8})$ .

## 10. Conclusion

In conclusion, we should summarize the results of a detailed analysis of experimental data on neutron decay, proton decay in the nucleus (the so-called $0^+ - 0^+$ transitions) and experimental data on oscillations of neutral mesons with CP-violation effects within the framework of the left-right asymmetry model of weak interaction.

This model assumes the presence of an an admixture of the right vector boson $W_R$ with the left vector boson $W_L$, where the sign of the mixing angle $W_R$ and $W_L$ reverses upon transition from $W^-$ to $W^+$, which leads to CP violation. Thus, this model inherently assumes 100% CP violation, since the sign of CP violation reverses upon transition from particles to antiparticles, or, in other words, the CP violation phase is $\pm\pi/2$ . Particle decay occurs through $W^-$, and antiparticle decay occurs through $W^+$. Thus, the lifetimes of neutral particles and antiparticles may differ. For example, the lifetimes of the neutron and antineutron may differ. And the oscillation processes of neutral mesons, such as $K^0 - \bar{K}^0$, $D^0 - \bar{D}^0$, $B^0 - \bar{B}^0$, $B_S^0 - \bar{B}_S^0$ will be accompanied by processes of 100% CP violation, i.e., a change in the sign of CP-violation upon transition from a meson to an antimeson. However, it is generally considered that a reconfiguration of quark flavor states occurs during oscillations, and that the process of CP violation arises precisely from quark mixing. However, in the left-right asymmetry model, CP-violation arises due to the presence of a right vector boson admixture, with the sign of CP violation being opposite for particles and antiparticles. Thus, the nature of CP-violation is associated with vector bosons. For this reason, in this paper, the analysis of experimental data on neutral meson oscillations was performed with a mixing matrix in which the effects of flavor oscillations are included in the off-diagonal matrix elements, and the effects of CP-violation are included in the diagonal matrix elements through a parameter $\delta\Gamma$ with opposite signs for mesons and antimesons.

Within the Standard Model, it is assumed that CP-violation is hidden in quark mixing via a single left quark $W_L$, and that introducing a complex mixing phase can solve all problems. However, if the nature of CP violation lies in the mixing of $W_R$ and $W_L$, then the mixing matrix proposed in this paper must be used. This was done in Section 3, and it was shown that the sign of the CP violation effect does indeed reverse during oscillations. This occurs especially clearly for $B^0 - \bar{B}^0$, $B_s^0 - \bar{B}_s^0$ and only for $D^0 - \bar{D}^0$ it is impossible to observe a second oscillation phase with the opposite sign of CP violation, because $D^0(\bar{D}^0)$ decays before it oscillates. Thus, it has been experimentally demonstrated that the left-right asymmetry model works successfully because it correctly guessed that the nature of CP violation lies in the mixing of vector bosons, not in the quark sector.

The CKM matrix determines the quark mixing angles in the weak interaction via vector bosons, but CP violation in the weak interaction is determined by $W_R$ and $W_L$ boson mixing. Currently, the CP phase in the SM is $\delta_{13}$ = 1.20 ± 0.08 radians = 68.8° ± 4.5°, which differs from $\pi/2$ by 4.7 $\sigma$. This raises the question of whether it is possible to extend the SM to account for mixing $W_R$ and $W_L$ with a CP phase $\pi/2$. This would require adjusting all elements of the CKM matrix. In the left-right asymmetry model, new parameters $\delta$ and $\zeta$ appear, and the "area" of the unitarity triangles can change due to new contributions. Ultimately, the maximum value of the Jarlskog invariant will be obtained, corresponding to maximum CP violation.

An experimental indication of the need to use the left-right model was the 3.7$\sigma$ discrepancy in neutron asymmetries. However, the need to modify the left-right manifest model to a left-right asymmetry model is due to the discrepancy between the values of $V_{ud}^{00}$ and $V_{ud}^{unt}$. Incidentally, this discrepancy between the values $V_{ud}^{00}$ and $V_{ud}^{unt}$ is an experimental indication of CP violation in baryons. It is CP violation in baryons that is most clearly manifested within the left-right asymmetry model for neutrons and antineutrons and leads to the possibility of explaining the baryon asymmetry of the Universe.

Section 2 analyzes the most recent, most accurate experimental data on neutron decay, which points to the need to extend the Standard Model by introducing a right vector boson admixture $W_R$. A left-right asymmetry model of the weak interaction is presented, in which the sign of the mixing angle $W_R$ and $W_L$ is reversed upon transition from $W^-$ to $W^+$. In this model, CP violation occurs due to the presence of a right vector boson admixture. The model parameters are the mixing angle $\zeta = -(2.7 \pm 0.9) \cdot 10^{-3}$ and the ratio of the squares of the masses $W_1$ and $W_2$ $\delta = (5.5 \pm 1.8) \times 10^{-2}$ extracted from neutron decay – lifetime and decay asymmetries, as well as from the study of proton decay in nuclei, the so-called superallowed $0^+ - 0^+$ transitions.

Section 5 examines the formation of the baryon asymmetry of the Universe in the left-right asymmetry model of weak interaction. The presence of an admixture of a right vector boson with a different sign of the mixing angle for $W^-$ and $W^+$ leads to a difference in the lifetimes of neutrons and antineutrons, which decay via $W^-$ and $W^+$. This difference gives rise to baryon asymmetry during the hadronization of quark-gluon plasma at temperatures below 150 MeV. During the phase transition from quark-gluon plasma to hadronic liquid, all three of A.D. Sakharov's conditions for the emergence of baryon asymmetry in the Universe are met: CP violation and non-stationarity of the process. As a result, baryon number violation occurs due to the difference in the decay probabilities of neutrons and antineutrons.

Section 7 examines the formation of the lepton asymmetry of the Universe in the left-right model. This process is associated with the presence of sterile (right) neutrinos, which do not thermalize and leave the cosmic plasma, carrying with them a lepton asymmetry with a sign opposite to the baryon asymmetry. Overall, a baryon-lepton asymmetry arises with the conservation of the difference between the baryon and lepton numbers, B - L . The mechanism for the formation of dark matter by sterile neutrinos and the mechanism for generating lepton asymmetry are presented.

Finally, it should be noted that the left-right asymmetry model of the weak interaction with CP -violation presented here requires a significant increase in experimental accuracy. The possibility of further improving the accuracy of neutron decay measurements exists. This goal, for example, is pursued by the PNPI NRC KI project "Neutron Beta Decay" for the PIK reactor [64-66], which plans to use a long-baseline superconducting solenoid for neutron decay to increase the statistics of decay events, and a magnetic mirror collimator to isolate the direction of electron emission. The magnetic field in the region of the uniform magnetic field is 0.34 T, and in the mirror region it is 0.86 T. This is a development of the 1998 PNPI RAS experiment [67] and is planned to achieve a relative measurement accuracy of $10^{-3}$ for neutrino and electron decay asymmetries, and, most importantly, the electron-neutrino asymmetry and neutrino asymmetry (a-small and B ), for which the largest discrepancy of 3.7 σ is observed . It is estimated that it will be possible to increase the accuracy by a factor of three and increase the reliability to 5σ and better [68]. The experimental scheme is presented below.

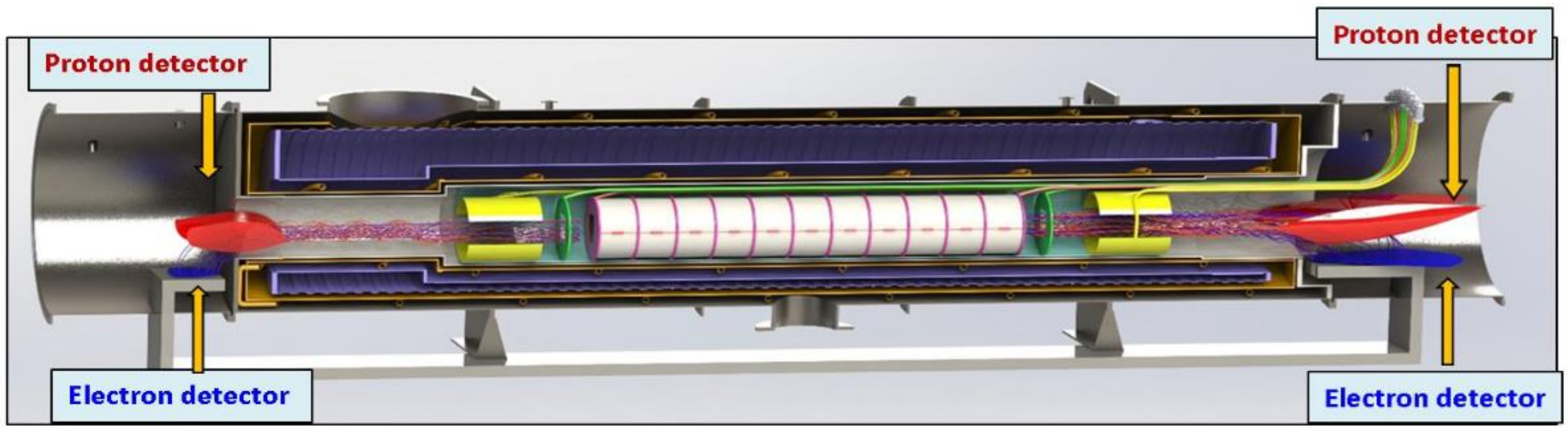

Fig. 10.1 . Detailed diagram of the setup for measuring neutron decay asymmetry. Electron trajectories are shown in blue, proton trajectories in red. The white cylinder is at a potential of +30 kV. The yellow half-cylinders are the plates of high-voltage capacitors with a potential of 20 kV.

It is also necessary to increase the accuracy of the Neutrino-4 experiment, which searches for light sterile neutrinos. To this end, a new, more sensitive setup has been created to continue searching for light sterile neutrinos. Therefore, conducting two experiments—a neutron experiment at the PIK reactor and a neutrino experiment at the SM-3 reactor—is fundamentally important for clarifying the cause of the baryon and lepton asymmetry of the Universe [69].

In conclusion, it should be noted that there is nothing surprising in the development of theoretical propositions in connection with the development of experiments. It is enough to recall the history of the revision of theoretical propositions regarding C, P and T invariants.

The first to be violated was the principle of spatial inversion P (Madame Wu's experiment was performed in 1956). But at that time, it was said that combined parity was conserved (Lev Landau, 1957). However, this principle was later refuted experimentally in 1964 by Cronin and Fitch in kaon decay. It was then said that if CP is violated, then T-invariance is also violated, because CPT invariance cannot be violated under any circumstances due to Lorentz invariance.

As we see, theoretical assumptions are being revised as the experiment progresses. We are currently at the next stage of experimental development, which requires a revision of theoretical assumptions. Currently, it is generally accepted that C-invariance is preserved, while CP and T-invariance are violated. However, an analysis of experimental data shows the opposite situation: C-invariance and CP-invariance are violated, while T-invariance is preserved.

This is precisely the focus of our article, which is essentially based on new experimental data. A new theoretical generalization of this situation is the task of theorists. However, it should be noted that the asymmetric left-right model works quite well. Our article is extremely important for particle physics and cosmology. It marks a turning point in understanding the nature of CP violation and the origin of the baryon asymmetry of the Universe.

**Acknowledgments**

The authors express their gratitude to the staff of the High Energy Physics Department of PNPI for discussing the work at the HEPD seminars of the NRC KI - PNPI.

**Financing**

This work was supported by the Russian Science Foundation (Project No. 24-12-00091 https://rscf.ru/project/24-12-00091/ ).

**Conflict of interest**

The authors declare no conflict of interest.